\documentclass[12pt]{article}
\usepackage[utf8]{inputenc}
\usepackage[svgnames,dvipsnames]{xcolor}
\usepackage[a4paper,left=1.5cm,right=1.5cm,top=1.5cm,bottom=1.5cm]{geometry}
\usepackage[bottom, multiple]{footmisc}
\usepackage{booktabs, multirow, nicefrac, float, xspace}
\usepackage{graphicx}
\usepackage{tikz}
\usetikzlibrary{decorations.markings}
\usetikzlibrary{decorations.pathmorphing}
\usepackage{framed}
\usepackage{csquotes}
\definecolor{shadecolor}{rgb}{0.90,0.90,0.90}
\usepackage{cite}
\usepackage{hyperref}
\usepackage{subcaption}
\usepackage{pifont}
\usepackage{mathtools}
\usepackage{mathtools}

\usepackage{setspace}
\usepackage{amsmath, amssymb, amsthm, float, graphicx,amsfonts,braket}
\numberwithin{equation}{section}
\usepackage{amsthm}

\theoremstyle{definition}

\usepackage{float}
\usepackage{multirow}
\usepackage{multicol}
\allowdisplaybreaks

\hypersetup{colorlinks=true, linkcolor=Maroon, citecolor=FireBrick,urlcolor=Green,linktocpage}
\definecolor{amber(sae/ece)}{rgb}{1.0, 0.49, 0.0}
\definecolor{deepsaffron}{rgb}{1.0, 0.6, 0.2}

\usepackage{multirow}

\usepackage{amstext} 
\usepackage{array}   
\newcolumntype{L}{>{$}l<{$}}
\newcolumntype{C}{>{$}c<{$}}

\def\beq{\begin{eqnarray}}\def\eeq{\end{eqnarray}}
\def\be{\begin{equation}}\def\ee{\end{equation}}

\def\s{\sigma}

\def\a{\alpha}

\def\D{\Delta}
\def\G{\Gamma}
\def\l{\lambda}

\def\G{\Gamma}

\usepackage{bm}
\usepackage{bm}

\begin{document}

\title{\bf Learning the open string spectrum \\ with neural networks}
\author{Debapriyo Chowdhury$^{a}$\footnote{debapriyoc@iisc.ac.in}\\
\it ${^a}$Centre for High Energy Physics,
\it Indian Institute of Science,\\ \it C.V. Raman Avenue, Bangalore 560012, India.}

\maketitle

\abstract{The tree-level open superstring has an integer mass-squared spectrum on linear Regge trajectories. To what extent is this fixed by consistency conditions: locality, analyticity, crossing, unitarity together with the requirement that $2\to2$ scattering minimize the entanglement it generates? Using the local two-channel crossing symmetric dispersion relation with one-parameter ambiguity ($\lambda$-CSDR) we set up a bootstrap treating both the mass-squared spectrum and the partial-wave coefficients as unknowns which lets the machine learning algorithm to learn the mass-squared spectrum and partial-wave coefficients simultaneously, going beyond earlier work in which the spectrum was fixed. Using the local three-channel crossing symmetric dispersion relation with one-parameter ambiguity we obtain, within the class of amplitudes with soft Regge behaviour and infinite higher-spin tower, a bound on the leading Wilson coefficients that we compare with existing bounds and trace to spin-0 dominance. We locate the Virasoro-Shapiro amplitude in the allowed region of the Wilson coefficients space. For amplitudes with two-channel crossing symmetry we find that a Non-dual amplitude exhibits a Polchinski-Strassler type transition from Regge to power-law behavior.} 

\newpage
\tableofcontents
\onehalfspacing

\section{Introduction}
\begin{figure}[H]
    \centering
    \includegraphics[width=0.5\linewidth]{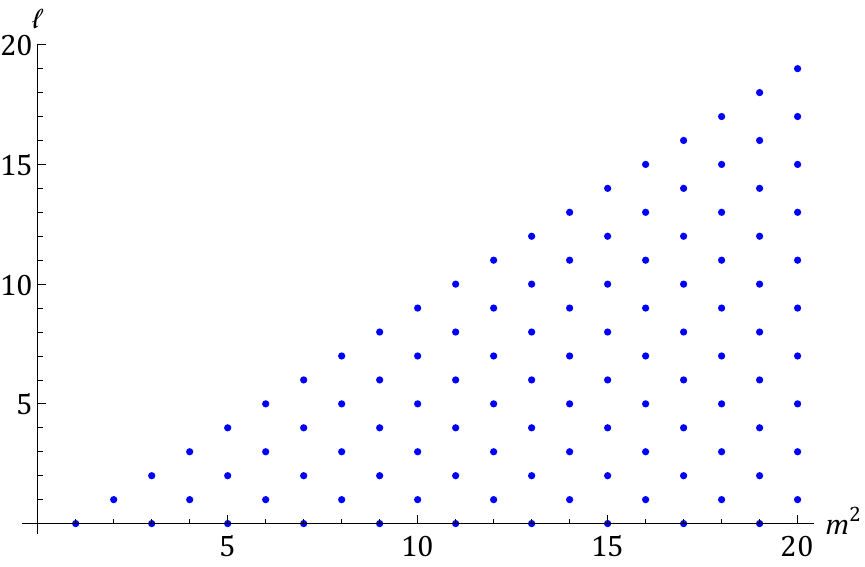}
    \caption{Open String Spectrum.}
    \label{fig:Open String Spectrum}
\end{figure}

In this work, we study the properties of the mass-squared spectrum of the open superstring amplitude using the fundamental physical principles of locality, analyticity, unitarity, and crossing symmetry. The four-point color-ordered scattering amplitude of gluons in Type-I open string theory at tree-level is given by 
\begin{equation}
    \mathcal{M}_{OS}(s,t)=g^2\mathcal{F}^4\frac{\Gamma(-s)\Gamma(-t)}{\Gamma(1-s-t)}
\end{equation}
Here, $g^2$ is the coupling and $\mathcal{F}^4$ is a form factor consisting of polarizations and momenta of the external particles coming from Type-I supersymmetry. The resonance mass-squared spectrum of the open string amplitude is integer valued due to the fact that the poles of the $\Gamma$ functions in the numerator are present at non-negative integer values of $s$ and $t$ that is for $s=n$ and $t=n$ with $n\in\mathbb{Z}_{\geq0}$. The residues of the open string amplitude at positive integer values of the variable $s$ that is at $s=n$ with $n\in\mathbb{Z}_{>0}$ are polynomials in the variable $t$. The residues can be expanded in terms of Gegenbauer polynomials
\begin{equation}
\label{res}
    \text{Res}_{s=n}\mathcal{M}_{OS}(s,t)=\sum_{\substack{\ell=0 \\ n-\ell \equiv 1 \ \mathrm{mod}\ 2}}^{n-1}c_\ell^{(n)}\mathcal{G}_\ell^{\left(\frac{D-3}{2}\right)}\left(1+\frac{2t}{n}\right)\;, \qquad n\in\mathbb{Z}_{>0}\;.
\end{equation}
In other words, the partial wave coefficients $c_\ell^{(n)}$ are non-zero only when the spin $\ell$ belongs to a particular set. At the level $n$ the following $\ell$ contribute:
\begin{equation}
\label{OSspectrum}
    0\leq\ell\leq n-1 \text{ and } n-\ell\equiv 1 \text{ mod } 2\;.
\end{equation}
This particular set is called the spectrum of the open string amplitude. Therefore, the spectrum of the open string amplitude looks as shown in figure \ref{fig:Open String Spectrum}.
In figure \ref{fig:Open String Spectrum}, $m$ and $\ell$ denote the masses and the spins respectively of the particles being exchanged in scattering. It can be clearly seen that the values of $m^2$ are located only at positive integers. Hence, it is clear that the open string amplitude has a unique positive integer mass-squared spectrum. 

Regge theory is the study of the angular momentum of particles as a function of their mass-squared. It is useful in the context of pion-nucleon scattering and enables us to plot the angular momentum $J$ versus mass-squared $M^2$ plots of hadronic resonances observed in pion-nucleon scattering data. These are known as Chew-Frautschi plots. These were introduced by Geoffrey Chew and Steven Frautschi in the early 1960s \cite{Chew:1962eu}.
\begin{figure}[H]
    \centering
    \includegraphics[width=0.55\linewidth]{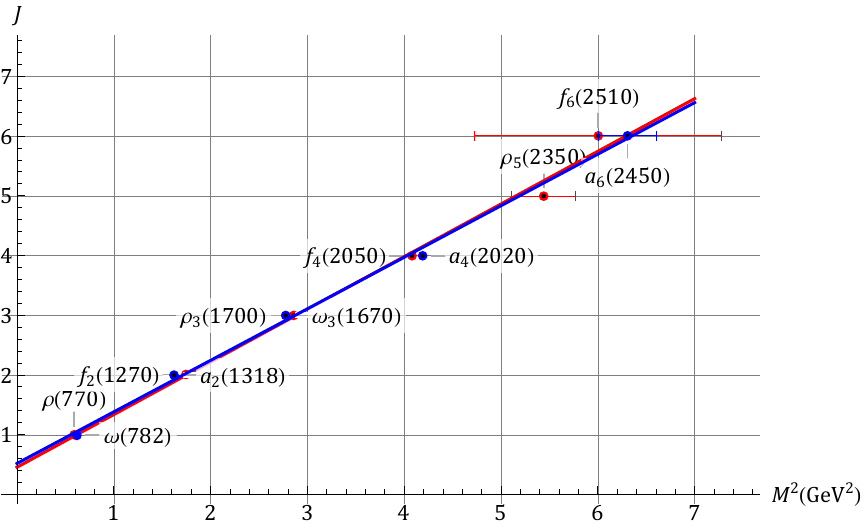}
    \caption{Chew-Frautschi plot of hadronic resonances showing their angular momentum ($J$) versus mass-squared ($M^2$). The red and blue plots denote the $\rho$ and $\omega$ family of light vector mesons. The masses of the particles have been indicated in parentheses in MeV. Data has been collected from \cite{Desgrolard:2000sf}.}
    \label{fig:Chew-Frautschi plot}
\end{figure}
In figure \ref{fig:Chew-Frautschi plot} we show a Chew-Frautschi plot for hadronic resonances with data collected from \cite{Desgrolard:2000sf}. It can be clearly seen that the resonances lie on straight lines. These striking straight trajectories are called Regge trajectories. The function behavior of these Regge trajectories can be written as
\begin{equation}
    J=\a(M^2)\simeq\a_0+\a_R'\,M^2
\end{equation}
with intercept $\alpha_0$ and Regge slope $\alpha_R'$. This led to the birth of string theory. The Veneziano amplitude (1968) was constructed to explain the linearly rising trajectories of hadronic resonances \cite{Veneziano:1968yb}.
It is also clear from figure \ref{fig:Open String Spectrum} that the spectrum of the open string amplitude consists of linear Regge trajectories. Recently a lot of progress has been made to constrain four and five-point string theory amplitudes \cite{Saha:2024qpt,Bhat:2024agd,Bhat:2025zex,Saha:2026ftv,gu1,gu2,Chiang:2023quf,Haring:2023zwu, Berman:2023jys, Berman:2024wyt, Berman:2024kdh, Berman:2024eid, Berman:2024owc, Berman:2025owb, Berman:2026ezk, Eckner:2024ggx, Eckner:2024pqt, Eckner:2025kve, Wan:2026pjq, Xu:2026kix, deRham:2026lvc}. Crossing Symmetric Dispersion Relations have emerged as powerful tools to constrain quantum field theories, gravitational effective field theories, and string theories \cite{ak, Mahoux:1974ej,Sinha:2020win,Gopakumar:2021dvg,bieb,Raman:2021pkf,Chowdhury:2021ynh,Bissi:2022mrs,Ghosh:2022net,Sinha:2022sdo, Bissi:2022fmj,az1,az2,Bhat:2023ekh,Song:2023quv,bhat,ak1,Saha:2024qpt,Bhat:2024agd,Bhat:2025zex,Bhat:2025ygi,Das:2026ckf,Alday:2022xwz,deRham:2022gfe,Li:2023qzs,Pasiecznik:2025eqc,Chang:2025cxc,EliasMiro:2025rqo}. In this paper we will look at what insights we can gather about the mass-squared spectrum of the open string amplitude using machine learning.  We will use the local two-channel crossing symmetric dispersion relation with one-parameter ambiguity ($\lambda$-CSDR) derived in \cite{Bhat:2024agd,Bhat:2025zex} for this purpose. We will also use linear programming to see how the open string mass-squared spectrum is associated with lower entanglement compared to general non-integer mass-squared spectra. For this we will use entanglement generated in $2\to2$ gluon scattering amplitudes as the physical quantity for minimization as was done in \cite{Bhat:2024agd}. Since, we use dispersive techniques to express the amplitude $\mathcal{M}(s_1,s_2)$ in terms of an integral over the $s_1$-channel imaginary part of the amplitude $\mathcal{A}^{(s_1)}(s_1,s_2)$, as a result the amplitude has to satisfy a few physical assumptions. We list the following set of physical assumptions for the amplitudes:
\begin{enumerate}
    \item Analyticity: the amplitude $\mathcal{M}(s_1,s_2)$ is analytic for all values of $s_1$ and $s_2$ except at the location of physical singularities.
    \item Locality: at tree-level the amplitude $\mathcal{M}(s_1,s_2)$ is meromorphic with simple poles in $s_1$ and $s_2$ as the only physical singularities of the amplitude.
    \item Crossing symmetry: the amplitude is crossing symmetric in $s_1$ and $s_2$
    \begin{equation}
        \mathcal{M}(s_1,s_2)=\mathcal{M}(s_2,s_1)\;. 
    \end{equation}
    \item Unitarity: we assume tree-level unitarity. Unitarity at tree-level implies that for the polynomial residues in \eqref{res}, the partial wave expansion is accompanied by positive partial wave coefficients, $c_\ell^{(n)}\geq 0$.   
    \item Maximal spin: at a particular mass-squared level $n$ we assume a hard spin cut-off similar to the open string amplitude. This is motivated by the fact that the amplitude has a leading Regge trajectory.
    \item \label{assump6} The amplitude $\mathcal{M}(s_1,s_2)$ goes to zero as $s_1\to\infty$ for $s_2<0$ faster than a constant. Mathematically, we have $\lim_{s_1\to\infty}\mathcal{M}(s_1,s_2)\sim \mathcal{O}(s_1^{-\epsilon})$, with $\epsilon>0$ for $s_2<0$.
\end{enumerate}
The outcomes of the paper are the following. a) Using these physical assumptions and the $\lambda$-CSDR we learn the open string mass-squared spectrum. b) Using entanglement minimization we provide evidence that an integer mass-squared spectrum is associated with lower entanglement than general non-integer mass-squared spectra. c) We use the local three-channel crossing symmetric dispersion relation with one-parameter ambiguity to carve out the space of allowed theories of three-channel crossing symmetric string theory amplitudes and identify the Virasoro-Shapiro amplitude in this space. We use machine learning to learn the partial-wave coefficients of the Virasoro-Shapiro amplitude. d) For an amplitude with two-channel crossing symmetry we show that a Non-dual amplitude shows a Polchinski-Strassler type transition from Regge behavior to power law behavior.  

We organize the paper as follows. In section \ref{CSPR} we write down the local two-channel crossing symmetric dispersion relation with one-parameter ambiguity. This will form our ansatz for the analysis that we perform in this paper. In section \ref{EP} we explain entanglement minimization. In section \ref{MLsetup} we give details about the machine learning algorithm to learn the mass-squared spectrum. In section \ref{results} we give details of our results about the open string mass-squared spectrum using machine learning and linear programming. In section \ref{Closed String} we move on to amplitudes with three channel crossing symmetry and give our results for the Virasoro-Shapiro amplitude. In section \ref{Regge} we compare the Regge trajectory of a Non-dual amplitude first found in \cite{Bhat:2024agd} with the Regge trajectory of the open string amplitude. We conclude and give future directions in section \ref{conclusions}.
\section{Crossing Symmetric Parametric Representations}
\label{CSPR}
The two-channel local crossing symmetric dispersion relation was derived in \cite{Saha:2024qpt}. It is given by
\begin{equation}
\label{LCSDR}
    \mathcal{M}(s_1,s_2)=\mathcal{M}(0,0)+\frac{1}{\pi}\int_{s_0}^{\infty}\text{d}\sigma\left[\frac{1}{\sigma-s_1}+\frac{1}{\sigma-s_2}-\frac{1}{\sigma}\right]\mathcal{A}^{(s_1)}\left(\sigma,\frac{y}{\sigma}\right)-\frac{1}{\pi}\int_{s_0}^{\infty}\frac{\text{d}\sigma}{\sigma}\mathcal{A}^{(s_1)}(\sigma,0)
\end{equation}
where $s_0$ is the location of the pole at the first massive exchange. Here, $\mathcal{A}^{(s_1)}(s_1,s_2)$ is the $s_1$-channel imaginary part of the amplitude 
\begin{equation}
    \mathcal{A}^{(s_1)}(s_1,s_2)=\lim_{\epsilon\rightarrow0^+}\frac{\mathcal{M}(s_1+i\epsilon,s_2)-\mathcal{M}(s_1-i\epsilon,s_2)}{2i}\;.
\end{equation}
Here we define
\begin{equation}
    x=s_1+s_2\;,\qquad y=s_1 s_2\;,
\end{equation}
which are the crossing symmetric variables. In writing \eqref{LCSDR} it is implicitly assumed that there is no massless pole and $\mathcal{M}(0,0)$ is finite. If the amplitude consists of a massless pole then $\mathcal{M}(0,0)$ has to be replaced using the following replacement rule
\begin{equation}
    \mathcal{M}(0,0)\to\text{Massless Pole}+W_{00}\;.
\end{equation}
The Massless Pole is the singular part of the amplitude at $s_1=0$ or $s_2=0$ and $W_{00}$ is finite. Wilson coefficients are defined as follows. One can expand the amplitude $\mathcal{M}(s_1,s_2)$ after subtracting out the Massless Pole as a Taylor series expansion around $x=y=0$\;,
\begin{equation}
    \mathcal{M}(s_1,s_2)-\text{Massless Pole}=\sum_{p,q=0}^\infty W_{pq}x^py^q\;.
\end{equation}
So, Wilson coefficients are the low energy Taylor expansion coefficients of the amplitude. Using shifted variables one can derive the two-channel local crossing symmetric dispersion relation with one-parameter ambiguity ($\lambda$-CSDR) \cite{Bhat:2024agd, Bhat:2025zex}
\begin{equation}
\label{lambdaLCSDR2channel}
\begin{split}
    \mathcal{M}(s_1,s_2)&=\mathcal{M}(-\l,-\l)-\frac{1}{\pi}\int_{s_0}^\infty\text{d}\s\frac{\mathcal{A}^{(s_1)}(\s,-\l)}{\s+\l}\\&+\frac{1}{\pi}\int_{s_0}^\infty\text{d}\s\left[\frac{1}{\s-s_1}+\frac{1}{\sigma-s_2}-\frac{1}{\s+\l}\right]\mathcal{A}^{(s_1)}\left(\s,\frac{(s_1+\l)(s_2+\l)}{\s+\l}-\l\right)
\end{split}
\end{equation}
Let us denote the Massless Pole by 
\begin{equation}
    \text{Massless Pole}=P\left(\frac{1}{s_1},\frac{1}{s_2}\right)\;.
\end{equation}
Then \eqref{LCSDR} can be written as 
\begin{equation}
\label{LCSDR2}
\begin{split}
    \mathcal{M}(s_1,s_2)&=P\left(\frac{1}{s_1},\frac{1}{s_2}\right)+W_{00}-\frac{1}{\pi}\int_{s_0}^{\infty}\frac{\text{d}\sigma}{\sigma}\mathcal{A}^{(s_1)}(\sigma,0)+\frac{1}{\pi}\int_{s_0}^{\infty}\text{d}\sigma\left[\frac{1}{\sigma-s_1}+\frac{1}{\sigma-s_2}-\frac{1}{\sigma}\right]\mathcal{A}^{(s_1)}\left(\sigma,\frac{y}{\sigma}\right)\;.
\end{split}
\end{equation}
Let us now focus on the open string amplitude. Since, the open string amplitude behaves as $s_1^{-1+s_2}$ as $s_1\rightarrow\infty$ this implies that in the forward limit $s_2\rightarrow0$ the open string amplitude falls off as $s_1^{-1}$ as $s_1\rightarrow\infty$. Therefore, using Cauchy's residue theorem we obtain \cite{Bhat:2024agd}
\begin{equation}
\label{w00}
    W_{00}=\frac{1}{\pi}\int_{s_0}^{\infty}\frac{\text{d}\sigma}{\sigma}\mathcal{A}^{(s_1)}(\sigma,0)
\end{equation}
and for the open string amplitude we also have 
\begin{equation}
P\left(\frac{1}{s_1},\frac{1}{s_2}\right)=\frac{1}{s_1s_2}\;.
\end{equation}
Therefore, for the open string amplitude \eqref{LCSDR} can be written as
\begin{equation}
\label{LSDROS}
    \mathcal{M}(s_1,s_2)=\frac{1}{s_1s_2}+\frac{1}{\pi}\int_{s_0}^{\infty}\text{d}\sigma\left[\frac{1}{\sigma-s_1}+\frac{1}{\sigma-s_2}-\frac{1}{\sigma}\right]\mathcal{A}^{(s_1)}\left(\sigma,\frac{y}{\sigma}\right).
\end{equation}
Using shifted variables on the above dispersion relation we obtain a one-parameter family of dispersion relations for the open string amplitude \cite{Bhat:2024agd}
\begin{equation}
\label{lambdaLCSDROS}
\begin{split}
    \mathcal{M}(s_1,s_2)&=\frac{1}{s_1s_2}+\frac{1}{\pi}\int_{s_0}^\infty\mathrm{d}\s\left[\frac{1}{\s-s_1}+\frac{1}{\sigma-s_2}-\frac{1}{\s+\l}\right]\mathcal{A}^{(s_1)}\left(\s,\frac{(s_1+\l)(s_2+\l)}{\s+\l}-\l\right)\;.
\end{split}
\end{equation}
We will now see what inference we can gather about the mass-squared spectrum of the open string amplitude using this dispersion relation. For this, we will use both machine learning and linear programming.

For an amplitude with simple poles and polynomial residues the $s_1$-channel imaginary part can be decomposed in a basis of Gegenbauer polynomials as a partial wave expansion
\begin{equation}\label{sdisc}
    \mathcal{A}^{(s_1)}(s_1,s_2)=\sum_{(\ell,R_n)\in\text{spectrum}}-\pi\, c_\ell^{(R_n)}\,\mathcal{G}_\ell^{\left(\frac{D-3}{2}\right)}\left(1+\frac{2s_2}{s_1}\right)\,\delta(s_1-R_n)\:.
\end{equation}
Here $R_n$ is a positive real number representing the location of the poles of the amplitude. The set of pairs $\left(\ell, R_n\right)$ is called the spectrum of the theory. Substituting \eqref{sdisc} into \eqref{lambdaLCSDROS} we obtain
\begin{equation}
\label{gendispsum}
\begin{split}
     &\mathcal{M}(s_1,s_2)=\frac{1}{s_1s_2}
    \\&+\sum_{(\ell,R_n)\in\text{spectrum}} c_\ell^{(R_n)}  \left[\frac{1}{s_1-R_n}+\frac{1}{s_2-R_n}+ \frac{1}{\lambda+R_n}\right] \mathcal{G}_\ell^{\left(\frac{D-3}{2}\right)}\left(1+\frac{2}{R_n}\left(\frac{(s_1+\lambda)(s_2 + \lambda)}{R_n + \lambda} - \lambda \right)\right)\:.
\end{split}
\end{equation}
It is clear that poles of the above amplitude are located at positive real values of $s$ and $t$ provided that the mass-squared spectrum is real and positive. Note that the total number of $(\ell, n)$ pairs in the open string spectrum is unbounded, but for the purpose of numerical simulations we have to truncate the $(\ell, R_n)$ pairs. We define $N_{max}$ as the number of distinct $R_n$ values in the mass-squared spectrum.
\section{Entanglement minimization}
\label{EP}
Here, we describe what quantity we minimize in our numerical algorithm. We minimize entanglement generated in $2\rightarrow2$ scattering of gluons and gravitons \cite{Bhat:2024agd,bose1,bose2,bell,PhysRevLett.125.181602,Beane:2021zvo,Aoude:2024xpx,Kowalska:2024kbs,Low:2024mrk,Low:2024hvn,McGinnis:2025iab,McGinnis:2025xgt,Carena:2025wyh,Hu:2025lua}. The entanglement generated in $2\to2$ scattering was studied in \cite{Aoude:2024xpx}. Let the initial states be $|p_1,a\rangle$ and $|p_2,b\rangle$ where $p_i$ denote the 4-momenta of the particles and $a$ and $b$ denote the internal quantum numbers of the particles if any. The general initial state is $|\Omega\rangle=\sum_{a,b}\Omega_{ab}|p_1,a\rangle_A\otimes|p_2,b\rangle_B$. Here $A$ and $B$ denote the two Hilbert spaces $\mathcal{H}_A$ and $\mathcal{H}_B$ of the two incoming particles. The linear entropy between these two subsystems is given by
\begin{equation}
    \mathcal{E}[\Omega]=1-\text{tr}_A\left[\rho_A^2\right]\;.
\end{equation}
$\rho_{AB}=|\Omega\rangle\langle\Omega|$ is the density matrix constructed from the initial state $|\Omega\rangle$ and $\rho_A=\text{tr}_B[\rho_{AB}]$. The initial state is normalized such that $\text{tr}_A[\rho_A]=1$. The linear entropy of the initial state is given by $\mathcal{E}_i[\Omega]=1-\text{tr}\left[(\Omega^\dagger\Omega)^2\right]$. If the initial state is a product state of the form
\begin{equation}
    |\Omega\rangle_\text{prod}=|p_1,a\rangle_A\otimes|p_2,b\rangle_B
\end{equation}
then $\text{tr}\left[(\Omega^\dagger\Omega)^2\right]=1$ and the initial linear entropy $\mathcal{E}_i$ vanishes. The difference in linear entropy between the final and initial states at the leading order in the coupling strength is given by
\begin{equation}
\label{EPgen}
    \Delta\mathcal{E}[\Omega]=4\,\mathcal{N}\,\text{Im}\left[\sum_{abij}\Omega_{ab}\mathcal{M}_{ij}^{ab}(p_1,p_2\rightarrow p_1,p_2)\left(\Omega^\dagger\cdot\Omega\cdot\Omega^\dagger\right)_{ij}\right]\;.
\end{equation}
Here $\mathcal{M}_{ij}^{ab}$ is the S-matrix element with initial states denoted by $a,b$ and final states denoted by $i,j$. For product initial state we have $\left(\Omega\cdot\Omega^\dagger\cdot\Omega\right)_{kl}=\Omega_{kl}$. Therefore, for product initial state \eqref{EPgen} reduces to
\begin{equation}\label{EPprod}
    \Delta\mathcal{E}\left[\Omega^{\text{prod}}\right] = 4\,\mathcal{N}\,\text{Im}\mathcal{M}_{\alpha\beta}^{\alpha\beta}\left(p_{1},p_{2}\rightarrow p_{1}, p_{2}\right)\;.
\end{equation}
Here, the normalization is $\mathcal{N} = \frac{T}{4E_{\Vec{p}_{1}}E_{\Vec{p}_{2}}V}$, with $T=2\pi\delta\left(0\right)$ coming from the energy conservation and $V=(2\pi)^3\delta^{(3)}\left(\vec{0}\right)$ coming from the 3-momenta conservation. Now let us calculate the entanglement generated in $2\to2$ gluon scattering. Consider the four-point color-ordered scattering amplitude of gluons $\mathcal{M}_4(1^{h_1},2^{h_2},3^{h_3},4^{h_4})$ in type-I open superstring theory. $h_i$ is the helicity of the $i$th external particle. The external particles can have linear combinations of helicities so it is convenient to work with linear combinations of helicity amplitudes $\mathcal{M}_{\varepsilon_1\varepsilon_2}^{\varepsilon_3\varepsilon_4}(p_1,p_2,p_3,p_4)$. We will follow the convention where all outgoing momenta are positive. Then,
\begin{equation}
    s=-(p_1+p_2)^2\;,\quad t=-(p_1+p_4)^2\;,\quad u=-(p_1+p_3)^2\;.
\end{equation}
The tree-level color-ordered four-gluon amplitude in type-I open superstring theory is given by
\begin{equation}
\label{gluonamp}
    \mathcal{M}_{OS}(s,t)=g^2\mathcal{F}^4\frac{\Gamma(-s)\Gamma(-t)}{\Gamma(1-s-t)}\;.
\end{equation}
This is the tree-level open superstring amplitude for $2\to2$ gluon scattering. We are using units where $\alpha'=1$. Here $g^2$ is the coupling strength and the prefactor $\mathcal{F}^4$ is given by 
\begin{eqnarray}
\label{gluonF4}
	\mathcal{F}^{4} & = & F_{1,\mu\nu}F_{2}^{\mu\nu}F_{3,\alpha\beta}F_{4}^{\alpha\beta} + F_{1,\mu\nu}F_{3}^{\mu\nu}F_{4,\alpha\beta}F_{2}^{\alpha\beta} + F_{1,\mu\nu}F_{4}^{\mu\nu}F_{2,\alpha\beta}F_{3}^{\alpha\beta} \nonumber\\
	&& -4\biggl\{F_{1,\mu\nu}F_{2}^{\nu\alpha}F_{3,\alpha\beta}F_{4}^{\beta\mu} + F_{1,\mu\nu}F_{3}^{\nu\alpha}F_{4,\alpha\beta}F_{2}^{\beta\mu} + F_{1,\mu\nu}F_{4}^{\nu\alpha}F_{2,\alpha\beta}F_{3}^{\beta\mu}\biggr\}\;.
\end{eqnarray}
Here trace over gauge theory indices is implied and $F_{i,\mu\nu} = p_{i,\mu}\varepsilon_{i,\nu}-p_{i,\nu}\varepsilon_{i,\mu}$ for the $i$th particle. In the forward limit $t\to0$, we can choose a frame such that the momenta of the incoming and outgoing particles are along the $z$-axis and the polarization vectors lie in the $xy$ plane. For such a configuration, $\varepsilon_i\cdot p_j=0$ for all $i,j$ (for $i=j$ this becomes $\varepsilon_i\cdot p_i=0$ which is the usual physical transversality condition). In this limit, the prefactor becomes 
\begin{equation}
\label{F4forwardlim}
	\lim_{t\rightarrow 0}\mathcal{F}^{4} = -2s^{2}(\varepsilon_{1}\cdot\varepsilon_{4})\; (\varepsilon_{2}\cdot\varepsilon_{3})\;.
\end{equation}
To apply bootstrap techniques to this problem we have to assume the following form for the four-point amplitude
\begin{equation}
    \mathcal{M}_{OS}(s,t)=g^2\mathcal{F}^4M(s,t)\,.
\end{equation}
Here $M(s,t)$ is a meromorphic form factor with simple poles at non-negative values of $s$ and $t$
\begin{equation}
    M(s,t)=\frac{\Gamma(-s)\Gamma(-t)}{\Gamma(1-s-t)}\;.
\end{equation}
Thus in the forward limit after dropping the coupling strength $g^2$, \eqref{EPprod} becomes  
\begin{equation}
\label{EPgluon}
    \Delta\mathcal{E}_{gluon} = 8\pi\mathcal{N}s^{2}(\varepsilon_{1}\cdot\varepsilon_{4})\; (\varepsilon_{2}\cdot\varepsilon_{3})\sum_{n=1}^{\infty}M_n\;\delta\left(s-R_n\right)\,.
\end{equation}
Here $M_n=\lim_{t\to0}\text{Res}_{s=R_n}M\left(s,t\right)$
and we have summed over massive spectrum values $R_n$. We restrict to the polarization-preserving elastic forward channel. In the all-outgoing convention, the forward momentum conditions and the corresponding polarization conditions (see \eqref{EPprod}) are
\begin{equation}
    \begin{split}
        p_4=-p_1\;,&\quad p_3=-p_2\;,\\
        \varepsilon_4=\varepsilon_1^*\;,&\quad\varepsilon_3=\varepsilon_2^*\;.
    \end{split}
\end{equation}
The complex conjugation appears because the outgoing polarization wavefunctions belong to the bra state. Therefore,
\begin{equation}
\label{EPgluonelaspol}
    \Delta\mathcal{E}_{gluon} = 8\pi\mathcal{N}s^{2}(\varepsilon_{1}\cdot\varepsilon_{1}^*)\; (\varepsilon_{2}\cdot\varepsilon_{2}^*)\sum_{n=1}^{\infty}M_n\;\delta\left(s-R_n\right)\,.
\end{equation}
For unit-normalized physical polarizations
\begin{equation}
    \varepsilon_i\cdot\varepsilon_i^*=1
\end{equation}
so the result becomes 
\begin{equation}
\label{EPgluonfinal}
    \Delta\mathcal{E}_{gluon} = 8\pi\mathcal{N}s^{2}\sum_{n=1}^{\infty}M_n\;\delta\left(s-R_n\right)\,.
\end{equation}
Notice the sum over delta functions. To make this expression finite we need to integrate over these delta functions. The normalization $\mathcal{N}\sim \frac{1}{s}$. We will be interested in amplitudes which satisfy $\lim_{s\to\infty}|\lim_{t\to0}\left(M(s,t)-\frac{1}{st}\right)|\sim s^{-\epsilon}\,,\; \epsilon\geq0$. For such amplitudes
\begin{equation}
\label{EPintgluon}
    \int_1^\Lambda\mathrm{d}s\frac{\Delta\mathcal{E}_{gluon}}{s^2}
\end{equation}
is finite. Substituting \eqref{EPgluonfinal} into \eqref{EPintgluon}, dropping the $8\pi$ factor and expanding the residue in terms of Gegenbauer polynomials we obtain
\begin{equation}
\label{phi}
    \Phi=\sum_{(\ell,R_n)\in\text{spectrum}}\frac{1}{R_n}c_\ell^{(R_n)}\mathcal{G}_\ell^{(\frac{D-3}{2})}(1)\;.
\end{equation}
For our numerical simulations we will use $R_n<\Lambda$ in \eqref{gendispsum} and \eqref{phi} for $1\leq n\leq N_{max}$. We call $\Phi$ the entanglement power. We minimize $\Phi$ in both machine learning and linear programming. Notice that from \eqref{w00} we obtain $\Phi=-W_{00}$ and hence minimizing $\Phi$ would mean maximizing $W_{00}$. If we had considered a tensor structure of the form 
\begin{equation}
    \tilde{\mathcal{F}}^4=c_1\left\{F_{1,\mu\nu}F_{2}^{\mu\nu}F_{3,\alpha\beta}F_{4}^{\alpha\beta}+\text{cyclic}(2\to3\to4)\right\}-c_2\left\{F_{1,\mu\nu}F_{2}^{\nu\alpha}F_{3,\alpha\beta}F_{4}^{\beta\mu}+\text{cyclic}(2\to3\to4)\right\}\;,
\end{equation}
then the leading order difference in entropy would be
\begin{equation}
    \Delta\mathcal{E}_{gluon} = -4\pi\mathcal{N}s^{2}\left\{\left(c_1-\frac{c_2}{4}\right)(\varepsilon_{1}\cdot\varepsilon_{2}\; \varepsilon_{1}^*\cdot\varepsilon_{2}^*+\varepsilon_{1}\cdot\varepsilon_{2}^*\; \varepsilon_{2}\cdot\varepsilon_{1}^*)-\frac{c_2}{2}\varepsilon_{1}\cdot\varepsilon_{1}^*\; \varepsilon_{2}\cdot\varepsilon_{2}^*\right\}\sum_{n=1}^{\infty}M_n\;\delta\left(s-R_n\right)\,.
\end{equation}
For the superstring case we have $c_1=1$ and $c_2=4$ and therefore the first term vanishes.

Now let us consider $2\to2$ scattering of gravitons. 
The four-point scattering amplitude of gravitons in type-II superstring theory is given by \cite{DHoker:2019blr}
\begin{equation}
\label{graviton}
\begin{split}
    &\mathcal{M}_{CS}(s,t,u)=\kappa^2\mathcal{R}^4\frac{\G(1-s)\G(1-t)\G(1-u)}{stu\G(1+s)\G(1+t)\G(1+u)}=-\kappa^2\mathcal{R}^4\frac{\G(-s)\G(-t)\G(-u)}{\G(1+s)\G(1+t)\G(1+u)}\;,\\& s+t+u=0\;.
\end{split}
\end{equation}
This is the tree-level closed string amplitude for $2\to 2$ graviton scattering. For the closed sring amplitude we work in units where $\frac{\alpha'}{4}=1$. Here $\kappa^2=8\pi G_N$, $G_N$ being Newton's constant. Similar to the gluon amplitude the graviton amplitude may be written as
\begin{equation}
    \mathcal{M}_{CS}(s,t,u)=-\kappa^2\mathcal{R}^4 M(s,t,u)\;.
\end{equation}
Here $M(s,t,u)$ is the meromorphic form factor for graviton scattering given by
\begin{equation}
    M(s,t,u)=\frac{\G(-s)\G(-t)\G(-u)}{\G(1+s)\G(1+t)\G(1+u)}\;.
\end{equation}
For a small perturbation around flat spacetime $g_{\mu\nu}=\eta_{\mu\nu}+h_{\mu\nu}$ the linearized Riemann tensor is
\begin{equation}
    R_{\mu\nu\rho\sigma}=\frac{1}{2}\left(\partial_\rho\partial_\nu h_{\mu\sigma}+\partial_\sigma\partial_\mu h_{\nu\rho}-\partial_\sigma\partial_\nu h_{\mu\rho}-\partial_\rho\partial_\mu h_{\nu\sigma}\right)\;.
\end{equation}
For scattering using plane wave gravitons we have 
\begin{equation}
    h_{\mu\nu}=\epsilon_{\mu\nu}e^{ip\cdot x}\;.
\end{equation}
Replacing $\partial_\mu\to ip_{\mu}$ and removing the common plane-wave factor we get 
\begin{equation}
     R_{\mu\nu\rho\sigma}=\frac{1}{2}\left(p_\mu p_\rho\epsilon_{\nu\sigma}+p_\nu p_\sigma\epsilon_{\mu\rho}-p_\mu p_\sigma\epsilon_{\nu\rho}-p_\nu p_\rho\epsilon_{\mu\sigma}\right)\;.
\end{equation}
The physical polarization conditions for the $i$th external particle are
\begin{equation}
    p_i^2=0\;,\quad \epsilon_{i,\mu\nu}=\epsilon_{i,\nu\mu}\;,\quad p_i^\mu\epsilon_{i,\mu\nu}=0\;,\quad \epsilon_i{}^\mu{}_\mu=0\;.
\end{equation}
The type-II four-graviton kinematic structure is 
\begin{equation}
\mathcal{R}^{4}
=
t_8^{\mu_1\cdots\mu_8}
t_8^{\nu_1\cdots\nu_8}
R_{1,\mu_1\mu_2\nu_1\nu_2}
R_{2,\mu_3\mu_4\nu_3\nu_4}
R_{3,\mu_5\mu_6\nu_5\nu_6}
R_{4,\mu_7\mu_8\nu_7\nu_8}\;.
\label{eq:graviton-kinematic-factor}
\end{equation}
This is the gravitational analogue of the $\mathcal{F}^4$ factor in \eqref{gluonamp}.  
$t_8t_8R^4$ means two copies of an 8-index tensor contracted with four Riemann tensors. The next step is to evaluate $\mathcal{R}^4$ in the forward limit, giving the graviton analogue of \eqref{F4forwardlim}. We use factorized polarizations as a computational basis. A physical graviton polarization is a symmetric, traceless rank-two tensor, and we expand it into products of transverse polarization vectors \cite{Bern:2019prr}. The general expansion is 
\begin{equation}
    \epsilon_{i,\mu\nu}=\sum_{a,b} C_{i,ab}\;\varepsilon_{i,\mu}^a\;\varepsilon_{i,\nu}^b\;.
\end{equation}
\(a\) and \(b\) label the transverse polarization basis vectors. Each runs over
\(a,b=1,\ldots,D-2.\). This is because a massless vector has $D-2$ independent physical polarization directions. Two directions are removed because of transversality and gauge invariance. The entire sum is symmetric, traceless and transverse. So, $C_{i,ab}=C_{i,ba}$ and $\sum_a C_{i,aa}=0$. We first calculate one term
\begin{equation}
    \epsilon_{i,\mu\nu}=\varepsilon_{i,\mu}^a\;\varepsilon_{i,\nu}^b\;.
\end{equation}
and then sum over the terms. This works because the scattering amplitude is linear in each external polarization tensor.
In the forward limit $t\to0$, we have $p_4=-p_1$ and $p_3=-p_2$. Since all external momenta lie along the $z$-axis and all polarization vectors lie in the $xy$ plane, we have
\begin{equation}
    \varepsilon_i^a\cdot p_j=\varepsilon_i^b\cdot p_j=0,\quad\forall \;i,j\;.
\end{equation}
Consequently, 
\begin{equation}
p_j^\mu\epsilon_{i,\mu\nu}=p_j^\nu\epsilon_{i,\mu\nu}=0\;,\quad\forall\;i,j\;.
\end{equation}
The condition for $j=i$ is the usual physical transversality condition. The condition for all $i,j$ is the additional forward-limit gauge choice needed to simplify the contractions in $t_8t_8R^4$.
We define
\begin{equation}
    F_{i,\mu\nu}^a=p_{i,\mu}\varepsilon_{i,\nu}^a-p_{i,\nu}\varepsilon_{i,\mu}^a\;,\quad F_{i,\mu\nu}^b=p_{i,\mu}\varepsilon_{i,\nu}^b-p_{i,\nu}\varepsilon_{i,\mu}^b\;.
\end{equation}
Then
\begin{equation}
    R_{i,\mu\nu\rho\sigma}=\frac{1}{2}F_{i,\mu\nu}^a F_{i,\rho\sigma}^b\;.
\end{equation}
Using the standard identity \cite{Peeters:2000qj} $t_8F_1F_2F_3F_4=-2\mathcal{F}^4$, where $\mathcal{F}^4$ is the gluon prefactor in \eqref{gluonF4} we get,
\begin{equation}
    \mathcal{R}^4=\frac{1}{16}(t_8F_1^aF_2^aF_3^aF_4^a)(t_8F_1^bF_2^bF_3^bF_4^b)=\frac{1}{4}\mathcal{F}^4[\varepsilon^a]\mathcal{F}^4[\varepsilon^b]\;.
\end{equation}
Now applying the gluon forward limit to each factor we get
\begin{equation}
\begin{split}
&\lim_{t\to0}\mathcal{F}^4[\varepsilon^a]=-2s^2(\varepsilon_1^a\cdot\varepsilon_4^a)(\varepsilon_2^a\cdot\varepsilon_3^a)\;,\\
&\lim_{t\to0}\mathcal{F}^4[\varepsilon^b]=-2s^2(\varepsilon_1^b\cdot\varepsilon_4^b)(\varepsilon_2^b\cdot\varepsilon_3^b)\;.
\end{split}
\end{equation}
Therefore, 
\begin{equation}
    \lim_{t\to0}\mathcal{R}^4=s^4(\varepsilon_1^a\cdot\varepsilon_4^a)(\varepsilon_1^b\cdot\varepsilon_4^b)(\varepsilon_2^a\cdot\varepsilon_3^a)(\varepsilon_2^b\cdot\varepsilon_3^b)\;.
\end{equation}
Now, $(\varepsilon_i^a\cdot\varepsilon_j^a)(\varepsilon_i^b\cdot\varepsilon_j^b)=\epsilon_i:\epsilon_j$ where, $\epsilon_i:\epsilon_j=\epsilon_{i,\mu\nu}\epsilon_{j}^{\mu\nu}$. Therefore,
\begin{equation}
    \lim_{t\to0}\mathcal{R}^4=s^4(\epsilon_1:\epsilon_4)(\epsilon_2:\epsilon_3)\;.
\end{equation}
Therefore, $\mathcal{R}^4$ contributes $s^4$ in the forward limit. Substituting \eqref{graviton} in \eqref{EPprod} and dropping the $\kappa^2$ factor we obtain 
\begin{equation}
\label{EPgraviton}
    \Delta\mathcal{E}_{graviton} = 4\pi\mathcal{N}s^{4}(\epsilon_1:\epsilon_4)(\epsilon_2:\epsilon_3)\sum_{n=1}^{\infty}M_n\;\delta\left(s-n\right)\,,
\end{equation}
where $M_n=\lim_{t\to0}\text{Res}_{s=n}M\left(s,t,-s-t\right)$. Using the polarization conditions we have
\begin{equation}
    \epsilon_4=\epsilon_1^*\;,\quad\epsilon_3=\epsilon_2^*\;,\quad\epsilon_i^*:\epsilon_i=1\;.
\end{equation}
Therefore, the final result becomes
\begin{equation}
\label{EPgravitonfinal}
    \Delta\mathcal{E}_{graviton} = 4\pi\mathcal{N}s^{4}\sum_{n=1}^{\infty}M_n\;\delta\left(s-n\right)\,.
\end{equation}
Notice that for the closed string we have assumed the mass-squared spectrum to be integer valued $n$. In the forward limit, the residue $M_n$ of the scalar closed string meromorphic form factor at the $n$th massive pole is $M_n=\frac{1}{n^2}$. Then
\begin{align}
\label{EPintgravitons3}
    &\int_1^\Lambda\mathrm{d}s\frac{\Delta\mathcal{E}_{graviton}}{s^3}\qquad\text{and}\\ 
\label{EPintgravitons4}
    &\int_1^\Lambda\mathrm{d}s\frac{\Delta\mathcal{E}_{graviton}}{s^4}
\end{align}
are finite. In this work, we will minimize \eqref{EPintgravitons4}, which gives the same expression for $\Phi$ as \eqref{phi}. Let us define
\begin{equation}
\begin{split}
&A[\varepsilon^a]=F_{1,\mu\nu}^aF_{2}^{a,\mu\nu}F_{3,\alpha\beta}^aF_{4}^{a,\alpha\beta}+\text{cyclic}(2\to3\to4)\;,\\
&B[\varepsilon^a]=F_{1,\mu\nu}^aF_{2}^{a,\nu\alpha}F_{3,\alpha\beta}^a F_{4}^{a,\beta\mu}+\text{cyclic}(2\to3\to4)\;,
\end{split}
\end{equation}
and define $A[\varepsilon^b], B[\varepsilon^b]$ using $F_i^b$. Using this we can write
\begin{equation}
    \tilde{\mathcal{F}}[\varepsilon^a]=c_1A[\varepsilon^a]-c_2B[\varepsilon^a]
\end{equation}
and the gravitational deformation is 
\begin{equation}
    \tilde{\mathcal{R}}^4=\frac{1}{4}(c_1A[\varepsilon^a]-c_2B[\varepsilon^a])(c_1A[\varepsilon^b]-c_2B[\varepsilon^b])\;.
\end{equation}
At $c_1=1$, $c_2=4$ this recovers $\mathcal{R}^4=t_8t_8R_4$\;. Using $p_4=-p_1$, $p_3=-p_2$ in the forward limit and $\varepsilon_4=\varepsilon_1^*\;,\;\varepsilon_3=\varepsilon_2^*$ we define
\begin{equation}
    Q[\varepsilon]=\left(c_1-\frac{c_2}{4}\right)(\varepsilon_{1}\cdot\varepsilon_{2}\; \varepsilon_{1}^*\cdot\varepsilon_{2}^*+\varepsilon_{1}\cdot\varepsilon_{2}^*\; \varepsilon_{2}\cdot\varepsilon_{1}^*)-\frac{c_2}{2}\varepsilon_{1}\cdot\varepsilon_{1}^*\; \varepsilon_{2}\cdot\varepsilon_{2}^*\;.
\end{equation}
Then
\begin{equation}
    \lim_{t\to 0}\tilde{\mathcal{F}}[\varepsilon]=s^2Q[\varepsilon]
\end{equation}
and
\begin{equation}
    \lim_{t\to 0}\tilde{\mathcal{R}}^4=\frac{s^4}{4}Q[\varepsilon^a]Q[\varepsilon^b]\;.
\end{equation}
Therefore, for the deformed factor $\tilde{\mathcal{R}}^4$ we have 
\begin{equation}
\label{EPgravitondeformed}
    \Delta\mathcal{E}_{graviton} = \pi\mathcal{N}s^{4}Q[\varepsilon^a]Q[\varepsilon^b]\sum_{n=1}^{\infty}M_n\;\delta\left(s-n\right)\;.
\end{equation}
\section{Machine Learning Set-up for Learning Spectrum}
\label{MLsetup}
Machine learning has emerged as a powerful technique for solving problems in physics with applications in a wide variety of areas. In theoretical physics machine learning has been used in the context of numerical learning of Calabi-Yau manifolds \cite{CYML1,CYML2,DeLuca:2024njg} and also to solve problems in conformal field theory and numerical conformal bootstrap \cite{CBootML1, CBootML2, CBootML3, CBootML4, CBootML5, Ghosh:2026xnp, Ghosh:2026jbw, Ghosh:2026efw}. Recently, machine learning has been used extensively in the context of quantum mechanical scattering and S-matrix bootstrap \cite{QMML, SbootML1, SbootML2, Bhat:2024agd, Gumus:2024lmj, Gumus:2026mhb}. 

In this section we will describe our machine learning set-up. The amplitude $\mathcal{M}(s_1,s_2)$ is given by \eqref{gendispsum}. The problem in machine learning is to learn the $R_n$ values and the corresponding $c_\ell^{(R_n)}$ values and to check what kind of amplitudes we get from machine learning. In machine learning our goal is to construct a loss function such that the minimization of the loss function results in the following bootstrap constraints getting satisfied:
\begin{enumerate}
    \item We want tree-level unitarity \cite{Arkani-Hamed:2022gsa}, which is given by the following condition
    \begin{equation}
        c_\ell^{(R_n)}\geq 0\;.
    \end{equation}
    \item We want the Wilson coefficients $W_{10}$ and $W_{01}$ to be equal to their corresponding values for the open string amplitude. Therefore, the machine learning set-up must output
\begin{equation}
    \left|W_{10}-(-\zeta(3))\right|\leq\epsilon\quad\text{and}\quad\left|W_{01}-\frac{7}{4}\zeta(4)\right|\leq \epsilon
\end{equation}
with a tolerance $0<\epsilon<<1$.
    \item According to assumption \ref{assump6}, for an amplitude $\mathcal{M}(s_1,s_2)$ that goes to zero as $s_1\to\infty$ for some value of $s_2<0$ then \eqref{gendispsum} converges to the amplitude and it is independent of $\lambda$ around $\lambda=-s_2$. For example, the open string amplitude goes to zero around $s_2=-14.6$ as $s_1\to \infty$. Therefore, for this value of $\lambda$ the machine learning set-up must output
\begin{equation}
\label{DerRatioCons}
   \left|\frac{\partial^k \mathcal{M}(s_1,s_2)}{\partial\lambda^k} \right| \leq \epsilon, \quad \text{for }k =1,2,3,\cdots,k_{max}
\end{equation}
with a tolerance $0<\epsilon<<1$. These $\lambda$-independence constraints are exactly equal to zero for the full infinite spectrum. In the numerical implementation we truncate the partial-wave expansion at $N_{max}$, so the resulting constraints are enforced approximately. We will use $\lambda=14.6$ throughout for the open string numerical simulations since the ansatz shows good convergence for this value. Although, these $\lambda$-independence constraints should hold for arbitrarily large $k$ but for numerical purposes we will impose them up to $k_{max}$. 
    \item We want to minimize the Entanglement Power ($\Phi$) in $2\rightarrow 2$ scattering of gluons in Type I open string theory. The expression for the Entanglement Power ($\Phi$) is given by 
\begin{equation}
    \Phi=\sum_{(\ell,R_n)\in\text{spectrum}}\frac{1}{R_n}c_\ell^{(R_n)}\mathcal{G}_\ell^{(\frac{D-3}{2})}(1)=-W_{00}\,.
\end{equation}
Hence, minimization of $\Phi$ would mean maximization of the leading Wilson coefficient $W_{00}$. 
\end{enumerate}
To solve this problem we will train a neural network \cite{HORNIK1989359} to learn the partial wave coefficients $c_\ell^{(R_n)}$. A neural network which finds the solution to a problem whose constraints are motivated by a physical scenario is called a Physics Informed Neural Network (PINN) \cite{RAISSI2019686}. The task of the PINN is to minimize $\Phi$ given these set of constraints. This can be achieved via the following Physics Informed Loss Function
\begin{equation}
    \label{Loss_Func}
    \mathcal{L} = \Phi + \beta_{10} \bigg{(}W_{10} - (-\zeta(3))\bigg{)}^2 + \beta_{01} \left(W_{01} - \frac{7\zeta(4)}{4}\right)^2 +   \sum_{k=1}^{k_{max}}\beta_k\tilde{\mathcal{L}}_k
\end{equation}
Here, $\tilde{\mathcal{L}}_k$ are $\lambda$-independence constraints imposed on a set $\mathcal{C}$ of $s_1, s_2$ values of length $C$:
\begin{equation}
\label{lambdaderivcons}
    \tilde{\mathcal{L}}_k=\frac{1}{C} \sum_{s_1,s_2 \in \mathcal{C}}\left(\frac{\partial^k\mathcal{M}(s_1,s_2)}{\partial\lambda^k}\right)^2\,.
\end{equation}
 Here, $\beta$s are positive weights that multiply the respective constraints in the loss function. They help push the constraints to smaller and smaller values depending on the tolerance ($\epsilon$) required to solve the problem. We can further see from \eqref{gendispsum} that because of the presence of $R_n$ in the denominator the constraints in the loss function become non-linear, which means that it is not possible to impose these constraints using linear programming techniques. Machine Learning becomes very helpful in this context. This problem is different from the problem described in \cite{Bhat:2024agd}. In \cite{Bhat:2024agd} the mass-squared spectrum was fixed to integer values and the neural network was learning the imaginary part of the partial wave coefficients. However, in this problem, the mass-squared spectrum $R_n$ and the imaginary part of the partial wave coefficients $c_\ell^{(R_n)}$ have to be learned simultaneously. To achieve this, we use a neural network to learn the $c_\ell^{(R_n)}$ and use gradient descent to learn the $R_n$ values. We use a fully-connected feed-forward neural network as shown in figure 14 of \cite{Bhat:2024agd} which is implemented with the help of \textit{PyTorch}\cite{Paszke:2019xhz}. From \eqref{gendispsum} we can see that, the argument of the Gegenbauer polynomial is a function of $R_n$. Therefore, it is necessary to pass gradient descent through all the functions i.e. $W_{10}$, $W_{01}$ and the $\tilde{\mathcal{L}}_k(s_1,s_2)$ which are derived from \eqref{gendispsum}. The key to do this is to use a Gegenbauer polynomial ($\mathcal{G}_\ell^{(\a)}(z)$) which is fully compatible with \textit{PyTorch}'s autograd with respect to its argument ($z$). We know that the Gegenbauer polynomials have the following representation
\begin{equation}
\label{geghyp2f1}
    \mathcal{G}_\ell^{(\a)}(z)=\frac{(2\a)_\ell}{\ell!} {}_2F_1\left(-\ell,2\a+\ell;\a+\frac{1}{2};\frac{1-z}{2}\right)
\end{equation}
Here, $(a)_k$ is the Pochhammer symbol: $(a)_0=1$, $(a)_k=(a)_{k-1}\times(a+k-1)$. The prefactor multiplying the hypergeometric function $\frac{(2\a)_\ell}{\ell!}$ is independent of $R_n$. But $z$ is a function of $R_n$. So, the last argument of the hypergeometric function $\frac{1-z}{2}$ is a function of $R_n$. Therefore, to learn the $R_n$ we need to differentiate the hypergeometric function using \textit{PyTorch}'s autograd. The standard hypergeometric functions that are available in Python for example \texttt{scipy} \texttt{hyp2f1}, \texttt{mpmath} \texttt{hyp2f1} and \texttt{sympy} \texttt{hyper} are not compatible with \textit{PyTorch}'s autograd. Therefore, we use the series representation of the hypergeometric function to make it \textit{PyTorch} autograd compatible
\begin{equation}
\label{hyp2f1ser}
    {}_2F_1(a,b;c;x)=\sum_{k=0}^\infty\frac{(a)_k(b)_k}{(c)_kk!}x^k\;.
\end{equation}
Moreover, from \eqref{geghyp2f1} we can see that the first argument of the hypergeometric function is $-\ell$. Now, $(-\ell)_k=0$ for $k>\ell$. Therefore, the series terminates!
\begin{equation}
\label{gegser}
    {}_2F_1\left(-\ell,2\a+\ell;\a+\frac{1}{2};\frac{1-z}{2}\right)=\sum_{k=0}^\ell\frac{(-\ell)_k(2\a+\ell)_k}{\left(\a+\frac{1}{2}\right)_kk!}\left(\frac{1-z}{2}\right)^k\;.
\end{equation}
This provides a \textit{PyTorch} autograd compatible series representation of the hypergeometric function ${}_2F_1$. 

 The input of our neural network is of size 2, one for the variable $\ell$ and the other for the variable $R_n$. The set of allowed pairs $(\ell,R_n)$ that are used as input to the neural network is called the spectrum of the theory. The neural network has 3 hidden layers of size 128 each. This is followed by the output layer which gives a scalar output i.e. the $c_\ell^{(R_n)}$. We use activation functions that add non-linearity to our network. We employ the commonly used (Rectified Linear Unit) activation function defined by $\text{ReLU}(x)=max(0,x)$ in the forward passes from the input layer to the first hidden layer, from the first hidden layer to the second hidden layer and from the second hidden layer to the third hidden layer. We use the Softplus activation function defined by $\gamma(x)=\log(1+e^x)$ in the forward pass from the third hidden layer to the output layer. The Softplus activation function enforces positivity of the $c_\ell^{(R_n)}$ thus maintaining unitarity of the theory.

We use both the standard optimizer Adam (Adaptive Moment Estimation) \cite{Kingma:2014vow} as well as the optimizer L-BFGS (Limited-memory Broyden–Fletcher–Goldfarb–Shanno) \cite{Liu:1989esw}. For the Adam optimizer, we use two different learning rates for the $c_\ell^{(R_n)}$ and the $R_n$ values. While the learning rate for $c_\ell^{(R_n)}$ is fixed to $10^{-5}$, the learning rate for the spectrum $R_n$ is varied from a large value for example $10^{-4}$ to a much smaller value for example $10^{-8}$ gradually depending on the value of the loss function and how close it is to the converged solution. The L-BFGS optimizer uses curvature information about the loss function and evaluates the loss over many iterations within one epoch of the training loop. The curvature information of the loss function is related to the second derivative of the loss function or the Hessian matrix
\begin{equation}
    H_{ij}=\frac{\partial^2\mathcal{L}(\theta_1,\theta_2,\cdots,\theta_n)}{\partial\theta_i\partial\theta_j}\;.
\end{equation}
The L-BFGS constructs an approximation to the inverse Hessian matrix using previous gradients and parameter updates.
It stores the computed gradients and the curvature information in memory with a specified History Size (how many past gradients and curvature updates are stored), and then it uses a Line Search Function to determine the step size for updating the parameters. For our set-up, we use a learning rate of $1.0$ for the L-BFGS optimizer for both $c_\ell^{(R_n)}$ and the $R_n$ values. We use Max-Iterations of 20 within each epoch and a History Size of 10. We use the Strong Wolfe Line Search Function to determine the final step size to update the parameters. Unlike ordinary gradient descent, which updates the parameters as follows 
\begin{equation}
    \theta_{t+1}=\theta_t-\eta\,\nabla_\theta\mathcal{L}\;,\qquad\eta=\text{Learning Rate}
\end{equation}
the L-BFGS optimizer updates the parameters as follows
\begin{equation}
    \theta_{t+1}=\theta_t-\eta\,\alpha_t\,H_t^{-1}\nabla_\theta\mathcal{L}\;,
\end{equation}
where $\a_t=\text{step size at step }t,\;H_t^{-1}=\text{approximation to the inverse Hessian Matrix at step }t$. We mainly use the Adam optimizer throughout the training. The L-BFGS optimizer may be used at various points during the training. The L-BFGS optimizer further decreases the loss and brings its value close to convergence depending on how close it is to the solution. The details of our machine learning algorithm are given in the table below.
\begin{table}[H]
    \centering
        \begin{tabular}{|c|c|} 
            \hline
             Input \textbar{} Hidden \textbar{} Output & 2 \textbar{} 128, 128, 128 \textbar{} 1  \\
            \hline
            Activation Function & ReLU \\
            \hline
            Final Layer & Softplus \\
            \hline
            Optimizers & Adam, L-BFGS \\
            \hline
            Adam Learning Rate for $c_\ell^{(R_n)}$ &  $10^{-5}$ \\
            \hline
            Adam Learning Rate for $R_n$ &  $10^{-4}\text{ to }10^{-8}$ \\
            \hline
            \multirow{4}{*}{ L-BFGS parameters for $c_\ell^{(R_n)}$ and $R_n$ } &  Learning Rate = 1.0,\\ 
            & Max-Iterations = 20, \\ 
            & History Size = 10, \\ 
            & Line Search Function = Strong Wolfe \\
            \hline
        \end{tabular}
    \caption{Hyperparameters for the network.}
    \label{tab:HyperTable}
\end{table}

\section{Results}
\label{results}
\subsection{Machine Learning}
\label{resultsML}
Here we present our results for learning the open string mass-squared spectrum using machine learning. We choose $\l=14.6$, $N_{max}=10$ and $k_{max}=5$. We run the machine learning simulations in $D=10$. We build the $(\ell,R_n)$ pairs as follows. At the level $R_n$ the following $\ell$ contribute: 
\begin{equation}
\label{MLspec}
\begin{split}
    0\leq\ell\leq \lfloor R_n\rceil-1 \text{ and } \lfloor R_n\rceil-\ell\equiv 1 \text{ mod } 2\;.
\end{split}
\end{equation} 
Here $\lfloor x\rceil$ represents the nearest integer function. We use the \texttt{torch.round}() operation in \textit{PyTorch} to achieve this. To learn the mass-squared spectrum $R_n$ we begin with an initial set of $R_n$ values and then use gradient descent to learn the $R_n$ values until we reach a converged solution thus obtaining the final set of $R_n$ values. The initial set of $R_n$ values are chosen such that $n< R_n< n+\delta$ where $0<\delta<0.5$. The spin template is reconstructed after every optimization epoch using the current values of $R_n$. In all optimization runs reported here, however, each learned pole satisfies $n<R_n<n+0.5$, so that \texttt{torch.round}$(R_n)=n$ identically throughout training. Consequently, the spin template remains unchanged during optimization, and the discontinuities associated with \texttt{torch.round}() are never encountered in practice. The machine learning problem then becomes the following. Given the open string spin template, can the $\lambda$-independence constraints reconstruct the integer mass-squared spacing? We choose the following set of points for the $\lambda$-independence constraints:
    \begin{equation}
        \mathcal{C} = \{ (s_1, s_2)\, |\, -9.1 \leq s_1 \leq 9.1 \, ,  \, -0.3 \leq s_2 \leq 0.3  \, , \, \D_{s_1} = 1\, , \, \D_{s_2} = 1/11\}\;.
    \end{equation} 
 Let us define
\begin{equation}
    \tilde{\mathcal{L}}_{\text{mean}}=\frac{1}{k_{max}}\sum_{k=1}^{k_{max}}\sqrt{\tilde{\mathcal{L}}_k}\;.
\end{equation}
\begin{itemize}
\item $R_1=1$\\
For this case we fix the level $R_1=1$ and learn the rest of the levels from $R_2$ to $R_{10}$. We use $\beta_{10}=10^6$, $\beta_{01}=10^6$,   $\beta_{1}=10^8$ and $\beta_2=\beta_3=\beta_4=\beta_{5}=10^6$. We mainly use the Adam optimizer throughout training. It is observed that in the middle of training the L-BFGS technique further decreases the loss by $86.81\%$. Then we continue to use Adam until the loss has converged. This leads to the following solution:
\begin{equation}
    \begin{split}
        \Phi = 1.590 \, &, \quad  \,  W_{10}-(-\zeta(3)) = 2.28 \times 10^{-4} \, , \\
        W_{01}-\frac{7\zeta(4)}{4} = - 1.86 \times 10^{-4} \, &, \quad  \, \tilde{\mathcal{L}}_{mean}  = 9.30 \times 10^{-5}\,. 
    \end{split}
\end{equation}
 Here the initial $R_n$ values were chosen such that $n< R_n< n+0.4$\;. The learned final $R_n$ values are given in the table below. 
\begin{table}[H]
    \centering
    \begin{tabular}{|c|c|c|c|c|c|c|c|c|c|c|}  
        \hline
        \multicolumn{11}{|c|}{Level 1 fixed} \\
        \hline
        Initial $R_n$ & 1 & 2.0964 & 3.2040 & 4.1564 & 5.0302 & 6.0430 & 7.2830 & 8.1194 & 9.0233 & 10.3838\\         
        \hline
        Final $R_n$ & 1 & 2.0128 & 3.0212 & 4.1816 & 5.0162 & 6.0902 & 7.2423 & 8.1491 & 9.0082 & 10.3758\\ 
        \hline
    \end{tabular}
    \label{tab:sample1}
\end{table}

\item $R_1=1, R_2=2$\\
For this case we fix the levels $R_1=1$, $R_2=2$ and learn the rest of the levels from $R_3$ to $R_{10}$. In the beginning of training we use $\beta_{10}=10^6$, $\beta_{01}=10^6$, $\beta_{1}=10^8$ and $\beta_2=\beta_3=\beta_4=\beta_{5}=10^6$. We mainly use the Adam optimizer throughout training. It is observed that near the end of training the L-BFGS technique further decreases the loss by $13.42\%$ when it is close to convergence. At the end of training we use $\beta_{10}=10^6$, $\beta_{01}=10^6$,  $\beta_1=\beta_2=\beta_3=\beta_4=\beta_{5}=10^8$ to push the constraints to smaller values. This leads to the following solution:
\begin{equation}
    \begin{split}
        \Phi = 1.608 \, &, \quad  \,  W_{10}-(-\zeta(3)) = 8.45 \times 10^{-6} \, , \\
        W_{01}-\frac{7\zeta(4)}{4} = - 4.63 \times 10^{-6} \, &, \quad  \, \tilde{\mathcal{L}}_{mean}  = 6.35 \times 10^{-5}\,. 
    \end{split}
\end{equation}
Here the initial $R_n$ values were chosen such that $n< R_n< n+0.5$\;. The learned final $R_n$ values are given in the table below.
\begin{table}[H]
    \centering
    \begin{tabular}{|c|c|c|c|c|c|c|c|c|c|c|}  
        \hline
        \multicolumn{11}{|c|}{Levels 1, 2 fixed} \\
        \hline
        Initial $R_n$ & 1 & 2 & 3.1186 & 4.1806 & 5.2318 & 6.0696 & 7.3018 & 8.3082 & 9.2454 & 10.1998\\         
        \hline
        Final $R_n$ & 1 & 2 & 3.0286 & 4.2386 & 5.1909 & 6.1685 & 7.2933 & 8.2840 & 9.1794 & 10.3163\\ 
        \hline
    \end{tabular}
    \label{tab:sample2}
\end{table}
\item $R_1=1, R_2=2, R_3=3$\\
For this case we fix the levels $R_1=1$, $R_2=2$, $R_3=3$ and learn the rest of the levels from $R_4$ to $R_{10}$. We use $\beta_{10}=10^6$, $\beta_{01}=10^6$, $\beta_{1}=10^8$ and $\beta_2=\beta_3=\beta_4=\beta_{5}=10^6$. We use the Adam optimizer throughout training. This leads to the following solution:
\begin{equation}
    \begin{split}
        \Phi = 1.560 \, &, \quad  \,  W_{10}-(-\zeta(3)) = 1.29 \times 10^{-5} \, , \\
        W_{01}-\frac{7\zeta(4)}{4} = 1.07 \times 10^{-6} \, &, \quad  \, \tilde{\mathcal{L}}_{mean}  = 2.50 \times 10^{-4}\,. 
    \end{split}
\end{equation}
Here the initial $R_n$ values were chosen such that $n< R_n< n+0.1$\;. The learned final $R_n$ values are given in the table below.
\begin{table}[H]
    \centering
    \begin{tabular}{|c|c|c|c|c|c|c|c|c|c|c|}  
        \hline
        \multicolumn{11}{|c|}{Levels 1, 2, 3 fixed} \\
        \hline
        Initial $R_n$ & 1 & 2 & 3 & 4.0299 & 5.0577 & 6.0710 & 7.0448 & 8.0431 & 9.0471 & 10.0281\\         
        \hline
        Final $R_n$ & 1 & 2 & 3 & 4.0367 & 5.0513 & 6.0768 & 7.0396 & 8.0479 & 9.0428 & 10.0321\\ 
        \hline
    \end{tabular}
    \label{tab:sample3}
\end{table}
\item $R_1=1, R_2=2, R_3=3, R_4=4$\\
For this case we fix the levels $R_1=1$, $R_2=2$, $R_3=3$, $R_4=4$ and learn the rest of the levels from $R_5$ to $R_{10}$. We use $\beta_{10}=10^6$, $\beta_{01}=10^6$, $\beta_{1}=10^8$ and $\beta_2=\beta_3=\beta_4=\beta_{5}=10^6$. We use the Adam optimizer throughout training. This leads to the following solution:
\begin{equation}
    \begin{split}
        \Phi = 1.543 \, &, \quad  \,  W_{10}-(-\zeta(3)) = 6.75 \times 10^{-5} \, , \\
        W_{01}-\frac{7\zeta(4)}{4} = -8.31 \times 10^{-5} \, &, \quad  \, \tilde{\mathcal{L}}_{mean}  = 9.57 \times 10^{-5}\,. 
    \end{split}
\end{equation}
Here the initial $R_n$ values were chosen such that $n< R_n< n+0.4$\;. The learned final $R_n$ values are given in the table below.
\begin{table}[H]
    \centering
    \begin{tabular}{|c|c|c|c|c|c|c|c|c|c|c|}  
        \hline
        \multicolumn{11}{|c|}{Levels 1, 2, 3, 4 fixed} \\
        \hline
        Initial $R_n$ & 1 & 2 & 3 & 4 & 5.3870 & 6.0412 & 7.3920 & 8.2010 & 9.1628 & 10.1048\\         
        \hline
        Final $R_n$ & 1 & 2 & 3 & 4 & 5.0781 & 6.2032 & 7.2887 & 8.2587 & 9.1089 & 10.1495\\ 
        \hline
    \end{tabular}
    \label{tab:sample4}
\end{table}

Notice that the machine learning simulations performed here require at least the first mass-squared level fixed to its open string value that is $R_1=1$ for the entanglement objective to converge to a value that is approximately close to the value $\Phi_{\text{string}}=1.5997$ \cite{Bhat:2024agd} corresponding to the open string solution and for the loss function to converge to the same order of magnitude as the entanglement objective so that the constraints are satisfied up to the required tolerance. We have not observed convergence to the open string solution when $R_1$ is left free. We perform separate optimization runs with the first one, two, three and four mass-squared levels fixed to their open string values, respectively: \(R_n=n\), with \(n=1,2,3,4\). In all four configurations, the entanglement objective converges approximately close to the open string value, while the remaining unfixed levels are optimized. However, the remaining unfixed levels do not approach the open string spectrum uniformly. For example, when only $R_1$ is fixed the levels $R_2,R_3,R_5,R_7,R_9,R_{10}$ improve and move closer to the integer values whereas the levels $R_4,R_6,R_8$ worsen and move away from the integer values. When $R_1, R_2$ are fixed the levels $R_3,R_5,R_7,R_8,R_9$ improve whereas the levels $R_4,R_6,R_{10}$ worsen. For our neural network architecture, the level $R_1$ is the most important. Within a specified finite meromorphic ansatz, with the first pole used to fix the mass scale and with a clearly stated spin template, $\lambda$-CSDR consistency and positive residues numerically favor approximately equally spaced mass-squared levels. However, the present optimization runs do not establish convergence of the unfixed levels to the exact open string integer values. It would be interesting to investigate whether a different neural network architecture could recover the open string integer mass-squared spectrum without fixing any initial level. This use of the first pole to fix the mass scale is similar to \cite{Wan:2026pjq} where splitting relations were used to uniquely analytically bootstrap the open string mass-squared spectrum. Let us look carefully at the entanglement objective $\Phi$. We will compare the converged solutions with \cite{Bhat:2024agd} where $\Phi_{\text{string}}=1.5997$ was obtained using linear programming with the spectrum fixed to the open string spectrum \eqref{OSspectrum}. In the first two cases when we fix $R_1$ and $R_1,R_2$ respectively we find that the converged solution has $\Phi\simeq\Phi_{\text{string}}$. But in the final two cases when we fix $R_1, R_2, R_3$ and $R_1, R_2, R_3, R_4$ respectively we find that $\Phi$ is slightly smaller than $\Phi_{\text{string}}$. This shows that in the first two cases when $R_1$ and $R_1,R_2$ are fixed, $\Phi\simeq\Phi_{\text{string}}$ because the final learned levels are approximately integer valued and approximately equispaced. As we fix more and more mass levels to integer values the open string solution obtained using machine learning has a converged entanglement objective $\Phi$ slightly smaller than $\Phi_{\text{string}}$.

In \cite{Gumus:2024lmj, Gumus:2026mhb} 2-2 scattering of identical scalar particles was studied. In \cite{Gumus:2024lmj} the double discontinuity of the amplitude was set to zero and a fully-connected neural network was used to learn the single discontinuity of the amplitude. The elastic unitarity condition was imposed using a loss function to obtain bounds on the leading Wilson coefficients of the amplitude. In \cite{Gumus:2026mhb} the double discontinuity of the amplitude is non-zero. Two different neural networks were used to learn both the single discontinuity and the double discontinuity of the amplitude. The neural network architectures in these works are similar to the approach used in this paper where we have used a fully-connected feed-forward neural network to learn the $c_\ell^{(R_n)}$. But we do not use skip connections which means that our neural network has a uniform width of 128 for the 3 hidden layers. Furthermore, we not only optimize the weights and biases of the neural network but also optimize the mass-squared spectrum values $R_n$ which means we had to build up a fully \textit{PyTorch} autograd compatible loss function including both the parameters of the neural network as well as our functional ansatz consisting of Gegenbauer polynomials and its derivatives. In \cite{Gumus:2024lmj, Gumus:2026mhb} the activation function $\text{CELU}(x) = max(0,x)+min(0,\alpha\,\text{exp}(\alpha \, x)-1)$ was used. For the problem stated in this paper we have used the ReLU activation function $\text{ReLU}(x) = max(0,x)$.
\end{itemize}
\subsection{Linear Programming}
\begin{figure}[H]
    \centering
    \begin{subfigure}{0.49\textwidth}
        \includegraphics[width=8cm]{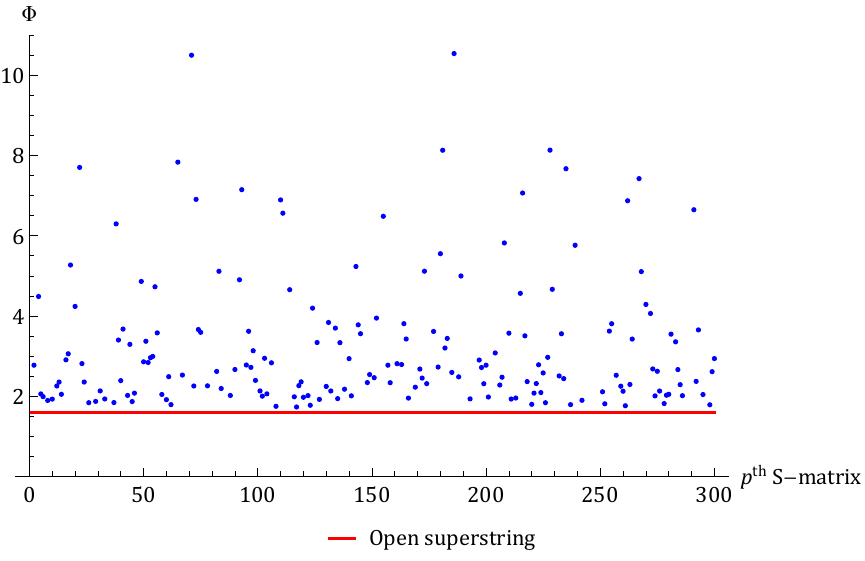}
        \caption{$(W_{10}, W_{01})=(-\zeta(3),\frac{7}{4}\zeta(4))$.}
        \label{fig:phirandOS}
    \end{subfigure}
    \begin{subfigure}{0.49\textwidth}
        \includegraphics[width=8cm]{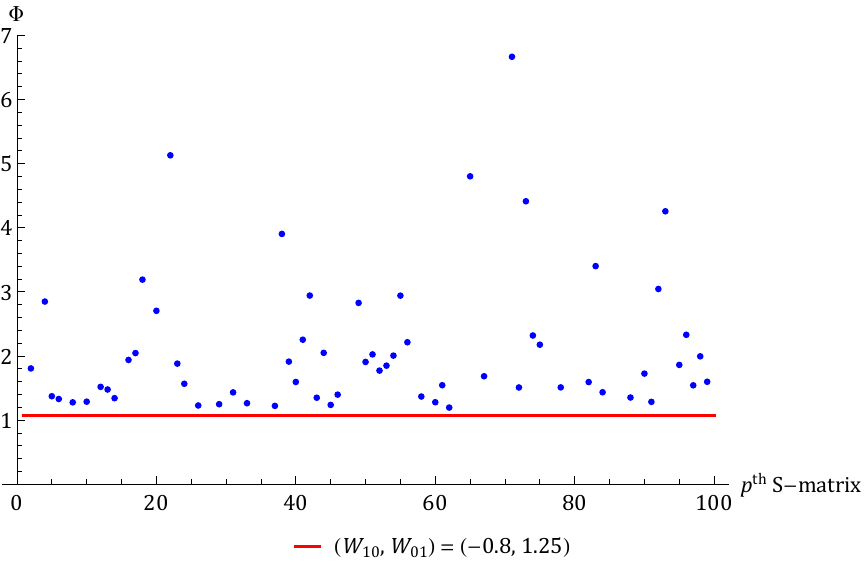}
        \caption{$(W_{10}, W_{01})=(-0.8,1.25)$.}
        \label{fig:phirandSL}
    \end{subfigure}
    \caption{We plot the $\Phi$ values generated after minimization for the open string point with $(W_{10}, W_{01})=(-\zeta(3),\frac{7}{4}\zeta(4))$ with 187 different general non-integer mass-squared spectra and compare with the $\Phi$ generated after minimization for the same point using the open string spectrum \eqref{OSspectrum}. We perform the same exercise for a different point located on the line of hypergeometric deformations of the open string amplitude namely $(W_{10},W_{01})=(-0.8,1.25)$ with 63 different general non-integer mass-squared spectra. In both the cases we find that the $\Phi$ values generated using general non-integer mass-squared spectra is always higher than the $\Phi$ values generated using the open string spectrum \eqref{OSspectrum}.} 
   \label{fig:phirandOSSL}
\end{figure}

Here we will show how linear programming helps us to show that the open string mass-squared spectrum is integer valued. First we will show that if we use a set of general non-integer $R_n$ then the entanglement generated in this process is always higher than the entanglement generated if we had used an integer mass-squared spectrum. For this we will refer to figure 6 of \cite{Bhat:2024agd}. Let us consider the open superstring point in this figure. We build the $(\ell,R_n)$ pairs as follows. At the level $R_n$ the following $\ell$ contribute: 
\begin{equation}
\label{spectrumsdpb}
    0\leq\ell\leq \lfloor R_n\rfloor\;.
\end{equation}
Here $\lfloor x \rfloor$ is the floor function that yields the greatest integer less than or equal to $x$. We use the semi-definite programming for bootstrap package SDPB to run our numerical simulations \cite{sdpb}. We use SDPB 2.5.1. For this exercise, we choose all the values of $R_n$ to be non-integer. We choose $N_{max}=30$ that is 30 distinct values for the mass levels $R_1, R_2, R_3,\cdots,R_{30}$ where none of them are integer valued and $0<R_1<R_2<R_3<\cdots<R_{30}<30$. The following code in Mathematica is used here
\begin{equation}
\label{tabn100}
    \texttt{tabn100 = Table[Sort[RandomReal[30, 30]], \{k, 1, 100\}];}
\end{equation}
which yields 100 lists of 30 random real numbers between 0 and 30. Each list thus has 30 $R_n$ values between 0 and 30 which are used as input spectrum to build the list of $(\ell, R_n)$ pairs according to \eqref{spectrumsdpb} for the linear programming set-up.  
We impose the following constraints
\begin{equation}
\begin{split}
    &W_{10}=-\zeta(3)\;,\qquad W_{01}=\frac{7}{4}\zeta(4)\;,\\&\left|\frac{\partial^k \mathcal{M}(s_1,s_2)}{\partial\lambda^k} \right| \leq \epsilon, \quad \text{for }k =1,2,\cdots, k_{max}\;.
\end{split}
\end{equation}
Similar to \cite{Bhat:2024agd} we choose $k_{max}=6$, $\lambda=14.6$ and $\epsilon=10^{-9}$. We perform the numerical simulations in $D=10$. The SDPB \texttt{dualityGapThreshold} is set to $10^{-15}$. The derivative constraints are imposed on a grid of 180 ordered pairs of $(s_1,s_2)$ values as shown in figure 2 of \cite{Bhat:2024agd}. We minimize the entanglement power ($\Phi$) subject to the above constraints 
\begin{equation}
    \Phi=\sum_{(\ell,R_n)\in\text{spectrum}}\frac{1}{R_n}c_\ell^{(R_n)}\mathcal{G}_\ell^{(\frac{D-3}{2})}(1)=-W_{00}\,.
\end{equation}
Our results are shown in figure \ref{fig:phirandOSSL}. In figure \ref{fig:phirandOS} we run \eqref{tabn100} three times, thus creating 300 distinct input spectra. We build the ($\ell,R_n$) pairs in accordance with \eqref{spectrumsdpb}. We minimize $\Phi$ for all 300 different input spectra. We find that after minimization using SDPB the $\Phi$ values for 187 input spectra are higher than $\Phi_{\text{string}}=1.5997$ \cite{Bhat:2024agd} which is the $\Phi$ value after minimization using the open string spectrum \eqref{OSspectrum} and for the other 113 input spectra, SDPB terminates with the status \texttt{maxComplementarity}. This means that for these 113 input spectra, the linear program could not satisfy all the constraints up to the desired tolerance $\epsilon=10^{-9}$ and are therefore excluded from quantitative comparison. These results provide strong evidence that the open string spectrum \eqref{OSspectrum} generates lower entanglement compared to general non-integer mass-squared spectra. We have also minimized $\Phi$ for an integer mass-squared spectrum $n$ where all spins $0\leq\ell\leq n$ contribute. This spectrum is the same as \eqref{spectrumsdpb} but for integer $n$. The $\Phi$ generated for this spectrum is $\Phi=1.5551$\,. 

Next, we repeat the same exercise with another point in figure 6 of \cite{Bhat:2024agd}, namely $(W_{10},W_{01})=(-0.8,1.25)$. This point is located on the black line that denotes the hypergeometric deformations of the open string amplitude \cite{Mansfield:2024wjc}. The results are shown in figure \ref{fig:phirandSL}. For this point, we minimize $\Phi$ for 100 different input spectra created using \eqref{tabn100}. We build the ($\ell,R_n$) pairs in accordance with \eqref{spectrumsdpb}. We find that after minimization using SDPB the $\Phi$ values for 63 input spectra are higher than $\Phi_{(-0.8,1.25)}=1.0745$ which is the $\Phi$ value after minimization using the open string spectrum \eqref{OSspectrum} for the point $(W_{10},W_{01})=(-0.8,1.25)$ and for the other 37 input spectra, SDPB terminates with the status \texttt{maxComplementarity}. Again we see strong evidence that the open string spectrum \eqref{OSspectrum} generates lower entanglement compared to general non-integer mass-squared spectra for the point $(W_{10},W_{01})=(-0.8,1.25)$. If we minimize $\Phi$ for an integer mass-squared spectrum $n$ where all spins $0\leq\ell\leq n$ contribute for this point $(W_{10},W_{01})=(-0.8,1.25)$, then $\Phi=1.0371$\,.

\begin{figure}[H]
    \centering
    \begin{subfigure}{0.47\textwidth}
        \includegraphics[width=8.5cm]{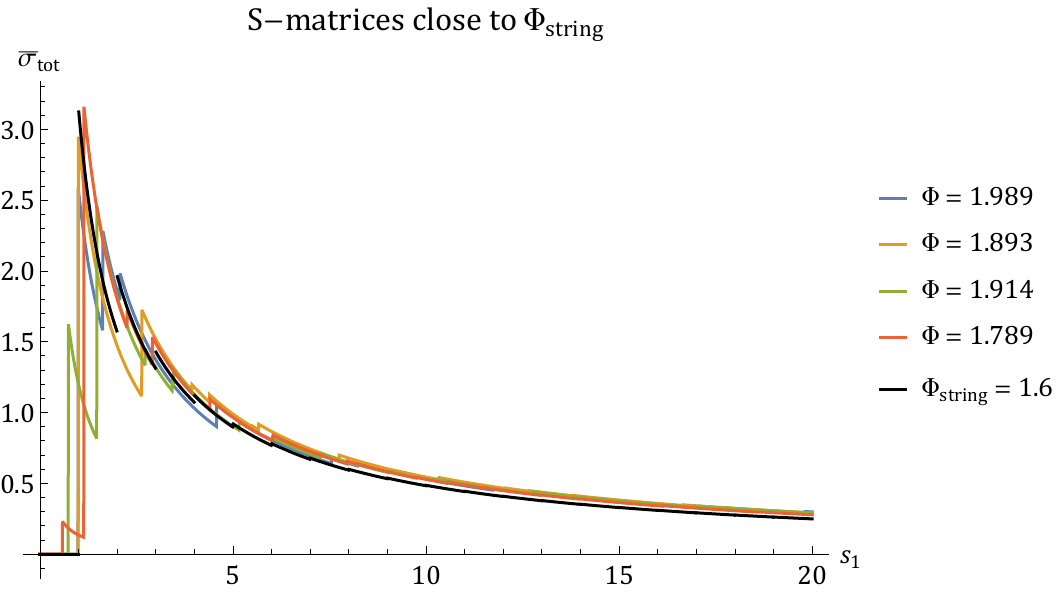}
        \caption{S-matrices with $\Phi$ close to $\Phi_{\text{string}}$.}
        \label{CSnearOS}
    \end{subfigure}
    \hspace{0.5cm}
    \begin{subfigure}{0.47\textwidth}
        \includegraphics[width=8.5cm]{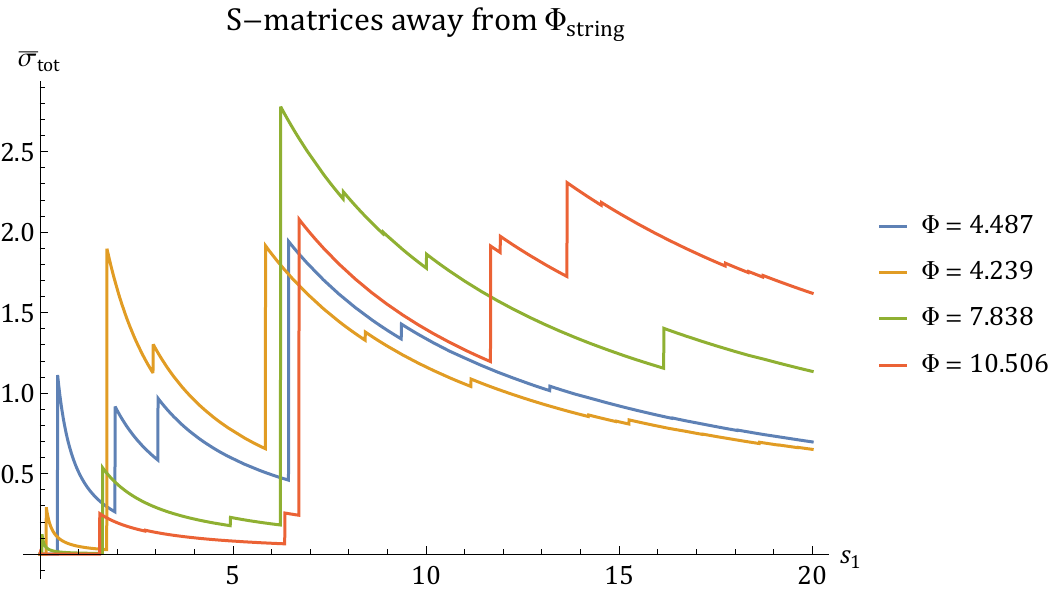}
        \caption{S-matrices with $\Phi$ away from $\Phi_{\text{string}}$.}
        \label{CSawayOS}
    \end{subfigure}
    \caption{Averaged total scattering cross-section ($\bar{\sigma}_{\text{tot}}(s_1)$) vs $s_1$ for S-matrices in figure \ref{fig:phirandOS}.}%
   \label{fig:CS}
\end{figure}

We plot the averaged total scattering cross-section of S-matrices close to and away from the open string S-matrix in terms of the entanglement power ($\Phi$). The averaged total scattering cross-section is given by 
\begin{equation*}
	\bar{\sigma}_{\text{tot}}(s_1)=\frac{1}{s_1}\int_{0}^{s_1}\mathrm{d}s'\frac{\text{Im}\mathcal{M}(s',s_2=0)}{s'}\;.
\end{equation*}
Our results are shown in figure \ref{fig:CS}. In figure \ref{CSnearOS} we plot $\bar{\sigma}_{\text{tot}}$ for four S-matrices in figure \ref{fig:phirandOS} with $\Phi$ slightly higher than $\Phi_{\text{string}}$. We find that the $\bar{\sigma}_{\text{tot}}$ for all four of these S-matrices are smoothly decaying and all four overlap with the $\bar{\sigma}_{\text{tot}}$ of the open string S-matrix. In figure \ref{CSawayOS} we plot $\bar{\sigma}_{\text{tot}}$ for four S-matrices in figure \ref{fig:phirandOS} with $\Phi$ much higher than $\Phi_{\text{string}}$. We find that the $\bar{\sigma}_{\text{tot}}$ for all four of these S-matrices do not show smooth behavior. 

Furthermore, it is observed that if we minimize $\Phi$ using the spin parameters in 
\begin{equation}
\label{spectrumsdpbfloor}
\begin{split}
    0\leq\ell\leq \lfloor R_n\rfloor-1 \text{ and } \lfloor R_n\rfloor-\ell\equiv 1 \text{ mod } 2
\end{split}
\end{equation}
instead of \eqref{spectrumsdpb} and use the same tolerance of $\epsilon=10^{-9}$ for the $W_{10}$, $W_{01}$ and $\lambda$-independence constraints then for this case SDPB terminates with the status \texttt{maxComplementarity}. This clearly means that a large number of spin parameters are required to satisfy all constraints up to the desired tolerance of $\epsilon=10^{-9}$ for a non-integer mass-squared spectrum $R_n$. These extra spin parameters in the ansatz are supplied by \eqref{spectrumsdpb} and not \eqref{spectrumsdpbfloor}. Finally, we have checked that the linear programming results do not change when we decrease $N_{max}$ to 25 or increase $N_{max}$ to $35$ without changing the number of $\lambda$-independence constraints. We have also checked that these results do not change when we change the number of $\lambda$-independence constraints from 180 to 210 ordered pairs of ($s_1,s_2$) values while fixing $N_{max}=30$.

An important difference between the machine learning and the linear programming approaches is that in the ML approach the tolerance ($\epsilon$) that we need the constraints to satisfy is set by the weights $(\beta s)$ in the loss function. This tolerance is achieved by the constraints as the loss gradually converges to the open string solution and the $\beta$s help in pushing the constraints to the required tolerance for the machine learning algorithm. On the other hand in the linear programming approach the desired tolerance ($\epsilon=10^{-9}$) is preset in the linear matrices that are used to impose the constraints.

\begin{figure}[H]
    \centering
    \begin{subfigure}{0.32\textwidth}
        \includegraphics[width=5.3cm]{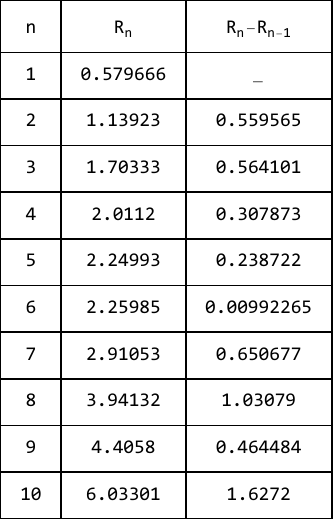}
        \caption{}
        \label{nRnDifftabn100R621to10}
    \end{subfigure}
    \begin{subfigure}{0.3\textwidth}
        \includegraphics[width=5cm]{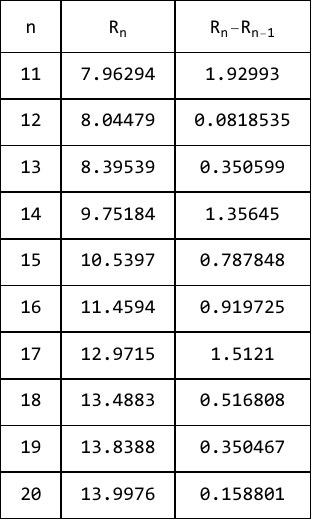}
        \caption{}
        \label{nRnDifftabn100R6211to20}
    \end{subfigure}
    \begin{subfigure}{0.3\textwidth}
        \includegraphics[width=5cm]{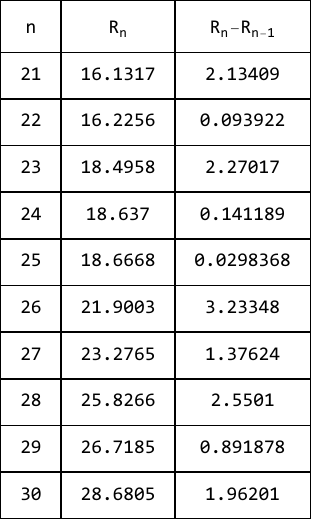}
        \caption{}
        \label{nRnDifftabn100R6221to30}
    \end{subfigure}
    \caption{$R_n-R_{n-1}$ for the S-matrix with $\Phi=1.789$ in figure \ref{CSnearOS}.}%
   \label{fig:nRnDifftabn100R621to30}
\end{figure}

Now let us reconcile the machine learning and the linear programming approaches. In the machine learning simulations the first level $R_1$ is fixed to its corresponding open string integer value $R_1=1$ whereas the rest of the levels are optimized using gradient descent. From the machine learning simulations we find that the learned levels are approximately integer valued but not exactly integer valued. Furthermore, the unfixed levels do not converge to the open string integer values in a uniform manner. Some of the unfixed levels improve and move closer to the integer values whereas the rest of the them worsen and move away from the integer values. The spacing between two consecutive learned levels is approximately close to one but not exactly one. The entanglement objective $\Phi$ also converges approximately close to $\Phi_{\text{string}}=1.5997$. All of these observations suggest that using machine learning the open string integer mass-squared spectrum yields $\Phi\simeq \Phi_{\text{string}}$ \cite{Bhat:2024agd} but it is possible to have other spectra that are not integer valued that have $\Phi\simeq\Phi_{\text{string}}$. In the linear programming approach we find similar conclusions. Consider figure \ref{fig:phirandOSSL}. In figure \ref{fig:phirandOS} we clearly see that the open string spectrum generates the lowest entanglement compared to 187 sampled non-integer spectra for the open string point $(W_{10},W_{01})=(-\zeta(3),\frac{7}{4}\zeta(4))$. But if we look closely we find a few S-matrices lying very close to $\Phi_{\text{string}}$. For example, consider the S-matrix with $\Phi=1.789$ in figure \ref{CSnearOS}. This S-matrix has $\Phi$ slightly higher than $\Phi_{\text{string}}$. Let us look carefully at the mass-squared spectrum of this S-matrix. In figure \ref{fig:nRnDifftabn100R621to30} we provide the $R_n$ values for this S-matrix. First of all none of the $R_n$ values are equal to $n$ which is a result of \eqref{tabn100}. Furthermore, the difference between consecutive $R_n$ values that is $R_n-R_{n-1}$ are not equal to one another. Rather, they are all very different from one another. We conclude that it is possible to have low values of $\Phi$ very close to $\Phi_{\text{string}}$ with non-integer $R_n$ which are not equally spaced. However, the open string integer spectrum does generate the lowest entanglement when compared with all of the 187 sampled non-integer spectra. We think this will remain true if we increase the sample space. This remains open to future investigations. If we consider the S-matrices in figure \ref{fig:phirandSL}, we find similar conclusions. 
\pagebreak
\section{Bounds on three-channel crossing symmetric scattering amplitudes}
\label{Closed String}
\begin{figure}[H]
    \centering
    \includegraphics[width=0.8\linewidth]{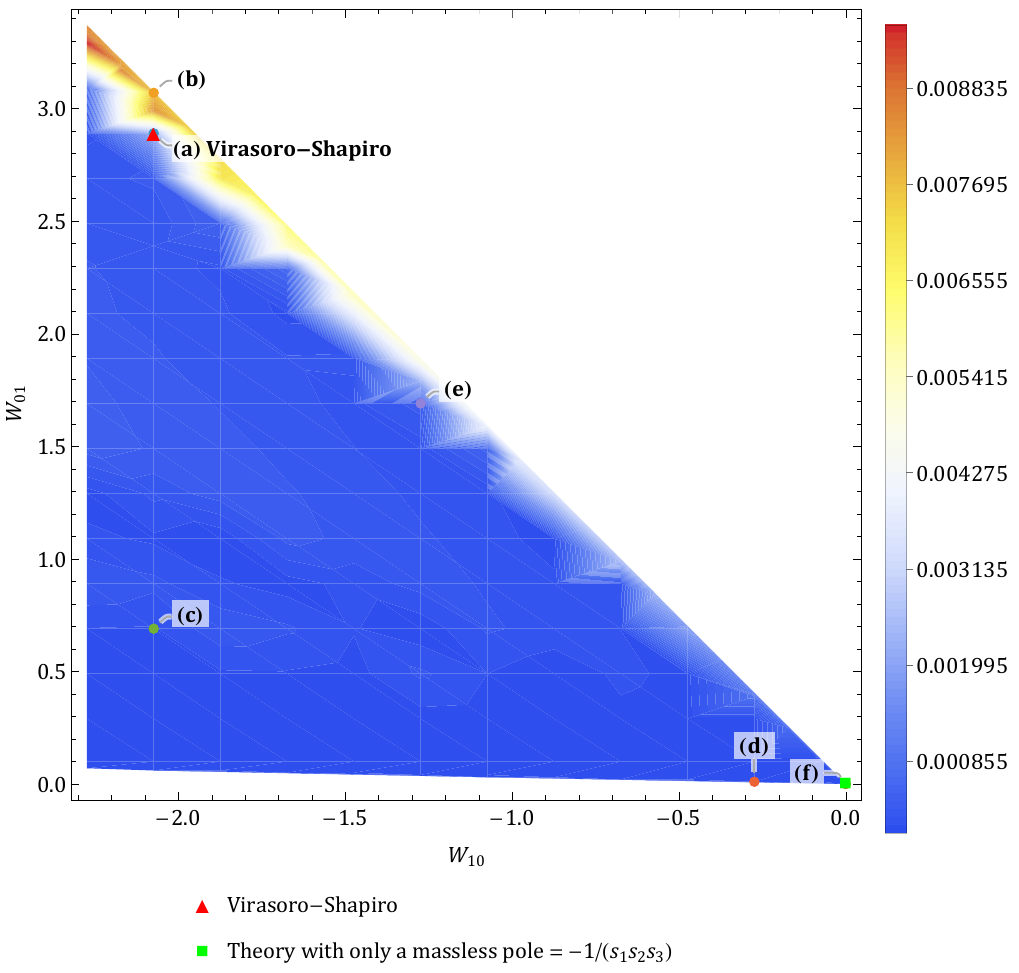}
    \caption{Allowed region in the space of $(W_{10},W_{01})$ for three channel crossing symmetric stringy amplitudes. The closed string amplitude has been marked with a red triangle.}
    \label{fig:ClosedStringContourPlot}
\end{figure}

\begin{figure}[H]
    \centering
    \includegraphics[width=0.55\linewidth]{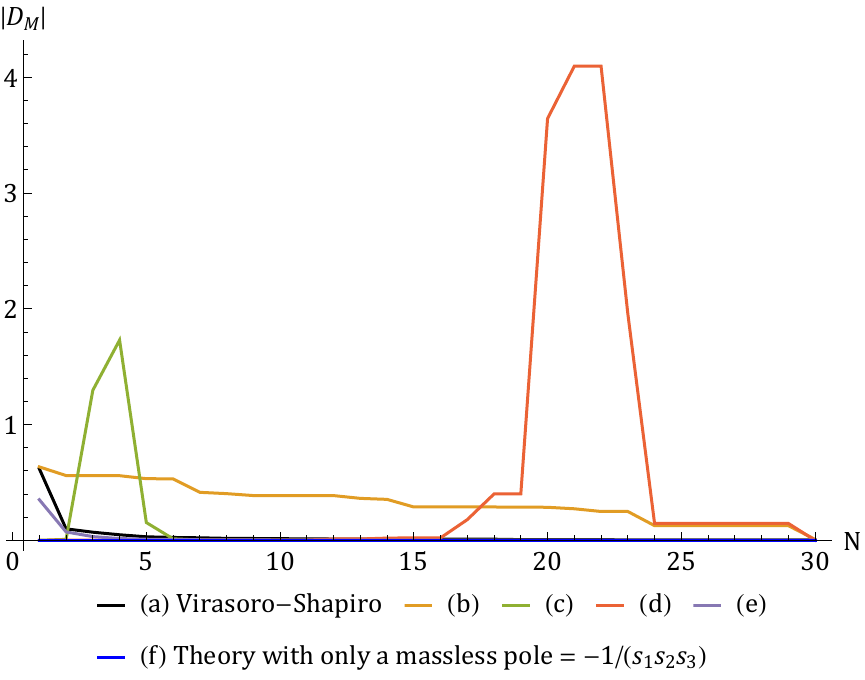}
    \caption{Convergence of the partial sums of $|D_M|$ as a function of $N$ for the S-matrices denoted by (a) to (f) in figure \ref{fig:ClosedStringContourPlot}.}
    \label{fig:ClosedStringPS}
\end{figure}

Let us now look at amplitudes with three-channel crossing symmetry.
For an amplitude that satisfies $\lim_{s_1\rightarrow\infty}|\mathcal{M}(s_1,s_2,s_3)|=0$ for some domain of $s_2\in\mathcal{D}$ and has discontinuities (poles and branch cuts) along some contour $\mathcal{C}_1$, the three-channel local crossing symmetric dispersion relation with one-parameter ambiguity was derived in \cite{Bhat:2025zex} and is given by
\begin{equation}
\label{3channellamCSDR}
    \mathcal{M}(s_1,s_2,s_3)=\frac{1}{\pi}\int_{\mathcal{C}_1}\mathrm{d\sigma}\;\mathcal{H}^{(\lambda)}(\s,s_1,s_2,s_3)\mathcal{A}^{(s_1)}(\s,\hat{s}_2^{(\lambda)}(\s,s_1,s_2,s_3),-\s-\hat{s}_2^{(\lambda)}(\s,s_1,s_2,s_3))
\end{equation}
for all $s_1, s_2$ and $-\lambda\in \mathcal{D}$. Here, $\mathcal{A}^{(s_1)}(s_1,s_2,s_3)$ is the imaginary part of the amplitude across $\mathcal{C}_1$ and
\begin{equation}
\begin{split}
    &\mathcal{H}^{(\lambda)}(s_1,s_2,s_3)=\frac{1}{\s-s_1}+\frac{1}{\s-s_2}+\frac{1}{\s-s_3}-\frac{1}{\s+\l}\;,\\
    &\hat{s}_2^{(\lambda)}(\s,s_1,s_2,s_3)=-\frac{\s}{2}\pm\sqrt{\frac{y+\lambda x}{\s+\l}+\frac{\s^3-3\l\s^2}{4(\s+\l)}}\;,\\&x=-(s_1s_2+s_2s_3+s_3s_1)\;,\quad y=-s_1s_2s_3\;,\\& s_1+s_2+s_3=0\;.
\end{split}
\end{equation}
We will use linear programming to carve out the space of three-channel crossing symmetric scattering amplitudes. The analytic expression for the meromorphic form factor of the amplitude for $2\to2$ scattering of gravitons in Type II superstring theory is obtained after stripping off factor of $-\kappa^2\mathcal{R}^4$ from \eqref{graviton} and is given by the Virasoro-Shapiro amplitude for graviton scattering
\begin{equation}
\label{VS}
    \mathcal{M}(s_1,s_2,s_3)=\frac{\G(-s_1)\G(-s_2)\G(-s_3)}{\Gamma(1+s_1)\Gamma(1+s_2)\Gamma(1+s_3)}\;.
\end{equation}
This amplitude has simple poles for $s_1,s_2,s_3\in\mathbb{Z}_{\geq0}$. After subtracting out the massless pole from this amplitude, we can expand the imaginary part in terms of partial waves 
\begin{equation}
\label{closedstringabs}
    \mathcal{A}^{(s_1)}(s_1,s_2)=-\pi\sum_{n=1}^{\infty}\sum_{\substack{\ell=0 \\ \ell \text{ even}}}^{2n-2}c_\ell^{(n)}\mathcal{G_\ell}^{\left(\frac{D-3}{2}\right)}\left(1+\frac{2s_2}{s_1}\right)\delta(s_1-n)\;.
\end{equation}
The spectrum of the Virasoro-Shapiro amplitude is such that at the level $n$, the following $\ell$ contribute:
\begin{equation}
\label{closedstringspec}
\begin{split}
 0 \leq\ell\leq 2n-2 \text{ and } \ell \text{ even}\;.
\end{split}
\end{equation}
Substituting this into \eqref{3channellamCSDR} we obtain the following
\begin{equation}
\label{3channellamCSDRsum}
\begin{split}
    \mathcal{M}(s_1,s_2,s_3)=-\frac{1}{s_1s_2s_3}+\sum_{n=1}^{\infty}\sum_{\substack{\ell=0 \\ \ell \text{ even}}}^{2n-2}&\left[\frac{1}{s_1-n}+\frac{1}{s_2-n}+\frac{1}{s_3-n}+\frac{1}{\l+n}\right]\\&\times c_\ell^{(n)}\mathcal{G}_\ell^{\left(\frac{D-3}{2}\right)}\left(\pm\frac{2}{n}\sqrt{\frac{y+\l x}{n+\l}+\frac{n^3-3\l n^2}{4(n+\l)}}\right)\;.
\end{split}
\end{equation}
We can use either sign in the argument of the Gegenbauer since only even $\ell$ contribute. For three-channel crossing symmetric scattering amplitudes that satisfy $\lim_{s_1\rightarrow\infty}|\mathcal{M}(s_1,s_2,s_3)|=0$ for some domain of $s_2\in\mathcal{D}$ and has discontinuities along $\mathcal{C}_1$ the fixed-$s_2$ dispersion relation is given by
\begin{equation}
\label{3channelfixedt}
    \mathcal{M}(s_1,s_2,s_3)=-\frac{1}{s_1s_2s_3}+\frac{1}{\pi}\int_{\mathcal{C}_1}\mathrm{d}\sigma\left[\frac{1}{\sigma-s_1}+\frac{1}{\sigma-s_3}\right]\mathcal{A}^{(s_1)}(\s,s_2,-\sigma-s_2)\;.
\end{equation}
Substituting \eqref{closedstringabs} into \eqref{3channelfixedt} we obtain
\begin{equation}
\label{3channelfixedtsum}
    \mathcal{M}(s_1,s_2,s_3)=-\frac{1}{s_1s_2s_3}+\sum_{n=1}^{\infty}\sum_{\substack{\ell=0 \\ \ell \text{ even}}}^{2n-2}\left[\frac{1}{s_1-n}+\frac{1}{s_3-n}\right]c_\ell^{(n)}\mathcal{G}_\ell^{(\frac{D-3}{2})}\left(1+\frac{2s_2}{n}\right)\;.
\end{equation}
After subtracting out the massless pole and setting $s_1=s_2=s_3=0$ in \eqref{3channellamCSDRsum} and \eqref{3channelfixedtsum} we obtain two different expressions for $W_{00}$ which are
\begin{align}
\label{w00lcsdr}
    \lambda\text{-CSDR}:&\qquad W_{00}=\sum_{n=1}^{\infty}\sum_{\substack{\ell=0 \\ \ell \text{ even}}}^{2n-2}\left[-\frac{3}{n}+\frac{1}{\lambda+n}\right]c_\ell^{(n)}\mathcal{G}_\ell^{\left(\frac{D-3}{2}\right)}\left(\sqrt{\frac{n-3\lambda}{n+\lambda}}\right)\;,\\
\label{w00fixeds2}
    \text{fixed-}s_2:&\qquad W_{00}=-\sum_{n=1}^{\infty}\sum_{\substack{\ell=0 \\ \ell \text{ even}}}^{2n-2}\frac{2}{n}c_\ell^{(n)}\mathcal{G}_\ell^{\left(\frac{D-3}{2}\right)}(1)\;.
\end{align}
Notice that the expression for $W_{00}$ using the fixed-$s_2$ dispersion relation we have
\begin{equation}
    W_{00}=-2\Phi
\end{equation}
where $\Phi$ is the fourth moment of the linear entropy generated in tree-level four-point graviton scattering \eqref{EPintgravitons4}. Below we also give the formulas for $W_{10}$ and $W_{01}$ obtained from \eqref{3channellamCSDRsum}. 

\begin{equation}
\label{W10W01}
\begin{split}
    &W_{10}=-\sum_{n=1}^{\infty}\sum_{\substack{\ell=0 \\ \ell \text{ even}}}^{2n-2}\frac{2c_\ell^{(n)}}{n^3}\left\{\mathcal{G}_\ell^{\left(\frac{D-3}{2}\right)}\left(\sqrt{\frac{n-3\lambda}{n+\lambda}}\right)+\frac{(D-3)\lambda(2n+3\lambda)}{\sqrt{(n-3\lambda)(n+\lambda)^3}}\mathcal{G}_{\ell-1}^{\left(\frac{D-1}{2}\right)}\left(\sqrt{\frac{n-3\lambda}{n+\lambda}}\right)\right\}\;,\\
    &W_{01}=\sum_{n=1}^{\infty}\sum_{\substack{\ell=0 \\ \ell \text{ even}}}^{2n-2}\frac{2c_\ell^{(n)}}{n^3}\left\{\frac{3}{2n}\mathcal{G}_\ell^{\left(\frac{D-3}{2}\right)}\left(\sqrt{\frac{n-3\lambda}{n+\lambda}}\right)-\frac{(D-3)(2n+3\lambda)}{\sqrt{(n-3\lambda)(n+\lambda)^3}}\mathcal{G}_{\ell-1}^{\left(\frac{D-1}{2}\right)}\left(\sqrt{\frac{n-3\lambda}{n+\lambda}}\right)\right\}\;.
\end{split}
\end{equation}
The $\mathcal{G}_{\ell-1}$ terms contribute from $\ell=2$ onward. Using \eqref{W10W01} we notice that 
\begin{equation}
    W_{10}-\lambda W_{01}=-\sum_{n=1}^{\infty}\sum_{\substack{\ell=0 \\ \ell \text{ even}}}^{2n-2}\frac{2c_\ell^{(n)}}{n^3}\left(1+\frac{3\lambda}{2n}\right)\mathcal{G}_\ell^{\left(\frac{D-3}{2}\right)}\left(\sqrt{\frac{n-3\lambda}{n+\lambda}}\right)\;.
\end{equation}
For even $\ell$, the Gegenbauer polynomial is always positive when its argument is real and $\text{argument}\geq 1$. Now, $\sqrt{\frac{n-3\lambda}{n+\lambda}}$ is real and $\sqrt{\frac{n-3\lambda}{n+\lambda}}\geq 1$ for $-1<\lambda\leq0$. $\left(1+\frac{3\lambda}{2n}\right)\geq0$ for $-\frac{2}{3}\leq\lambda\leq 0$. The upper limit $\lambda\to0$ gives the bound $W_{10}\leq 0$ and the lower limit $\lambda\to -\frac{2}{3}$ gives the rigorous upper bound 
\begin{equation}
    \label{W10W01bound}
    W_{01}\leq-\frac{3}{2}W_{10}\;.
\end{equation}
Using the local 3-channel $\lambda$-CSDR we obtain \eqref{W10W01bound}, to be compared with the CSDR bound of \cite{Raman:2021pkf}. \cite{Raman:2021pkf} derive an explicit two-sided bound $ -\frac{9}{16} \leq \frac{W_{01}}{W_{10}}\leq \frac{9}{8} $ by the same technique used here: tune the CSDR ambiguity parameter over the range where the Gegenbauers stay positive, with low-spin dominance fixing the boundary. We will discuss spin-0 dominance and reconcile these two approaches to obtain the bounds in detail in \ref{spin-0 dominance}. In \cite{Bhat:2024agd} the Virasoro-Shapiro amplitude was obtained using linear programming for minimization of both \eqref{EPintgravitons3} and \eqref{EPintgravitons4} after fixing $W_{10}=-2\zeta(5)$ and $W_{01}=2\zeta(3)^2$ which are the values of $W_{10}$ and $W_{01}$ for the Virasoro-Shapiro amplitude. In this paper we will carve out the entire allowed space of three-channel crossing symmetric scattering amplitudes that satisfy \eqref{3channellamCSDR} in the $(W_{10},W_{01})$ space. We will locate the Virasoro-Shapiro amplitude in this space. We use linear programming \cite{sdpb} to achieve this. The input spectrum is fixed to \eqref{closedstringspec}. We truncate the ansatz in \eqref{3channellamCSDRsum} up to $N_{max}=30$. We use $\lambda=1.6$ as the seed value in the ansatz \eqref{3channellamCSDRsum}. We impose 
\begin{equation}
    \left|\frac{\partial^k\mathcal{M}(s_1,s_2,s_3)}{\partial\lambda^k}\right|\leq\epsilon\;,\qquad k=1,2,\cdots,k_{max}\;.
\end{equation}
We choose $k_{max}=6$ and $\epsilon=10^{-6}$. We impose the $\lambda$-independence constraints on the same grid of $(s_1,s_2)$ values as in figure 2 of \cite{Bhat:2024agd}. For amplitudes satisfying the dual resonance condition \cite{Bhat:2024agd,DHS,Green:1987sp}, \eqref{w00lcsdr} and \eqref{w00fixeds2} must be equivalent. Thus we can define a duality measure for three-channel crossing symmetric scattering amplitudes
\begin{equation}
\label{DM}
    D_{M}\equiv\sum_{n=1}^{\infty}\sum_{\substack{\ell=0 \\ \ell \text{ even}}}^{2n-2}c_\ell^{(n)}\left\{\frac{2}{n}\mathcal{G}_\ell^{\left(\frac{D-3}{2}\right)}(1)+\left(-\frac{3}{n}+\frac{1}{\lambda+n}\right)\mathcal{G}_\ell^{\left(\frac{D-3}{2}\right)}\left(\sqrt{\frac{n-3\lambda}{n+\lambda}}\right)\right\}=0\;.
\end{equation}
It is clear that $D_M=0$ identically for $\lambda=0$. For $\lambda\neq0$, $D_M=0$ for amplitudes that satisfy the dual resonance condition. However, for amplitudes that do not satisfy the dual resonance condition $D_M\neq 0$. For $\lambda\neq0$, $D_M=0$ only if infinite higher masses and spins are included in the sum. We will discuss this in detail in \ref{spin-0 dominance}. In figure \ref{fig:ClosedStringContourPlot} we show the allowed region of three-channel crossing symmetric scattering amplitudes in the $(W_{10},W_{01})$ space. The shaded region is the allowed region whereas the white region is the disallowed region. The color coding indicates the value of $|D_M|$. Notice that the analytic bound in \eqref{W10W01bound} is satisfied by the upper boundary of the allowed region in the $(W_{10}, W_{01})$ space in figure \ref{fig:ClosedStringContourPlot}. The Virasoro-Shapiro marked with a red triangle sits at $W_{01}/W_{10} = -\zeta(3)^2/\zeta(5) \simeq -1.394$, inside the $-3/2$ edge. We see that for almost all of the S-matrices in figure \ref{fig:ClosedStringContourPlot} the dual resonance condition is satisfied very well. Next we plot the partial sums of $|D_M|$ as a function of $N$ in figure \ref{fig:ClosedStringPS} for various amplitudes which are marked in figure \ref{fig:ClosedStringContourPlot}. Here $N$ is the number of terms included in the $n$-sum in \eqref{DM}. For the S-matrices on the lower boundary of the allowed region we observe that the partial sums of $|D_M|$ peak when we cut off the sum around $N=20$ to $N=25$ and then quickly drops to zero when all the terms in the sum are included at $N=N_{max}=30$. For the S-matrices on the upper boundary we observe that the partial sums of $|D_M|$ gradually fall off to zero as we increase $N$. 

\subsection{Spin-0 dominance and the existence of infinite higher masses and spins}
\label{spin-0 dominance}
We have discussed about the upper boundary in figure \ref{fig:ClosedStringContourPlot} and that it respects the upper bound $W_{01}\leq -\frac{3}{2}W_{10}$. Now let us discuss about the lower boundary. We can see from figure \ref{fig:ClosedStringContourPlot} that $W_{01}>0$. We would like to know the reason for this. Taking derivative of $D_M$ with respect to $\lambda$ and then setting $\lambda=0$ we get
\begin{equation}
    \frac{\partial D_M}{\partial\lambda}\bigg|_{\lambda=0}=\sum_{n=1}^{\infty}\sum_{\substack{\ell=0 \\ \ell \text{ even}}}^{2n-2}\frac{c_\ell^{(n)}}{n^2}\left\{4(D-3)\mathcal{G}_{\ell-1}^{\left(\frac{D-1}{2}\right)}(1)-\mathcal{G}_\ell^{\left(\frac{D-3}{2}\right)}(1)\right\}=0\;.
\end{equation}
The $\mathcal{G}_{\ell-1}$ terms contribute from $\ell=2$ onward. This equation can be written as a sum of two parts: one part is the spin-0 contribution and another part is the contribution from higher spins:
\begin{equation}
	\frac{\partial D_M}{\partial \lambda}\bigg|_{\lambda=0}=\underbrace{-\sum_{n=1}^{\infty}\frac{c_0^{(n)}}{n^2}}_{\text{spin-0}}+\underbrace{
	\sum_{n=2}^{\infty}\sum_{\substack{\ell=2 \\ \ell \text{ even}}}^{2n-2}\frac{c_\ell^{(n)}}{n^2}\left\{4(D-3)\mathcal{G}_{\ell-1}^{\left(\frac{D-1}{2}\right)}\left(1\right)-\mathcal{G}_{\ell}^{\left(\frac{D-3}{2}\right)}\left(1\right)\right\}}_{\text{higher spins}} = 0\;.
\end{equation} 
The spin-0 contribution is always negative and the contribution from higher spins is always positive. This means that spin-0 must exist and that the spin-0 contribution must be greater than the contribution from any particular spin in the higher spin sum. This is called spin-0 dominance. Assuming exact \(\lambda\)-independence and that the sum converges well to expand it term by term around $\lambda=0$ we find that the spectrum must contain unbounded spins. The Taylor coefficient of $D_M$ at the $p$th order around $\lambda=0$ is given by $\frac{1}{p!}\frac{\partial^p D_M}{\partial\lambda^p}\Big|_{\lambda=0}$. For any finite spin cutoff, sufficiently high-order Taylor coefficients of all retained spin contributions have the same sign. These contributions cannot be canceled because $c_\ell^{(n)}\geq 0$ contradicting $\frac{\partial^p D_M}{\partial\lambda^p}\Big|_{\lambda=0}=0$. The relation \(\ell\le2n-2\) then also requires unbounded masses. 

The same conclusion holds true for $W_{01}$ as well. From \eqref{W10W01} we find that at $\lambda=0$, the spin-0 contribution is positive and all other spins contribute negatively. Therefore, spin-0 dominance would mean $W_{01}>0$ which is exactly what we find in figure \ref{fig:ClosedStringContourPlot}. Let us check this numerically. Consider the S-matrix marked as (d) in figure \ref{fig:ClosedStringContourPlot}. This S-matrix lies exactly on the lower boundary of the allowed region. Numerically we find that 
\begin{equation}
    \frac{\left(W_{01}\right)_{\text{spin-0}}}{\left(W_{01}\right)_{\text{higher spins}}}=-2.784
\end{equation}
which is a check of spin-0 dominance on the lower boundary. We have checked that spin-0 dominance holds true for all the S-matrices of figure \ref{fig:ClosedStringContourPlot}. 

This paper uses the local three-channel $\lambda$-CSDR of \cite{Bhat:2025zex}, not the original CSDR of \cite{Raman:2021pkf}. The positivity window is $\lambda \in [-2/3, 0]$, giving a different carving: $-\frac{3}{2} \leq \frac{W_{01}}{W_{10}} < 0$, which excludes $\frac{W_{01}}{W_{10}} > 0$ entirely, unlike \cite{Raman:2021pkf}'s $ -\frac{9}{16} \leq \frac{W_{01}}{W_{10}} \leq \frac{9}{8}$. Its quoted interval uses a dispersive threshold $\Lambda^2=8/3$. After rescaling to unit threshold, the endpoints become  $ -\frac{3}{2} \leq \frac{W_{01}}{W_{10}} \leq 3$. So the bound is a re-derivation via the local $\lambda$-CSDR, compared against \cite{Raman:2021pkf}, with the boundary tied to spin-0 dominance, not a first-time bound. \cite{Tolley:2020gtv, Caron-Huot:2020cmc} bound the same coefficient space. Spin-0 dominance is a sharp instance of
the low-spin dominance already used in \cite{Raman:2021pkf,Chowdhury:2021ynh}.

For a fully crossing-symmetric amplitude without a massless pole, the quadratic coefficient $a_2$ (coefficient of $s^2$) is sign-definite by ordinary positivity. With a massless spin-2 (graviton) $t$-channel pole, it is not. The pole contributes $\sim \frac{-s^2}{M_{Pl}^{D-2}t}$ piece that pushes the leading coefficient to either sign ($M_{Pl}^{D-2}=\frac{1}{8\pi G_N}$). The negative lower edge scales like $\sim\frac{-1}{M_{Pl}^{D-2}M_{Regge}^2}$ that closes only as $M_{Pl} \to \infty$. Here $M_{Regge}^2=1/\alpha'_R$ is inverse of the Regge slope. Consequently $a_2$ takes both signs in the presence of gravity. This is the massless-spin-2-pole
result \cite{Hamada:2018dde,Alberte:2020jsk,Tokuda:2020mlf,Caron-Huot:2021rmr}. This paper presents the bounds $W_{10}\leq 0$ and $0<W_{01}\leq -\frac{3}{2}W_{10}$. Notice that 
\begin{equation}
    M(s,t,u)=-\frac{1}{stu}+W_{00}+W_{10} x+W_{01}y+\cdots\;,\quad x=-(st+tu+us)\;,\;y=-stu\;.
\end{equation}
The closed string amplitude is given by 
\begin{equation}
    \mathcal{M}_{CS}(s,t,u)=-\kappa^2\mathcal{R}^4 M(s,t,u)\;.
\end{equation}
Then $\lim_{t\to 0}\;x=s^2$ and $\lim_{t\to 0}\mathcal{R}^4=s^4$ and the forward contribution becomes
\begin{equation}
    \lim_{t\to0}\mathcal{M}_{CS}(s,t,u)=-\frac{\kappa^2s^2}{t}-\kappa^2W_{00}s^4-\kappa^2W_{10}s^6-\cdots
\end{equation}
Thus, $W_{10}$ controls an $s^6$ contribution and is distinct from the quadratic coefficient affected by the graviton-pole obstruction to ordinary forward positivity. These bounds apply to the reduced crossing-symmetric meromorphic form factor $M$ under the assumed high-energy falloff and positivity of its partial-wave residues. The condition $\lim_{s_1\to\infty}|M| = 0$ is a strong assumption that holds true for the meromorphic form factor excluding Regge-growing graviton amplitudes. For example, in the Virasoro–Shapiro case, \(M\sim s^{2t-2}\), whereas restoring the gravitational kinematic factor \(\mathcal R^4\sim s^4\) gives \(\mathcal M_{\mathrm{CS}}\sim s^{2t+2}\). Thus, reduced-form-factor decay and physical Regge growth coexist for \(-1<t<0\). After restoring the gravitational kinematic factor, \(W_{10}\) contributes at order \(s^6\), whereas the standard graviton-pole obstruction concerns the quadratic coefficient of the physical amplitude. This distinction is consistent with \cite{deRham:2022gfe}, who derive positivity bounds for higher forward coefficients to which the quadratic graviton-pole term does not contribute. Comparing coefficients gives 
\begin{equation}
    a_6=-\kappa^2W_{10}\;.
\end{equation}
Since, $\kappa^2>0$, 
\begin{equation}
a_6\geq0 \Leftrightarrow W_{10}\leq 0\;.
\end{equation}
Thus \(W_{10}\leq 0\) corresponds to positivity of the physical \(s^6\) coefficient, not \(a_2\). For our specified type-II \(\mathcal R^4\) amplitude, the regular tree-level expansion actually begins at \(s^4\), so \(a_2=0\). The coefficient \(W_{01}\) requires a separate nonforward analysis. Its physical contribution is \(-\kappa^2\mathcal R^4W_{01}y\), with \(y=-stu\), and therefore vanishes in the strict forward limit. Near \(t=0\), this contribution takes the form
\[
\mathcal M_{W_{01}}
=-\kappa^2W_{01}s^6t+O(t^2).
\]
Consequently, the forward positivity argument above does not determine its sign. The bound \(W_{01}\le-\frac32W_{10}\) follows from our reduced-amplitude sum rules, while \(W_{01}>0\) follows from spin-0 dominance.
\subsection{Use of machine learning}
\begin{figure}[H]
    \centering
    \begin{subfigure}{0.47\textwidth}
        \includegraphics[width=8cm]{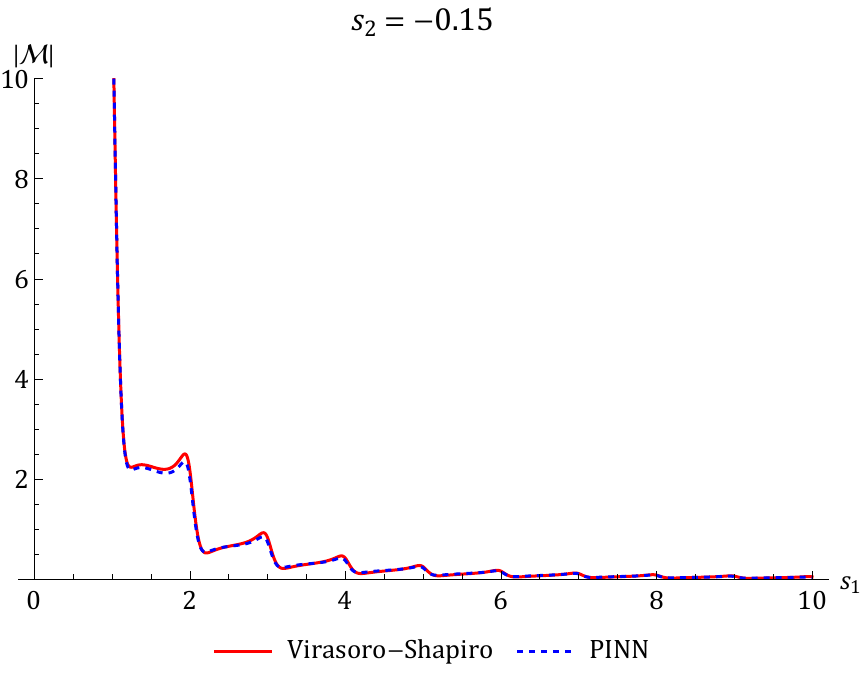}
        \caption{}
        \label{ClosedStringPINNs2m15by100}
    \end{subfigure}
    \hspace{0.5cm}
    \begin{subfigure}{0.47\textwidth}
        \includegraphics[width=8cm]{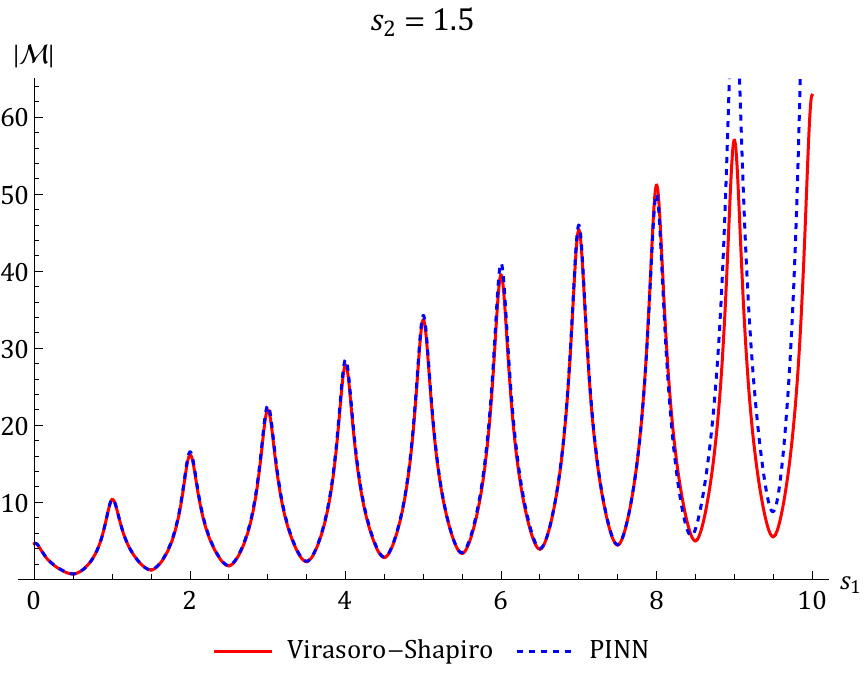}
        \caption{}
        \label{ClosedStringPINNs215by10}
    \end{subfigure}
    \caption{ $|\mathcal{M}(s_1+\frac{i}{10},s_2)|$ versus $s_1$ for the amplitude obtained using the PINN compared with the Virasoro-Shapiro amplitude. Left: We plot for $s_2=-0.15$. Right: We plot for $s_2=1.5$.}%
   \label{fig:ClosedStringPINN}
\end{figure}

Here we use a neural network to bootstrap the Virasoro-Shapiro amplitude. We use the same neural network architecture as in section \ref{MLsetup}. The input spectrum is fixed to the closed string spectrum \eqref{closedstringspec}. Therefore, the neural network learns the partial wave coefficients $c_\ell^{(n)}$ for integer $n$. The loss function is motivated by the Wilson coefficients of the Virasoro-Shapiro amplitude $(W_{10}=-2\zeta(5),W_{01}=2\zeta(3)^2)$ and the local three-channel crossing symmetric dispersion relation with one-parameter ambiguity \eqref{3channellamCSDRsum} gives the amplitude $\mathcal{M}(s,t)$. 
\begin{equation}
    \label{Loss_Func_CS}
    \mathcal{L} = \Phi + \beta_{10} \bigg{(}W_{10} - (-2\zeta(5))\bigg{)}^2 + \beta_{01} \bigg{(}W_{01} - 2\zeta(3)^2\bigg{)}^2 + \sum_{k=1}^{k_{max}}\beta_k\tilde{\mathcal{L}}_k\;.
\end{equation}
Here $\tilde{\mathcal{L}}_k$ are the $\lambda$-independence constraints as defined in \eqref{lambdaderivcons}. We choose $\lambda=1.6, N_{max}=10, k_{max}=3$ and run the machine learning simulations in $D=10$. We choose the following set of points for the $\lambda$-independence constraints:
\begin{equation}
        \mathcal{C} = \{ (s_1, s_2)\, |\, -9.1 \leq s_1 \leq 9.1 \, ,  \, -0.5 \leq s_2 \leq 0.5  \, , \, \D_{s_1} = 1\, , \, \D_{s_2} = 1/11\}\;.
\end{equation} 
We use $\beta_{10}=\beta_{01}=\beta_1=\beta_2=\beta_3=10^6$. This gives the following solution:
\begin{equation}
    \begin{split}
        \Phi = 1.184 \, &, \quad  \,  W_{10}-(-2\zeta(5)) = 1.27 \times 10^{-6} \, , \\
        W_{01}-2\zeta(3)^2 = 3.47 \times 10^{-6} \, &, \quad  \, \tilde{\mathcal{L}}_{mean}  = 3.94 \times 10^{-4}\,. 
    \end{split}
\end{equation}
The partial wave coefficients $c_\ell^{(n)}$ on the leading Regge trajectory obtained from the solution are given in the tables below:
\begin{table}[H]
    \centering
    \begin{tabular}{|c|c|c|c|c|c|c|}
        \hline
         & $c_0^{(1)}$ & $c_2^{(2)}$ & $c_4^{(3)}$ & $c_6^{(4)}$ & $c_8^{(5)}$ & $c_{10}^{(6)}$\\
        \hline
        Virasoro-Shapiro & 1 & 0.00794 & 0.000375 & 0.0000348 & $4.661\times10^{-6}$ & $7.842\times10^{-7}$\\    
        \hline
        PINN & 1.0025 & 0.00826 & 0.000397 & 0.0000374 & $4.994\times10^{-6}$ & $8.743\times10^{-7}$\\ 
        \hline
    \end{tabular}
    \label{tab:c16CS}
\end{table}
\begin{table}[H]
    \centering
    \begin{tabular}{|c|c|c|c|c|}
        \hline
         & $c_{12}^{(7)}$ & $c_{14}^{(8)}$ & $c_{16}^{(9)}$ & $c_{18}^{(10)}$\\
        \hline
        Virasoro-Shapiro & $1.544\times10^{-7}$ & $3.4097\times10^{-8}$ & $8.2173\times10^{-9}$ & $2.1216\times10^{-9}$\\    \hline
        PINN & $1.624\times10^{-7}$ &  $3.3818\times10^{-8}$ & $1.8423\times10^{-8}$ & $1.3882\times10^{-8}$\\    \hline
    \end{tabular}
    \label{tab:c710CS}
\end{table}
In figure \ref{fig:ClosedStringPINN} we plot $|\mathcal{M}(s_1+\frac{i}{10},s_2)|$ versus $s_1$ for the solution obtained using PINN and compared with the analytic form of the Virasoro-Shapiro amplitude \eqref{VS}.

\section{Regge trajectory of a Non-dual amplitude}
\label{Regge}
\begin{figure}[H]
    \centering
    \includegraphics[width=0.7\linewidth]{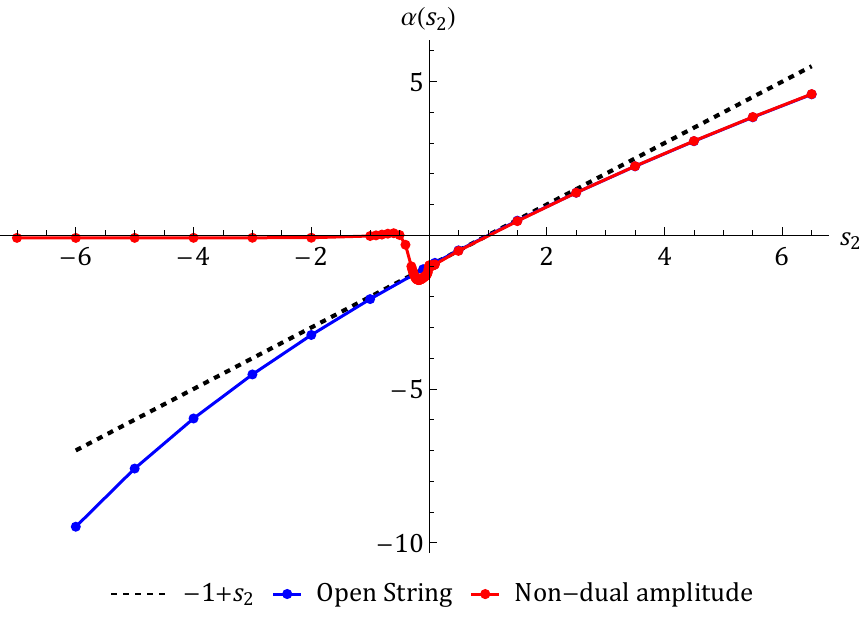}
    \caption{Plotted in red is the Regge trajectory of the Non-dual amplitude which was first found in \cite{Bhat:2024agd} using \eqref{reggefit}. The Regge trajectory of the open string amplitude obtained using \eqref{reggefit} has been plotted in blue for comparison. We also plot $-1+s_2$ which is the Regge trajectory of the open string amplitude obtained using \eqref{reggeos} with a dashed black straight line.}
    \label{fig:ReggeTrajectoryOSBS}
\end{figure}

In this section we will explore the Regge trajectory of the black star amplitude first discovered in \cite{Bhat:2024agd}. We named this the black star amplitude as marked in figure 9a of \cite{Bhat:2024agd}. This amplitude demonstrates a Polchinski-Strassler type transition \cite{polchinskistrassler} from Regge behavior to power-law behavior for $s_1>>1$ and $s_2<0$. The black star amplitude goes to a constant or in other words it behaves as $s_1^0$ for $s_1>>1$ and $s_2< 0$. Hence, it violates the dual resonance model \cite{DHS,Green:1987sp} for sufficiently negative $s_2$. Thus, it can simply be called a Non-dual amplitude. Here we will analyze the Regge trajectory of this Non-dual amplitude. To do this, we make a table of $\left|\mathcal{M}(s_1+\frac{i}{10},s_2)\right|$ versus $s_1$ for $s_1\in\{12,13,14,15,16,17,18\}$ for fixed $s_2$. For the open string amplitude we know that in the Regge limit $s_1\to\infty$ and fixed $s_2$ the amplitude behaves as
\begin{equation}
\label{reggeos}
    \lim_{s_1\to\infty}\mathcal{M}_{OS}(s_1,s_2)=\lim_{s_1\to\infty}\frac{\Gamma(-s_1)\Gamma(-s_2)}{\Gamma(1-s_1-s_2)}\simeq-\Gamma(-s_2)\,\frac{\sin(\pi(s_1+s_2))}{\sin(\pi s_1)}\,s_1^{-1+s_2}\;.
\end{equation}
Now notice that
\begin{equation}
\begin{split}
    &\left|-\Gamma(-s_2)\frac{\sin(\pi(s_1+\frac{i}{10}+n+s_2))}{\sin(\pi (s_1+\frac{i}{10}+n))}\left(s_1+\frac{i}{10}+n\right)^{-1+s_2}\right|\\=&\left|-\Gamma(-s_2)\frac{\sin(\pi(s_1+\frac{i}{10}+s_2))\cos(\pi n)+\cos(\pi(s_1+\frac{i}{10}+s_2))\sin(\pi n)}{\sin(\pi (s_1+\frac{i}{10}))\cos(\pi n)+\cos(\pi (s_1+\frac{i}{10}))\sin(\pi n)}\left(s_1+\frac{i}{10}+n\right)^{-1+s_2}\right|\\=&\left|-\Gamma(-s_2)\frac{\sin(\pi(s_1+\frac{i}{10}+s_2))}{\sin(\pi (s_1+\frac{i}{10}))}\left(s_1+\frac{i}{10}+n\right)^{-1+s_2}\right|\;.
\end{split}
\end{equation}
In other words the proportionality factor $\left|\frac{\sin(\pi(s_1+\frac{i}{10}+s_2))}{\sin(\pi (s_1+\frac{i}{10}))}\right|$ is invariant when $s_1$ is varied by an integer step. Motivated by this we use the following fit function to approximate the data table
\begin{equation}
\label{reggefit}
    \mathcal{M}_{\text{fit}}(s_1,s_2)=k(s_2)\,s_1^{\alpha(s_2)}\;,\qquad\text{for } s_1\in\{12,13,14,15,16,17,18\}\;.
\end{equation}
Here $k(s_2)$ is a proportionality factor and $\alpha(s_2)$ is called the Regge trajectory of the amplitude. We use the \texttt{FindFit} function in Mathematica with \texttt{WorkingPrecision -> 200} and \texttt{MaxIterations -> 10000} to determine $k(s_2)$ and $\alpha(s_2)$ for different values of $s_2$. In figure \ref{fig:ReggeTrajectoryOSBS} we plot the Regge trajectory $\alpha(s_2)$ versus $s_2$ for the Non-dual amplitude obtained using the above algorithm \eqref{reggefit}. We also plot the Regge trajectory of the open string amplitude obtained using \eqref{reggefit} on the same plot for comparison. The Regge trajectory of the open string amplitude is a straight line $\alpha_{OS}(s_2)=-1+s_2$, however, from the plot we see a slight bend when computed using \eqref{reggefit}. This is because we have limited the analysis to finite $s_1\in\{12,13,14,15,16,17,18\}$. We find that for $s_2\geq0$ the Regge trajectories of the Non-dual amplitude and the open string amplitude overlap with each other. But for $s_2<0$ the Regge trajectory of the Non-dual amplitude curves upwards and saturates to zero. Therefore, the Non-dual amplitude shows a transition from Regge behavior to power-law behavior, $s_1^0$ for $s_1>>1$ and $s_2<0$. This transition is similar to QFT amplitudes for example glueball scattering amplitudes as predicted by Polchinski-Strassler in \cite{polchinskistrassler}. See also \cite{Caron-Huot:2016icg} for related discussions. 

\section{Conclusions and Future Directions}
\label{conclusions}
In this paper, we studied the mass-squared spectrum of the open superstring amplitude using the local two-channel crossing symmetric dispersion relation with one-parameter ambiguity ($\lambda$-CSDR). We used machine learning techniques to show that the integer mass-squared spectrum of the open string amplitude favors the bootstrap constraints and minimizes the entanglement generated using a Physics-Informed Neural Network (PINN) and gradient descent simultaneously. PINN was used to learn the $c_\ell^{(R_n)}$ values and gradient descent was used to learn the $R_n$ values. We used a \textit{PyTorch} autograd compatible version of the hypergeometric function ${}_2F_1$ to optimize $c_\ell^{(R_n)}$ and $R_n$ simultaneously. The non-linear nature of the $R_n$ in the ansatz in \eqref{gendispsum} makes machine learning most suitable to handle this problem. We found that fixing $R_1=1$ was absolutely crucial, whereas the rest of $R_n$ were initialized to $n<R_n<n+\delta$ with $0<\delta<0.5$ with a small learning rate to reach the open string solution. Our method of using fully \textit{PyTorch} autograd compatible loss functions could be very useful in the future. Using linear programming, we found that an integer mass-squared spectrum is associated with lower entanglement when compared with general non-integer mass-squared spectra. 

Using the local three-channel crossing symmetric dispersion relation with one-parameter ambiguity we found analytic bounds on $W_{10}$ and $W_{01}$. We carved out the allowed region of S-matrices in the ($W_{10},W_{01}$) space and this allowed region respects the analytic bounds. We used our neural network architecture to obtain the Virasoro-Shapiro amplitude.

Finally we compared the Regge trajectories of QFT like amplitudes that show transition from Regge behavior to power-law behavior with the Regge trajectory of the open string amplitude.

Our work opens up many interesting directions that can be pursued in the future. We list a few of them below:
\begin{itemize}
    \item It would be great to reach the open string solution with all the $R_n$ levels unfixed or learn the integer values of the open string mass-squared spectrum from first principles using machine learning. Achieving this would require advanced neural network architectures which are more sophisticated than the one used in this work. 
    \item It would be fascinating to learn the analytic forms of the open string and the closed string amplitudes instead of the partial wave coefficients. It will be interesting to see what sort of neural network architectures enable this. One would need to dress the neural network with low energy and high energy behavior of the respective amplitude to achieve this.  
    \item It would be interesting to learn the $2\to2$ pion scattering amplitudes obtained in \cite{bhat} using machine learning and the $\lambda$-CSDR. As we know these pion scattering amplitudes have a branch cut for $s\geq4m^2$ where $m$ is the mass of a $\pi^0$ massive scalar particle. Thus, one would require an ansatz with a branch cut for $s\geq 4m^2$ for example the $\rho$ ansatz first proposed in \cite{Paulos:2017fhb}. Two recent works, \cite{Gumus:2024lmj,Gumus:2026mhb} have initiated progress in this direction. It would be fascinating to see what sort of amplitudes we get using the $\lambda$-CSDR. 
    \item Another research direction is the exploration of CFT amplitudes using machine learning. What light can machine learning shed on the numerical conformal bootstrap and the analytic functional bootstrap? Is it possible to learn the OPE coefficients and scaling dimensions of the 3D Ising model for example using the bootstrap principle and machine learning? This is a very important and challenging problem.
\end{itemize}
\section*{Acknowledgements}
We are immensely grateful to Aninda Sinha for suggesting the problem, for his continuous encouragement, and for numerous insightful and helpful discussions. We thank Faizan Bhat, Kausik Ghosh, Apratim Kaviraj, Arnab Priya Saha, Soumen Saren, and Ahmadullah Zahed for numerous useful discussions and comments on the draft. We acknowledge the use of ChatGPT 4, ChatGPT 5, and its higher versions for numerous useful ideas and helpful coding tips. We take sole responsibility for all the results presented in this work. We thank the participants of Asian Winter School 2026 held at IISER, Bhopal and the participants of Students' Talks on Trending Topics in Theory 2026 held at CMI, Chennai where this work was presented.

\appendix
\section{Convergence with $N_{max}$, constraints and $\lambda$}
Here we will check the convergence of the machine learning results for learning the open string mass-squared spectrum with a change in $N_{max}$, $\lambda$-independence constraints and the value of the seed $\lambda$. We will fix $R_1=1$ and train the rest of the levels $R_2,R_3,\cdots,R_{10}$. First, let us check the convergence of these results with a change in $N_{max}$.

\begin{itemize}
    \item $N_{max}=9$
    
    Let us check with $N_{max}=9$. We choose $\lambda=14.6$, $k_{max}=5$, $D=10$. We choose the following grid of ($s_1,s_2$) values for the $\lambda$-independence constraints: 
    \begin{equation}
        \mathcal{C} = \{ (s_1, s_2)\, |\, -8.1 \leq s_1 \leq 8.1 \, ,  \, -0.3 \leq s_2 \leq 0.3  \, , \, \D_{s_1} = 1\, , \, \D_{s_2} = 1/11\}\;.
    \end{equation}
    We use $\beta_{10}=10^6$, $\beta_{01}=10^6$, $\beta_{1}=10^8$ and $\beta_2=\beta_3=\beta_4=\beta_{5}=10^6$. This leads to the following solution:
    \begin{equation}
    \begin{split}
        \Phi = 1.529 \, &, \quad  \,  W_{10}-(-\zeta(3)) = 1.66 \times 10^{-5} \, , \\
        W_{01}-\frac{7\zeta(4)}{4} = 6.31 \times 10^{-6} \, &, \quad  \, \tilde{\mathcal{L}}_{mean}  = 1.53 \times 10^{-4}\,. 
    \end{split}
    \end{equation}
    Here the initial $R_n$ values were chosen such that $n< R_n< n+0.1$\;. The learned final $R_n$ values are given in the table below.
    \begin{table}[H]
    \centering
    \begin{tabular}{|c|c|c|c|c|c|c|c|c|c|}  
        \hline
        \multicolumn{10}{|c|}{Level 1 fixed} \\
        \hline
        Initial $R_n$ & 1 & 2.0310 & 3.0738 & 4.0262 & 5.0399 & 6.0792 & 7.0948 & 8.0746 & 9.0763 \\         
        \hline
        Final $R_n$ & 1 & 2.0279 & 3.0787 & 4.0206 & 5.0453 & 6.0745 & 7.0992 & 8.0700 & 9.0811 \\ 
        \hline
    \end{tabular}
    \label{tab:Nmax9}
\end{table}
    \item $N_{max}=11$
    
    Let us check with $N_{max}=11$. We choose $\lambda=14.6$, $k_{max}=5$, $D=10$. We choose the following grid of ($s_1,s_2$) values for the $\lambda$-independence constraints: 
    \begin{equation}
        \mathcal{C} = \{ (s_1, s_2)\, |\, -10.1 \leq s_1 \leq 10.1 \, ,  \, -0.3 \leq s_2 \leq 0.3  \, , \, \D_{s_1} = 1\, , \, \D_{s_2} = 1/11\}\;.
    \end{equation}
    We use $\beta_{10}=10^6$, $\beta_{01}=10^6$, $\beta_{1}=10^8$ and $\beta_2=\beta_3=\beta_4=\beta_{5}=10^6$. This leads to the following solution:
    \begin{equation}
    \begin{split}
        \Phi = 1.596 \, &, \quad  \,  W_{10}-(-\zeta(3)) = 4.61 \times 10^{-6} \, , \\
        W_{01}-\frac{7\zeta(4)}{4} = 2.13 \times 10^{-6} \, &, \quad  \, \tilde{\mathcal{L}}_{mean}  = 1.23 \times 10^{-4}\,. 
    \end{split}
    \end{equation}
    Here the initial $R_n$ values were chosen such that $n< R_n< n+0.1$\;. The learned final $R_n$ values are given in the table below.
    \begin{table}[H]
    \centering
    \begin{tabular}{|c|c|c|c|c|c|c|c|c|c|c|c|}  
        \hline
        \multicolumn{12}{|c|}{Level 1 fixed} \\
        \hline
        Initial $R_n$ & 1 & 2.0986 & 3.0496 & 4.0494 & 5.0754 & 6.0860 & 7.0510 & 8.0103 & 9.0077 & 10.0158 & 11.0615 \\         
        \hline
        Final $R_n$ & 1 & 2.0662 & 3.0172 & 4.0454 & 5.0768 & 6.0838 & 7.0537 & 8.0112 & 9.0041 & 10.0164 & 11.0645 \\ 
        \hline
    \end{tabular}
    \label{tab:Nmax11}
\end{table}
    
\item more $\lambda$-independence constraints

Next, let us check the convergence of these results with a change in the number of $\lambda$-independence constraints imposed. Let us choose the following grid of ($s_1,s_2$) values for the $\lambda$-independence constraints:
    \begin{equation}
        \mathcal{C} = \{ (s_1, s_2)\, |\, -9.1 \leq s_1 \leq 9.1 \, ,  \, -0.5 \leq s_2 \leq 0.5  \, , \, \D_{s_1} = 1\, , \, \D_{s_2} = 1/11\}\;.
    \end{equation}
Notice that we have changed the range of $s_2$ from $-0.3\leq s_2\leq0.3$ to $-0.5\leq s_2\leq 0.5$. We choose $\lambda=14.6$, $N_{max}=10$, $k_{max}=5$, $D=10$. We use $\beta_{10}=10^6$, $\beta_{01}=10^6$, $\beta_{1}=10^8$ and $\beta_2=\beta_3=\beta_4=\beta_{5}=10^6$. This leads to the following solution:
\begin{equation}
    \begin{split}
        \Phi = 1.571 \, &, \quad  \,  W_{10}-(-\zeta(3)) = 1.65 \times 10^{-5} \, , \\
        W_{01}-\frac{7\zeta(4)}{4} = - 2.02 \times 10^{-5} \, &, \quad  \, \tilde{\mathcal{L}}_{mean}  = 2.06 \times 10^{-4}\,. 
    \end{split}
\end{equation}
Here the initial $R_n$ values were chosen such that $n< R_n< n+0.1$\;. The learned final $R_n$ values are given in the table below. 
\begin{table}[H]
    \centering
    \begin{tabular}{|c|c|c|c|c|c|c|c|c|c|c|}  
        \hline
        \multicolumn{11}{|c|}{Level 1 fixed} \\
        \hline
        Initial $R_n$ & 1 & 2.0064 & 3.0928 & 4.0954 & 5.0684 & 6.0837 & 7.0372 & 8.0479 & 9.0518 & 10.0336\\         
        \hline
        Final $R_n$ & 1 & 2.0152 & 3.0869 & 4.0997 & 5.0645 & 6.0871 & 7.0351 & 8.0511 & 9.0418 & 10.0519\\ 
        \hline
    \end{tabular}
    \label{tab:moreC}
\end{table}
\item $\lambda=9.5$

Next, let us check the convergence of these results with a change in the value of the seed $\lambda$. Let us choose $\lambda=9.5$\;. We choose $N_{max}=10$, $k_{max}=5$, $D=10$. We choose the following grid of ($s_1,s_2$) values for the $\lambda$-independence constraints:
    \begin{equation}
        \mathcal{C} = \{ (s_1, s_2)\, |\, -9.1 \leq s_1 \leq 9.1 \, ,  \, -0.3 \leq s_2 \leq 0.3  \, , \, \D_{s_1} = 1\, , \, \D_{s_2} = 1/11\}\;.
    \end{equation}
We use $\beta_{10}=10^6$, $\beta_{01}=10^6$, $\beta_{1}=10^8$ and $\beta_2=\beta_3=\beta_4=\beta_{5}=10^6$. This leads to the following solution:
\begin{equation}
    \begin{split}
        \Phi = 1.544 \, &, \quad  \,  W_{10}-(-\zeta(3)) = 1.71 \times 10^{-5} \, , \\
        W_{01}-\frac{7\zeta(4)}{4} = 3.56 \times 10^{-6} \, &, \quad  \, \tilde{\mathcal{L}}_{mean}  = 1.11 \times 10^{-4}\,. 
    \end{split}
\end{equation}
Here the initial $R_n$ values were chosen such that $n< R_n< n+0.1$\;. The learned final $R_n$ values are given in the table below. 
\begin{table}[H]
    \centering
    \begin{tabular}{|c|c|c|c|c|c|c|c|c|c|c|}  
        \hline
        \multicolumn{11}{|c|}{Level 1 fixed} \\
        \hline
        Initial $R_n$ & 1 & 2.0499 & 3.0675 & 4.0906 & 5.0649 & 6.0845 & 7.0848 & 8.0579 & 9.0053 & 10.0737\\         
        \hline
        Final $R_n$ & 1 & 2.0216 & 3.0556 & 4.0844 & 5.0695 & 6.0838 & 7.0885 & 8.0639 & 9.0084 & 10.0768\\ 
        \hline
    \end{tabular}
    \label{tab:lambda95}
\end{table}
\end{itemize}
\section{Checks for $\lambda$-independence}
\begin{figure}[H]
    \centering
    \begin{subfigure}{0.47\textwidth}
        \includegraphics[width=8cm]{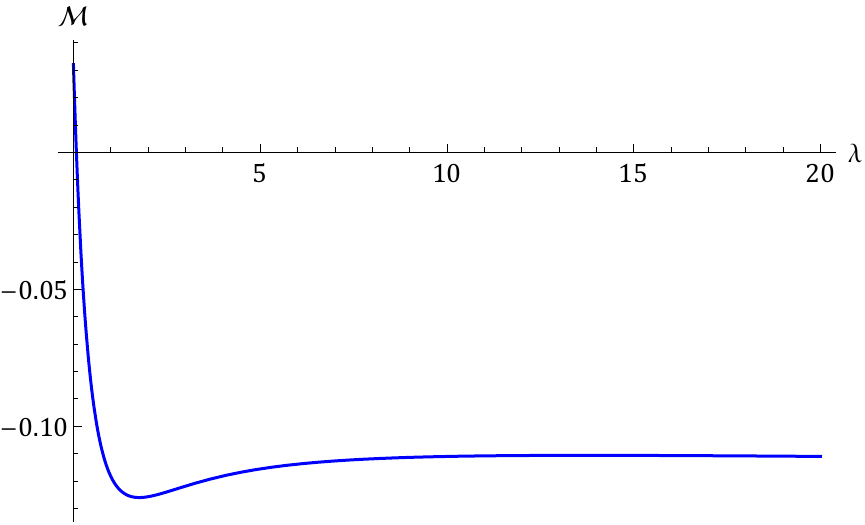}
        \caption{$s_1=3.1,s_2=-0.1$}
        \label{lConvergenceOSMLR1fixeds131by10s2m1by10}
    \end{subfigure}
    \hspace{0.5cm}
    \begin{subfigure}{0.47\textwidth}
        \includegraphics[width=8cm]{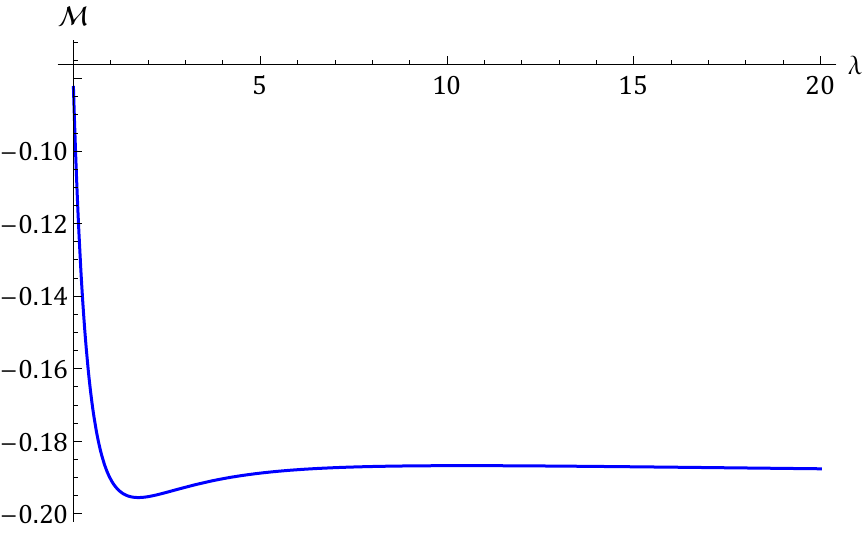}
        \caption{$s_1=-0.5,s_2=4.6$}
        \label{lConvergenceOSMLR1fixeds1m5by10s246by10}
    \end{subfigure}
    \caption{We plot $\mathcal{M}(s_1, s_2)$ versus $\lambda$ for the open string solution found with $R_1=1$ fixed in \ref{resultsML} for different values of $s_1, s_2$ and find nice plateaus around the seed value of $\lambda = 14.6$.}%
   \label{fig:lConvergenceOSMLR1fixed}
\end{figure}

\begin{figure}[H]
    \centering
    \begin{subfigure}{0.47\textwidth}
        \includegraphics[width=8cm]{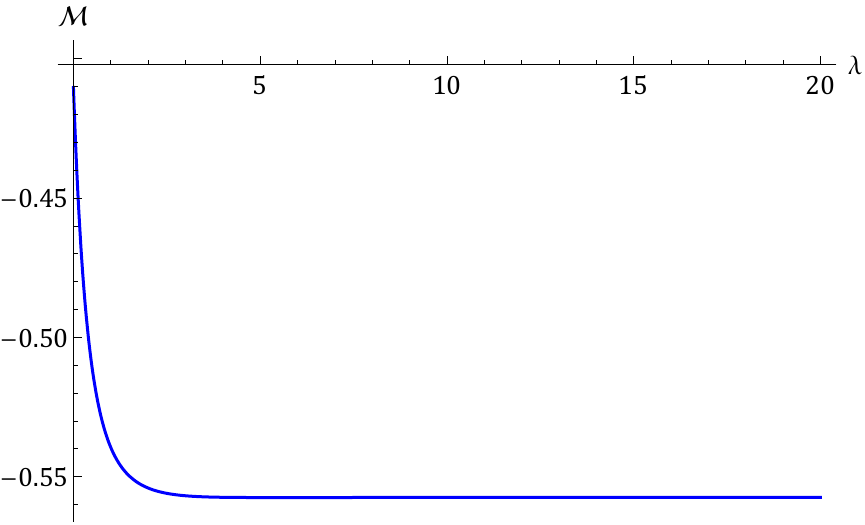}
        \caption{$s_1=3.1,s_2=-0.1$}
        \label{lConvergenceOSSPCSs131by10s2m1by10}
    \end{subfigure}
    \hspace{0.5cm}
    \begin{subfigure}{0.47\textwidth}
        \includegraphics[width=8cm]{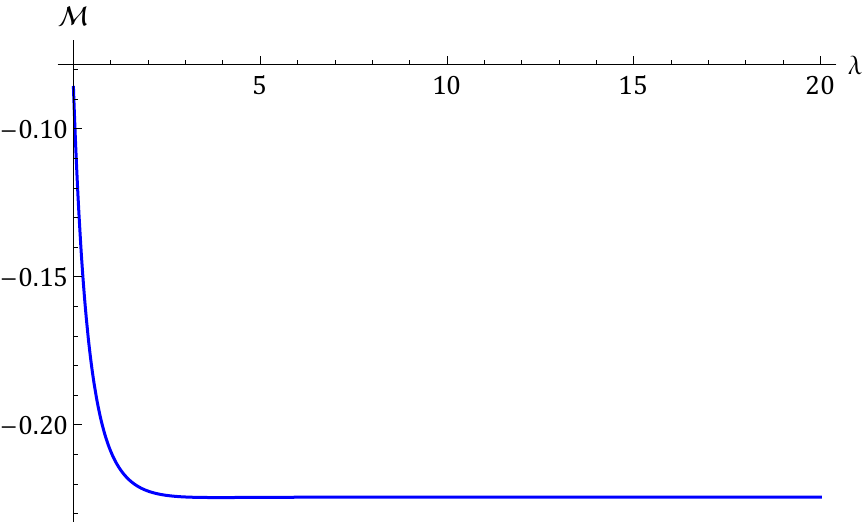}
        \caption{$s_1=-2.5,s_2=4.6$}
        \label{lConvergenceOSSPCSs1m25by10s246by10}
    \end{subfigure}
    \caption{We plot $\mathcal{M}(s_1, s_2)$ versus $\lambda$ for the S-matrix with $\Phi=1.789$ in figure \ref{CSnearOS} for different values of $s_1, s_2$ and find nice plateaus around the seed value of $\lambda = 14.6$.}%
   \label{fig:lConvergenceOSSPCS}
\end{figure}

\begin{figure}[H]
    \centering
    \begin{subfigure}{0.47\textwidth}
        \includegraphics[width=8cm]{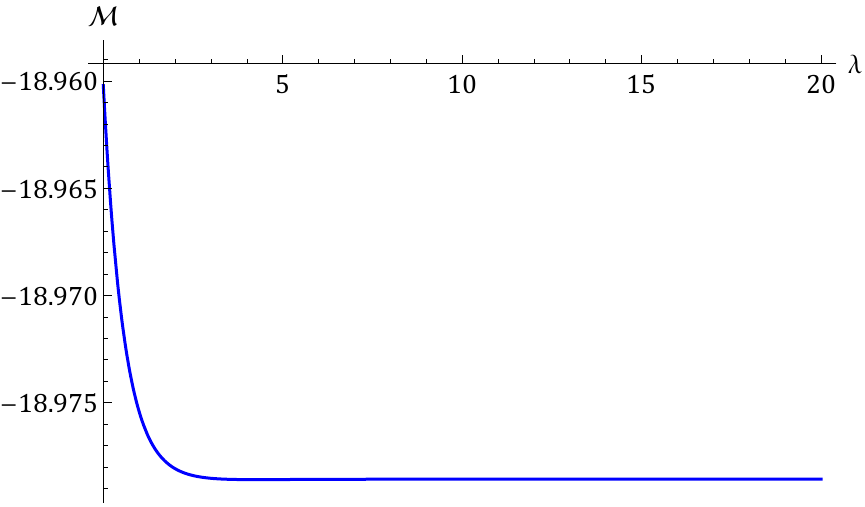}
        \caption{$s_1=3.1,s_2=-0.1$}
        \label{lConvergenceOSSPASs131by10s2m1by10}
    \end{subfigure}
    \hspace{0.5cm}
    \begin{subfigure}{0.47\textwidth}
        \includegraphics[width=8cm]{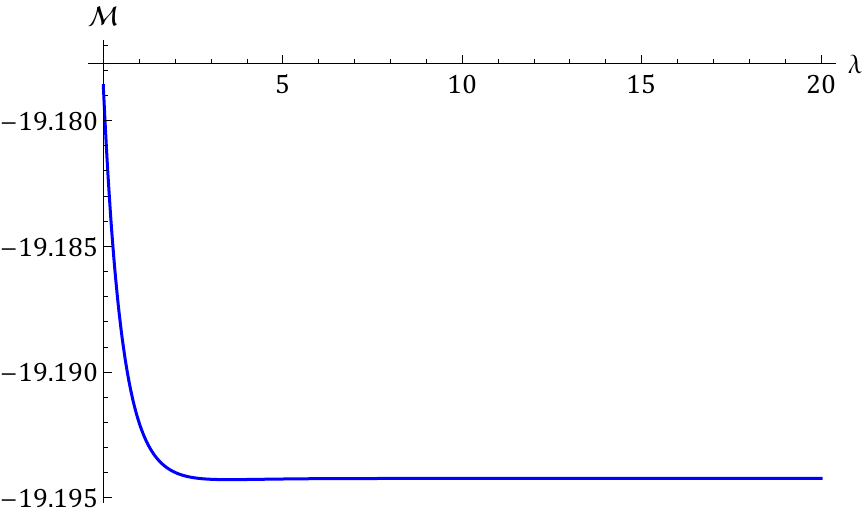}
        \caption{$s_1=-2.5,s_2=4.6$}
        \label{lConvergenceOSSPASs1m25by10s246by10}
    \end{subfigure}
    \caption{We plot $\mathcal{M}(s_1, s_2)$ versus $\lambda$ for the S-matrix with $\Phi=10.506$ in figure \ref{CSawayOS} for different values of $s_1, s_2$ and find nice plateaus around the seed value of $\lambda = 14.6$.}%
   \label{fig:lConvergenceOSSPAS}
\end{figure}

Here we check the $\lambda$-independence of various solutions obtained in the main text using machine learning and linear programming in section \ref{results}. In figure \ref{fig:lConvergenceOSMLR1fixed} we plot $\mathcal{M}(s_1,s_2)$ versus $\lambda$ with $R_1=1$ fixed for the open string solution obtained using machine learning in \ref{resultsML} and observe that it is $\lambda$-independent around the seed value $\lambda=14.6$. In figure \ref{fig:lConvergenceOSSPCS} we plot $\mathcal{M}(s_1,s_2)$ versus $\lambda$ for the S-matrix with $\Phi=1.789$ in figure \ref{CSnearOS} obtained using linear programming. In figure \ref{fig:lConvergenceOSSPAS} we plot $\mathcal{M}(s_1,s_2)$ versus $\lambda$ for the S-matrix with $\Phi=10.506$ in figure \ref{CSawayOS} obtained using linear programming. Again we find that both S-matrices obtained using linear programming are $\lambda$-independent around the seed value $\lambda=14.6$.
\section{Machine Learning in Wolfram Mathematica with arbitrary precision}
\label{MLmathematica}
In this section we will implement our fully-connected feed-forward network in Mathematica with arbitrary precision to learn the $c_{\ell}^{(n)}$ of the open string amplitude and compare our results with \cite{Bhat:2024agd}. To do this we need a custom built Adam optimizer in Mathematica at arbitrary precision.

\begin{figure}[H]
    \centering
    \begin{subfigure}{0.47\textwidth}
        \includegraphics[width=8cm]{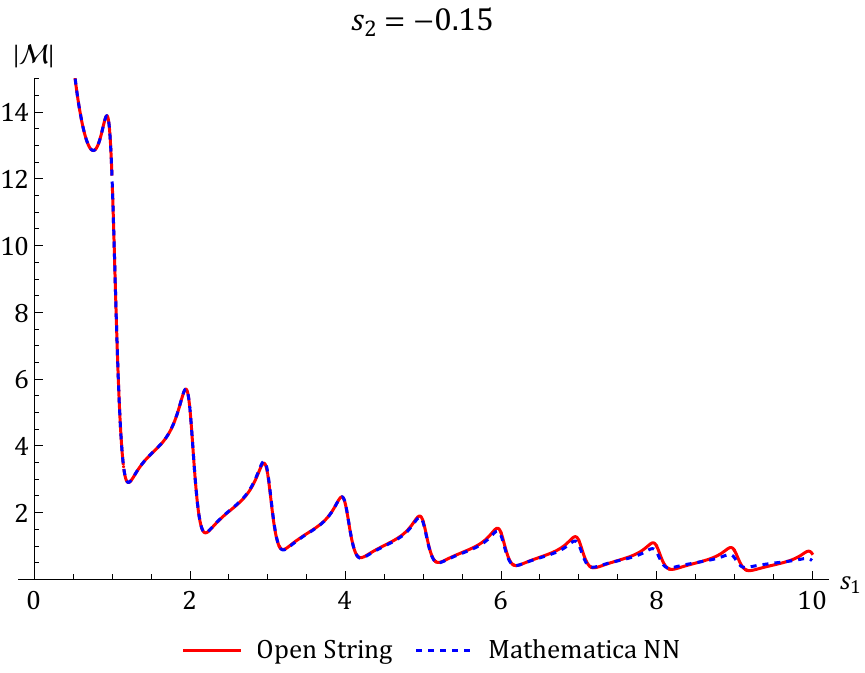}
        \caption{}
        \label{OpenStringMathematicaNNs2m15by100}
    \end{subfigure}
    \hspace{0.5cm}
    \begin{subfigure}{0.47\textwidth}
        \includegraphics[width=8cm]{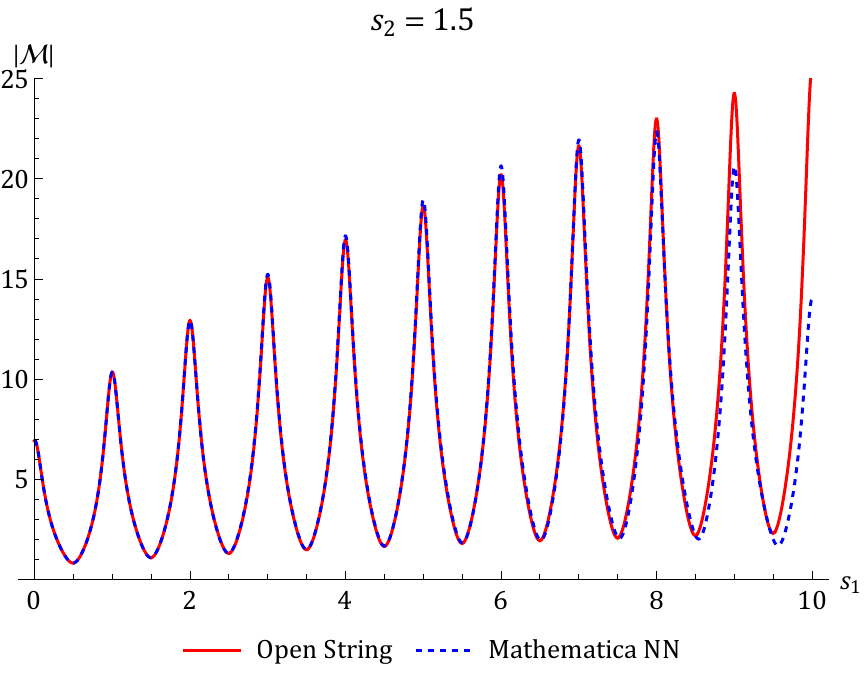}
        \caption{}
        \label{OpenStringMathematicaNNs215by10}
    \end{subfigure}
    \caption{ $|\mathcal{M}(s_1+\frac{i}{10},s_2)|$ versus $s_1$ for the amplitude obtained using the Mathematica NN compared with the open string amplitude. Left: We plot for $s_2=-0.15$. Right: We plot for $s_2=1.5$.}%
   \label{fig:OpenStringMathematicaNN}
\end{figure}

Before moving on to the open string problem we will solve a small toy problem with our custom built neural network in Mathematica. Consider the following problem:
\begin{equation}
\label{prototype}
\begin{split}
    \text{Find the maximum of}\quad&(-x-2y-5z)\\
    \text{subject to the constraints:}\quad &x\geq\frac{1}{2},\;y\geq\frac{1}{2},\;z\geq\frac{1}{2},\\ &x+y-z=10,\;2x-y+2z=10\;.
\end{split}
\end{equation}
If we solve this problem in Mathematica using \texttt{FindMaximum}, we obtain the maximum to be $-17$ and the solution is $x\to6.5,\;y\to4,\;z\to0.5$. We would like to solve this problem using a neural network and gradient descent with a precision that will be set beforehand in Mathematica. To achieve this we build a neural network in Mathematica. The neural network has an input layer of size 1, a hidden layer of size 8, and an output layer of size 1. We use custom made functions in Mathematica to build and run the neural network. For example, we use custom built \texttt{ReLU} and \texttt{Softplus} activation functions. We use \texttt{ReLU} from the first layer to the second layer. We use \texttt{Softplus} from the second layer to the output layer. We add $\frac{1}{2}$ to the usual \texttt{Softplus} function, that is, we use $\log(1+e^x)+\frac{1}{2}$ so that all the outputs of the neural network are greater than or equal to $\frac{1}{2}$. We define the function \texttt{InitMLP} which stores the initial weights and biases of the neural network (the full form of MLP is Multi Layer Perceptron). We define the function \texttt{LossAndGrad} which stores the loss function and the gradient of the loss function. For this problem we define the loss function as
\begin{equation}
    \mathcal{L}=-(-x-2y-5z)+\beta_1(x+y-z-10)^2+\beta_2(2x-y+2z-10)^2\;.
\end{equation}
We choose $\beta_1=\beta_2=100$. We define the \texttt{MLPForwardList} function that generates the forward pass of the neural network. We define the \texttt{MLPBackwardList} function that generates the gradients to update the weights and biases of the neural network. We also build a custom \texttt{AdamUpdate} function and a training loop. We build our customized functions using the inbuilt Mathematica function \texttt{SetPrecision} so that the inputs and outputs of the customized functions are at a predetermined precision. So, the entire training loop and all the variables run with the same precision. For example purposes, we provide the code for the \texttt{AdamUpdate} function below \cite{Kingma:2014vow}.

\begin{equation*}
\begin{split}
    &\texttt{AdamUpdate[params\_, grads\_, state\_, alpha\_, b1\_, b2\_, eps\_] :=}\\ &\texttt{Module[\{m, v, t, newM, newV, mHat, vHat, newParams\},}\\ &\texttt{\{m, v, t\} = state;}\\ &\texttt{newM = b1*m + (1 - b1)*grads;}\\ &\texttt{newV = b2*v + (1 - b2)*(grads}^\texttt{2}\texttt{);}\\ &\texttt{mHat = newM/(1 - b1}^\texttt{(t + 1)}\texttt{);}\\ &\texttt{vHat = newV/(1 - b2}^\texttt{(t + 1)}\texttt{);}\\ &\texttt{newParams = params - alpha*mHat/(Sqrt[vHat] + eps);}\\ &\texttt{\{newParams, \{newM, newV, t + 1\}\}];}
\end{split}
\end{equation*}
For the \texttt{AdamUpdate} we use $\texttt{b1}=0.9$, $\texttt{b2}=0.999$ and $\texttt{eps}=10^{-8}$. We set the initial Adam state to be constant arrays of zeros. 
\begin{equation*}
\begin{split}
    &\texttt{InitAdamState[paramArray\_] :=}\\
    &\texttt{\{ConstantArray[SetPrecision[0, prec], Dimensions[paramArray]],}\\
    &\texttt{ConstantArray[SetPrecision[0, prec], Dimensions[paramArray]],}\\
    &\texttt{0\}}
\end{split}
\end{equation*}
We find that our neural network in Mathematica solves the problem defined in \eqref{prototype} quite easily! For this problem the learning rate for \texttt{AdamUpdate} denoted by \texttt{alpha} is set to $10^{-4}$. We use $x=1,y=2,z=3$ as fixed inputs for the neural network. We run the training loop for $10^5$ epochs after which we find that the loss saturates. We find that the predetermined precision does not play a major role in the solution. Precision affects the solution after the fifth decimal place. We provide the details of the solutions obtained after the end of training in the table below.
\begin{table}[H]
\label{prototypesol}
    \centering
    \begin{tabular}{|c|c|c|c|c|}
        \hline
        Precision & $x$ & $y$ & $z$ & Loss\\
        \hline
        10 & 6.498 & 3.994 & 0.500 & 16.993\\
        \hline
        100 & 6.498 & 3.994 & 0.500 & 16.993\\
        \hline
        200 & 6.498 & 3.994 & 0.500 & 16.993\\
        \hline
    \end{tabular}
    \caption{Solution using Mathematica NN for the prototype problem defined in \eqref{prototype}.}
\end{table}
Next we move on to the open string problem. The Mathematica set-up is the same as for the previous problem. The neural network architecture is similar to \cite{Bhat:2024agd}. It has an input layer of size 2, two hidden layers of size 64 each, and an output layer of size 1. We use the \texttt{ReLU} activation function for the two hidden layers and the \texttt{Softplus} activation function for the output layer of the neural network. Here we learn the partial wave coefficients $c_\ell^{(n)}$ and the input spectrum is the same as the open string spectrum \eqref{OSspectrum}. The \texttt{Softplus} function in the output layer maintains unitarity of the $c_\ell^{(n)}$. The loss function is defined in \eqref{Loss_Func}. The learning rate of the \texttt{AdamUpdate} is set to $10^{-5}$. We use a precision of 100 in the \texttt{SetPrecision} functions throughout for the open string problem. We choose $\lambda=14.6$, $N_{max}=10$ and $k_{max}=3$. We impose the $\lambda$-independence constraints on the following set of points
\begin{equation}
        \mathcal{C} = \{ (s_1, s_2)\, |\, -9.1 \leq s_1 \leq 9.1 \, ,  \, -0.3 \leq s_2 \leq 0.3  \, , \, \D_{s_1} = 1\, , \, \D_{s_2} = 1/11\}\;.
\end{equation}
We use $\beta_{10}=10^6$, $\beta_{01}=10^6$, $\beta_{1}=10^8$ and $\beta_2=\beta_3=10^6$. We run for $2.5\times 10^5$ epochs and this leads to the following solution:
\begin{equation}
    \begin{split}
        \Phi = 1.533 \, &, \quad  \,  W_{10}-(-\zeta(3)) = -3.82 \times 10^{-6} \, , \\
        W_{01}-\frac{7\zeta(4)}{4} = -7.06 \times 10^{-7} \, &, \quad  \, \tilde{\mathcal{L}}_{mean}  = 2.07\times 10^{-4}\,. 
    \end{split}
\end{equation}
The values of $c_\ell^{(n)}$ on the leading Regge trajectory are given in the tables below.
\begin{table}[H]
    \centering
    \begin{tabular}{|c|c|c|c|c|c|c|}
        \hline
         & $c_0^{(1)}$ & $c_1^{(2)}$ & $c_2^{(3)}$ & $c_3^{(4)}$ & $c_4^{(5)}$ & $c_5^{(6)}$\\
        \hline
        Open String & 1 & 0.0714 & 0.0119 & 0.002886 & 0.000867 & 0.0002997\\    
        \hline
        Mathematica NN & 0.9998 & 0.0714 & 0.0120 & 0.002921 & 0.000893 & 0.0003188\\ 
        \hline
    \end{tabular}
    \label{tab:c16}
\end{table}
\begin{table}[H]
    \centering
    \begin{tabular}{|c|c|c|c|c|}
        \hline
         & $c_6^{(7)}$ & $c_7^{(8)}$ & $c_8^{(9)}$ & $c_9^{(10)}$\\
        \hline
        Open String & 0.0001143 & 0.0000469 & 0.0000204 & $9.264\times10^{-6}$\\    \hline
        Mathematica NN & 0.0001254 & 0.0000520 & 0.0000211 & $6.528\times10^{-6}$\\    \hline
    \end{tabular}
    \label{tab:c710}
\end{table}
In figure \ref{fig:OpenStringMathematicaNN} we plot $|\mathcal{M}(s_1+\frac{i}{10},s_2)|$ versus $s_1$ for the open string amplitude and the amplitude obtained using the Mathematica NN for $(W_{10},W_{01})=\left(-\zeta(3),\frac{7}{4}\zeta(4)\right)$.

\bibliographystyle{utphys}
\bibliography{Dispersion}

\providecommand{\href}[2]{#2}\begingroup\raggedright\begin{thebibliography}{100}

\bibitem{Chew:1962eu}
G.~F. Chew and S.~C. Frautschi, ``{Regge Trajectories and the Principle of
  Maximum Strength for Strong Interactions},''
  \href{https://dx.doi.org/10.1103/PhysRevLett.8.41}{{\em Phys. Rev. Lett.}
  {\bfseries 8} (1962) 41--44}.

\bibitem{Desgrolard:2000sf}
P.~Desgrolard, M.~Giffon, E.~Martynov, and E.~Predazzi, ``{Exchange degenerate
  Regge trajectories: A Fresh look from resonance and forward scattering
  regions},'' \href{https://dx.doi.org/10.1007/s100520100548}{{\em Eur. Phys.
  J. C} {\bfseries 18} (2001) 555--561},
  \href{https://arxiv.org/abs/hep-ph/0006244}{{\ttfamily
  arXiv:hep-ph/0006244}}.

\bibitem{Veneziano:1968yb}
G.~Veneziano, ``{Construction of a crossing - symmetric, Regge behaved
  amplitude for linearly rising trajectories},''
  \href{https://dx.doi.org/10.1007/BF02824451}{{\em Nuovo Cim. A} {\bfseries
  57} (1968) 190--197}.

\bibitem{Saha:2024qpt}
A.~P. Saha and A.~Sinha, ``{Field Theory Expansions of String Theory
  Amplitudes},'' \href{https://dx.doi.org/10.1103/PhysRevLett.132.221601}{{\em
  Phys. Rev. Lett.} {\bfseries 132} no.~22, (2024) 221601},
  \href{https://arxiv.org/abs/2401.05733}{{\ttfamily arXiv:2401.05733
  [hep-th]}}.

\bibitem{Bhat:2024agd}
F.~Bhat, D.~Chowdhury, A.~P. Saha, and A.~Sinha, ``{Bootstrapping string models
  with entanglement minimization and machine learning},''
  \href{https://dx.doi.org/10.1103/PhysRevD.111.066013}{{\em Phys. Rev. D}
  {\bfseries 111} no.~6, (2025) 066013},
  \href{https://arxiv.org/abs/2409.18259}{{\ttfamily arXiv:2409.18259
  [hep-th]}}.

\bibitem{Bhat:2025zex}
F.~Bhat, A.~P. Saha, and A.~Sinha, ``{Stringy dispersion relation for field
  theories},'' \href{https://dx.doi.org/10.1103/ljzx-q254}{{\em Phys. Rev. D}
  {\bfseries 113} no.~6, (2026) 066016},
  \href{https://arxiv.org/abs/2506.03862}{{\ttfamily arXiv:2506.03862
  [hep-th]}}.

\bibitem{Saha:2026ftv}
A.~P. Saha and A.~Sinha, ``{Five-point partial waves, splitting constraints and
  hidden zeros},'' \href{https://dx.doi.org/10.1007/JHEP06(2026)092}{{\em JHEP}
  {\bfseries 06} (2026) 092},
  \href{https://arxiv.org/abs/2601.15088}{{\ttfamily arXiv:2601.15088
  [hep-th]}}.

\bibitem{gu1}
A.~Guerrieri, J.~Penedones, and P.~Vieira, ``{Where Is String Theory in the
  Space of Scattering Amplitudes?},''
  \href{https://dx.doi.org/10.1103/PhysRevLett.127.081601}{{\em Phys. Rev.
  Lett.} {\bfseries 127} no.~8, (2021) 081601},
  \href{https://arxiv.org/abs/2102.02847}{{\ttfamily arXiv:2102.02847
  [hep-th]}}.

\bibitem{gu2}
A.~Guerrieri, H.~Murali, J.~Penedones, and P.~Vieira, ``{Where is M-theory in
  the space of scattering amplitudes?},''
  \href{https://dx.doi.org/10.1007/JHEP06(2023)064}{{\em JHEP} {\bfseries 06}
  (2023) 064}, \href{https://arxiv.org/abs/2212.00151}{{\ttfamily
  arXiv:2212.00151 [hep-th]}}.

\bibitem{Chiang:2023quf}
L.-Y. Chiang, Y.-t. Huang, and H.-C. Weng, ``{Bootstrapping string theory
  EFT},'' \href{https://dx.doi.org/10.1007/JHEP05(2024)289}{{\em JHEP}
  {\bfseries 05} (2024) 289},
  \href{https://arxiv.org/abs/2310.10710}{{\ttfamily arXiv:2310.10710
  [hep-th]}}.

\bibitem{Haring:2023zwu}
K.~H\"aring and A.~Zhiboedov, ``{The Stringy S-matrix Bootstrap: Maximal Spin
  and Superpolynomial Softness},''
  \href{https://arxiv.org/abs/2311.13631}{{\ttfamily arXiv:2311.13631
  [hep-th]}}.

\bibitem{Berman:2023jys}
J.~Berman, H.~Elvang, and A.~Herderschee, ``{Flattening of the EFT-hedron:
  supersymmetric positivity bounds and the search for string theory},''
  \href{https://dx.doi.org/10.1007/JHEP03(2024)021}{{\em JHEP} {\bfseries 03}
  (2024) 021}, \href{https://arxiv.org/abs/2310.10729}{{\ttfamily
  arXiv:2310.10729 [hep-th]}}.

\bibitem{Berman:2024wyt}
J.~Berman and H.~Elvang, ``{Corners and Islands in the S-matrix Bootstrap of
  the Open Superstring},'' \href{https://arxiv.org/abs/2406.03543}{{\ttfamily
  arXiv:2406.03543 [hep-th]}}.

\bibitem{Berman:2024kdh}
J.~Berman, ``{Analytic bounds on the spectrum of crossing symmetric
  S-matrices},'' \href{https://dx.doi.org/10.1007/JHEP08(2025)066}{{\em JHEP}
  {\bfseries 08} (2025) 066},
  \href{https://arxiv.org/abs/2410.01914}{{\ttfamily arXiv:2410.01914
  [hep-th]}}.

\bibitem{Berman:2024eid}
J.~Berman, H.~Elvang, N.~Geiser, and L.~L. Lin, ``{Bootstrapping extremal
  scalar amplitudes with and without supersymmetry},''
  \href{https://dx.doi.org/10.1007/JHEP05(2026)149}{{\em JHEP} {\bfseries 05}
  (2026) 149}, \href{https://arxiv.org/abs/2412.13368}{{\ttfamily
  arXiv:2412.13368 [hep-th]}}.

\bibitem{Berman:2024owc}
J.~Berman and N.~Geiser, ``{Analytic bootstrap bounds on masses and spins in
  gravitational and non-gravitational scalar theories},''
  \href{https://dx.doi.org/10.1007/JHEP06(2025)168}{{\em JHEP} {\bfseries 06}
  (2025) 168}, \href{https://arxiv.org/abs/2412.17902}{{\ttfamily
  arXiv:2412.17902 [hep-th]}}.

\bibitem{Berman:2025owb}
J.~Berman, H.~Elvang, and C.~Figueiredo, ``{Splitting regions and shrinking
  islands from higher point constraints},''
  \href{https://dx.doi.org/10.1007/JHEP10(2025)226}{{\em JHEP} {\bfseries 10}
  (2025) 226}, \href{https://arxiv.org/abs/2506.22538}{{\ttfamily
  arXiv:2506.22538 [hep-th]}}.

\bibitem{Berman:2026ezk}
J.~Berman, S.~Caron-Huot, A.~V. Chandra, H.~Elvang, A.~Herderschee, L.~L. Lin,
  and R.~Morales, ``{Gravitational Effective Theories with Maximal
  Supersymmetry and a Peculiar Parity},''
  \href{https://arxiv.org/abs/2607.14230}{{\ttfamily arXiv:2607.14230
  [hep-th]}}.

\bibitem{Eckner:2024ggx}
C.~Eckner, F.~Figueroa, and P.~Tourkine, ``{The Regge bootstrap, from linear to
  non-linear trajectories},''
  \href{https://arxiv.org/abs/2401.08736}{{\ttfamily arXiv:2401.08736
  [hep-th]}}.

\bibitem{Eckner:2024pqt}
C.~Eckner, F.~Figueroa, and P.~Tourkine, ``{On the number of Regge trajectories
  for dual amplitudes},'' \href{https://arxiv.org/abs/2405.21057}{{\ttfamily
  arXiv:2405.21057 [hep-th]}}.

\bibitem{Eckner:2025kve}
C.~Eckner, F.~Figueroa, S.~Metayer, and P.~Tourkine, ``{Regge trajectories for
  UV completions of graviton scattering from polynomial boundedness},''
  \href{https://arxiv.org/abs/2512.17828}{{\ttfamily arXiv:2512.17828
  [hep-th]}}.

\bibitem{Wan:2026pjq}
S.-L. Wan and S.-Y. Zhou, ``{Analytic Bootstrap of the Veneziano Amplitude},''
  \href{https://arxiv.org/abs/2605.11084}{{\ttfamily arXiv:2605.11084
  [hep-th]}}.

\bibitem{Xu:2026kix}
Y.~Xu, ``{A Dispersive Bootstrap for the Virasoro-Shapiro Amplitude},''
  \href{https://arxiv.org/abs/2606.19283}{{\ttfamily arXiv:2606.19283
  [hep-th]}}.

\bibitem{deRham:2026lvc}
C.~de~Rham, A.~J. Tolley, Z.-H. Wang, and S.-Y. Zhou, ``{Unitary Dual-Resonance
  S-matrices},'' \href{https://arxiv.org/abs/2607.24922}{{\ttfamily
  arXiv:2607.24922 [hep-th]}}.

\bibitem{ak}
G.~Auberson and N.~N. Khuri, ``{Rigorous parametric dispersion representation
  with three-channel symmetry},''
  \href{https://dx.doi.org/10.1103/PhysRevD.6.2953}{{\em Phys. Rev. D}
  {\bfseries 6} (1972) 2953--2966}.

\bibitem{Mahoux:1974ej}
G.~Mahoux, S.~M. Roy, and G.~Wanders, ``{Physical pion pion partial-wave
  equations based on three channel crossing symmetry},''
  \href{https://dx.doi.org/10.1016/0550-3213(74)90480-5}{{\em Nucl. Phys. B}
  {\bfseries 70} (1974) 297--316}.

\bibitem{Sinha:2020win}
A.~Sinha and A.~Zahed, ``{Crossing Symmetric Dispersion Relations in Quantum
  Field Theories},''
  \href{https://dx.doi.org/10.1103/PhysRevLett.126.181601}{{\em Phys. Rev.
  Lett.} {\bfseries 126} no.~18, (2021) 181601},
  \href{https://arxiv.org/abs/2012.04877}{{\ttfamily arXiv:2012.04877
  [hep-th]}}.

\bibitem{Gopakumar:2021dvg}
R.~Gopakumar, A.~Sinha, and A.~Zahed, ``{Crossing Symmetric Dispersion
  Relations for Mellin Amplitudes},''
  \href{https://dx.doi.org/10.1103/PhysRevLett.126.211602}{{\em Phys. Rev.
  Lett.} {\bfseries 126} no.~21, (2021) 211602},
  \href{https://arxiv.org/abs/2101.09017}{{\ttfamily arXiv:2101.09017
  [hep-th]}}.

\bibitem{bieb}
P.~Haldar, A.~Sinha, and A.~Zahed, ``{Quantum field theory and the Bieberbach
  conjecture},'' \href{https://dx.doi.org/10.21468/SciPostPhys.11.1.002}{{\em
  SciPost Phys.} {\bfseries 11} (2021) 002},
  \href{https://arxiv.org/abs/2103.12108}{{\ttfamily arXiv:2103.12108
  [hep-th]}}.

\bibitem{Raman:2021pkf}
P.~Raman and A.~Sinha, ``{QFT, EFT and GFT},''
  \href{https://dx.doi.org/10.1007/JHEP12(2021)203}{{\em JHEP} {\bfseries 12}
  (2021) 203}, \href{https://arxiv.org/abs/2107.06559}{{\ttfamily
  arXiv:2107.06559 [hep-th]}}.

\bibitem{Chowdhury:2021ynh}
S.~D. Chowdhury, K.~Ghosh, P.~Haldar, P.~Raman, and A.~Sinha, ``{Crossing
  Symmetric Spinning S-matrix Bootstrap: EFT bounds},''
  \href{https://dx.doi.org/10.21468/SciPostPhys.13.3.051}{{\em SciPost Phys.}
  {\bfseries 13} no.~3, (2022) 051},
  \href{https://arxiv.org/abs/2112.11755}{{\ttfamily arXiv:2112.11755
  [hep-th]}}.

\bibitem{Bissi:2022mrs}
A.~Bissi, A.~Sinha, and X.~Zhou, ``{Selected topics in analytic conformal
  bootstrap: A guided journey},''
  \href{https://dx.doi.org/10.1016/j.physrep.2022.09.004}{{\em Phys. Rept.}
  {\bfseries 991} (2022) 1--89},
  \href{https://arxiv.org/abs/2202.08475}{{\ttfamily arXiv:2202.08475
  [hep-th]}}.

\bibitem{Ghosh:2022net}
S.~Ghosh, P.~Raman, and A.~Sinha, ``{Celestial insights into the S-matrix
  bootstrap},'' \href{https://dx.doi.org/10.1007/JHEP08(2022)216}{{\em JHEP}
  {\bfseries 08} (2022) 216},
  \href{https://arxiv.org/abs/2204.07617}{{\ttfamily arXiv:2204.07617
  [hep-th]}}.

\bibitem{Sinha:2022sdo}
A.~Sinha, ``{Dispersion relations, knots polynomials, and the q-deformed
  harmonic oscillator},''
  \href{https://dx.doi.org/10.1103/PhysRevD.106.126019}{{\em Phys. Rev. D}
  {\bfseries 106} no.~12, (2022) 126019},
  \href{https://arxiv.org/abs/2204.13986}{{\ttfamily arXiv:2204.13986
  [hep-th]}}.

\bibitem{Bissi:2022fmj}
A.~Bissi and A.~Sinha, ``{Positivity, low twist dominance and CSDR for CFTs},''
  \href{https://dx.doi.org/10.21468/SciPostPhys.14.4.083}{{\em SciPost Phys.}
  {\bfseries 14} no.~4, (2023) 083},
  \href{https://arxiv.org/abs/2209.03978}{{\ttfamily arXiv:2209.03978
  [hep-th]}}.

\bibitem{az1}
A.~Zahed, ``{Positivity and geometric function theory constraints on pion
  scattering},'' \href{https://dx.doi.org/10.1007/JHEP12(2021)036}{{\em JHEP}
  {\bfseries 12} (2021) 036},
  \href{https://arxiv.org/abs/2108.10355}{{\ttfamily arXiv:2108.10355
  [hep-th]}}.

\bibitem{az2}
D.~Chowdhury, P.~Haldar, and A.~Zahed, ``{Locality and analyticity of the
  crossing symmetric dispersion relation},''
  \href{https://dx.doi.org/10.1007/JHEP10(2022)180}{{\em JHEP} {\bfseries 10}
  (2022) 180}, \href{https://arxiv.org/abs/2205.13762}{{\ttfamily
  arXiv:2205.13762 [hep-th]}}.

\bibitem{Bhat:2023ekh}
F.~Bhat and A.~Zahed, ``{A Celestial route to AdS bulk locality},''
  \href{https://dx.doi.org/10.1007/JHEP08(2023)112}{{\em JHEP} {\bfseries 08}
  (2023) 112}, \href{https://arxiv.org/abs/2304.02003}{{\ttfamily
  arXiv:2304.02003 [hep-th]}}.

\bibitem{Song:2023quv}
C.~Song, ``{Crossing-Symmetric Dispersion Relations without Spurious
  Singularities},''
  \href{https://dx.doi.org/10.1103/PhysRevLett.131.161602}{{\em Phys. Rev.
  Lett.} {\bfseries 131} no.~16, (2023) 161602},
  \href{https://arxiv.org/abs/2305.03669}{{\ttfamily arXiv:2305.03669
  [hep-th]}}.

\bibitem{bhat}
F.~Bhat, D.~Chowdhury, A.~Sinha, S.~Tiwari, and A.~Zahed, ``{Bootstrapping
  high-energy observables},''
  \href{https://dx.doi.org/10.1007/JHEP03(2024)157}{{\em JHEP} {\bfseries 03}
  (2024) 157}, \href{https://arxiv.org/abs/2311.03451}{{\ttfamily
  arXiv:2311.03451 [hep-th]}}.

\bibitem{ak1}
A.~Kaviraj, ``{Crossing antisymmetric Polyakov blocks + dispersion relation},''
  \href{https://dx.doi.org/10.1007/JHEP01(2022)005}{{\em JHEP} {\bfseries 01}
  (2022) 005}, \href{https://arxiv.org/abs/2109.02658}{{\ttfamily
  arXiv:2109.02658 [hep-th]}}.

\bibitem{Bhat:2025ygi}
F.~Bhat and A.~Sinha, ``{Ramanujan{\textquoteright}s 1/{\ensuremath{\pi}}
  Series and Conformal Field Theories},''
  \href{https://dx.doi.org/10.1103/c38g-fd2v}{{\em Phys. Rev. Lett.} {\bfseries
  135} no.~23, (2025) 231602},
  \href{https://arxiv.org/abs/2503.21656}{{\ttfamily arXiv:2503.21656
  [hep-th]}}.

\bibitem{Das:2026ckf}
D.~Das and A.~Sinha, ``{Bootstrapping black holes at low impact parameter},''
  \href{https://arxiv.org/abs/2607.05503}{{\ttfamily arXiv:2607.05503
  [hep-th]}}.

\bibitem{Alday:2022xwz}
L.~F. Alday, T.~Hansen, and J.~A. Silva, ``{AdS Virasoro-Shapiro from
  single-valued periods},''
  \href{https://dx.doi.org/10.1007/JHEP12(2022)010}{{\em JHEP} {\bfseries 12}
  (2022) 010}, \href{https://arxiv.org/abs/2209.06223}{{\ttfamily
  arXiv:2209.06223 [hep-th]}}.

\bibitem{deRham:2022gfe}
C.~de~Rham, S.~Jaitly, and A.~J. Tolley, ``{Constraints on Regge behavior from
  IR physics},'' \href{https://dx.doi.org/10.1103/PhysRevD.108.046011}{{\em
  Phys. Rev. D} {\bfseries 108} no.~4, (2023) 046011},
  \href{https://arxiv.org/abs/2212.04975}{{\ttfamily arXiv:2212.04975
  [hep-th]}}.

\bibitem{Li:2023qzs}
Y.-Z. Li, ``{Effective field theory bootstrap, large-N {\ensuremath{\chi}}PT
  and holographic QCD},''
  \href{https://dx.doi.org/10.1007/JHEP01(2024)072}{{\em JHEP} {\bfseries 01}
  (2024) 072}, \href{https://arxiv.org/abs/2310.09698}{{\ttfamily
  arXiv:2310.09698 [hep-th]}}.

\bibitem{Pasiecznik:2025eqc}
C.~Pasiecznik, ``{Bootstrapping Gravity with Crossing Symmetric Dispersion
  Relations},'' \href{https://arxiv.org/abs/2506.09884}{{\ttfamily
  arXiv:2506.09884 [hep-th]}}.

\bibitem{Chang:2025cxc}
C.-H. Chang and J.~Parra-Martinez, ``{Graviton loops and negativity},''
  \href{https://dx.doi.org/10.1007/JHEP08(2025)175}{{\em JHEP} {\bfseries 08}
  (2025) 175}, \href{https://arxiv.org/abs/2501.17949}{{\ttfamily
  arXiv:2501.17949 [hep-th]}}.

\bibitem{EliasMiro:2025rqo}
J.~Elias~Mir{\'o}, A.~Guerrieri, M.~A. G{\"u}m{\"u}s, and A.~Zahed, ``{A
  Geometric View on Crossing-Symmetric Dispersion Relations},''
  \href{https://arxiv.org/abs/2509.14170}{{\ttfamily arXiv:2509.14170
  [hep-th]}}.

\bibitem{bose1}
A.~Bose, P.~Haldar, A.~Sinha, P.~Sinha, and S.~S. Tiwari, ``{Relative entropy
  in scattering and the S-matrix bootstrap},''
  \href{https://dx.doi.org/10.21468/SciPostPhys.9.5.081}{{\em SciPost Phys.}
  {\bfseries 9} (2020) 081}, \href{https://arxiv.org/abs/2006.12213}{{\ttfamily
  arXiv:2006.12213 [hep-th]}}.

\bibitem{bose2}
A.~Bose, A.~Sinha, and S.~S. Tiwari, ``{Selection rules for the S-Matrix
  bootstrap},'' \href{https://dx.doi.org/10.21468/SciPostPhys.10.5.122}{{\em
  SciPost Phys.} {\bfseries 10} no.~5, (2021) 122},
  \href{https://arxiv.org/abs/2011.07944}{{\ttfamily arXiv:2011.07944
  [hep-th]}}.

\bibitem{bell}
A.~Sinha and A.~Zahed, ``{Bell inequalities in 2-2 scattering},''
  \href{https://dx.doi.org/10.1103/PhysRevD.108.025015}{{\em Phys. Rev. D}
  {\bfseries 108} no.~2, (2023) 025015},
  \href{https://arxiv.org/abs/2212.10213}{{\ttfamily arXiv:2212.10213
  [hep-th]}}.

\bibitem{PhysRevLett.125.181602}
R.~Aoude, M.-Z. Chung, Y.-t. Huang, C.~S. Machado, and M.-K. Tam, ``Silence of
  binary kerr black holes,''
  \href{https://dx.doi.org/10.1103/PhysRevLett.125.181602}{{\em Phys. Rev.
  Lett.} {\bfseries 125} (Oct, 2020) 181602}.
  \url{https://link.aps.org/doi/10.1103/PhysRevLett.125.181602}.

\bibitem{Beane:2021zvo}
S.~R. Beane, R.~C. Farrell, and M.~Varma, ``{Entanglement minimization in
  hadronic scattering with pions},''
  \href{https://dx.doi.org/10.1142/S0217751X21502055}{{\em Int. J. Mod. Phys.
  A} {\bfseries 36} no.~30, (2021) 2150205},
  \href{https://arxiv.org/abs/2108.00646}{{\ttfamily arXiv:2108.00646
  [hep-ph]}}.

\bibitem{Aoude:2024xpx}
R.~Aoude, G.~Elor, G.~N. Remmen, and O.~Sumensari, ``{Positivity in Amplitudes
  and Quantum Entanglement},''
  \href{https://dx.doi.org/10.1002/prop.70113}{{\em Fortsch. Phys.} {\bfseries
  74} (2026) e70113}, \href{https://arxiv.org/abs/2402.16956}{{\ttfamily
  arXiv:2402.16956 [hep-th]}}.

\bibitem{Kowalska:2024kbs}
K.~Kowalska and E.~M. Sessolo, ``{Entanglement in flavored scalar
  scattering},'' \href{https://dx.doi.org/10.1007/JHEP07(2024)156}{{\em JHEP}
  {\bfseries 07} (2024) 156},
  \href{https://arxiv.org/abs/2404.13743}{{\ttfamily arXiv:2404.13743
  [hep-ph]}}.

\bibitem{Low:2024mrk}
I.~Low and Z.~Yin, ``{An Area Law for Entanglement Entropy in Particle
  Scattering},'' \href{https://arxiv.org/abs/2405.08056}{{\ttfamily
  arXiv:2405.08056 [hep-th]}}.

\bibitem{Low:2024hvn}
I.~Low and Z.~Yin, ``{Elastic cross section is entanglement entropy},''
  \href{https://dx.doi.org/10.1103/PhysRevD.111.065027}{{\em Phys. Rev. D}
  {\bfseries 111} no.~6, (2025) 065027},
  \href{https://arxiv.org/abs/2410.22414}{{\ttfamily arXiv:2410.22414
  [hep-th]}}.

\bibitem{McGinnis:2025iab}
N.~McGinnis, ``{Quantum Computational Structure of $SU(N)$ Scattering},''
  \href{https://arxiv.org/abs/2511.10550}{{\ttfamily arXiv:2511.10550
  [quant-ph]}}.

\bibitem{McGinnis:2025xgt}
N.~McGinnis, ``{Crossing Symmetry and Entanglement},''
  \href{https://arxiv.org/abs/2511.10559}{{\ttfamily arXiv:2511.10559
  [hep-th]}}.

\bibitem{Carena:2025wyh}
M.~Carena, G.~Coloretti, W.~Liu, M.~Littmann, I.~Low, and C.~E.~M. Wagner,
  ``{Entanglement maximization and mirror symmetry in two-Higgs-doublet
  models},'' \href{https://dx.doi.org/10.1007/JHEP08(2025)016}{{\em JHEP}
  {\bfseries 08} (2025) 016},
  \href{https://arxiv.org/abs/2505.00873}{{\ttfamily arXiv:2505.00873
  [hep-ph]}}.

\bibitem{Hu:2025lua}
T.-R. Hu, K.~Sone, F.-K. Guo, T.~Hyodo, and I.~Low, ``{Entanglement
  suppression, quantum statistics, and symmetries in spin-3/2 baryon
  scatterings},'' \href{https://dx.doi.org/10.1103/x52w-4rbs}{{\em Phys. Rev.
  Res.} {\bfseries 7} no.~4, (2025) 043306},
  \href{https://arxiv.org/abs/2506.08960}{{\ttfamily arXiv:2506.08960
  [hep-ph]}}.

\bibitem{DHoker:2019blr}
E.~D'Hoker and M.~B. Green, ``{Exploring transcendentality in superstring
  amplitudes},'' \href{https://dx.doi.org/10.1007/JHEP07(2019)149}{{\em JHEP}
  {\bfseries 07} (2019) 149},
  \href{https://arxiv.org/abs/1906.01652}{{\ttfamily arXiv:1906.01652
  [hep-th]}}.

\bibitem{Bern:2019prr}
Z.~Bern, J.~J. Carrasco, M.~Chiodaroli, H.~Johansson, and R.~Roiban, ``{The
  duality between color and kinematics and its applications},''
  \href{https://dx.doi.org/10.1088/1751-8121/ad5fd0}{{\em J. Phys. A}
  {\bfseries 57} no.~33, (2024) 333002},
  \href{https://arxiv.org/abs/1909.01358}{{\ttfamily arXiv:1909.01358
  [hep-th]}}.

\bibitem{Peeters:2000qj}
K.~Peeters, P.~Vanhove, and A.~Westerberg, ``{Supersymmetric higher derivative
  actions in ten-dimensions and eleven-dimensions, the associated superalgebras
  and their formulation in superspace},''
  \href{https://dx.doi.org/10.1088/0264-9381/18/5/307}{{\em Class. Quant.
  Grav.} {\bfseries 18} (2001) 843--890},
  \href{https://arxiv.org/abs/hep-th/0010167}{{\ttfamily
  arXiv:hep-th/0010167}}.

\bibitem{CYML1}
L.~B. Anderson, M.~Gerdes, J.~Gray, S.~Krippendorf, N.~Raghuram, and F.~Ruehle,
  ``{Moduli-dependent Calabi-Yau and SU(3)-structure metrics from Machine
  Learning},'' \href{https://dx.doi.org/10.1007/JHEP05(2021)013}{{\em JHEP}
  {\bfseries 05} (2021) 013},
  \href{https://arxiv.org/abs/2012.04656}{{\ttfamily arXiv:2012.04656
  [hep-th]}}.

\bibitem{CYML2}
M.~R. Douglas, S.~Lakshminarasimhan, and Y.~Qi, ``{Numerical Calabi-Yau metrics
  from holomorphic networks},''
  \href{https://arxiv.org/abs/2012.04797}{{\ttfamily arXiv:2012.04797
  [hep-th]}}.

\bibitem{DeLuca:2024njg}
G.~B. De~Luca, ``{Machine Learning gravity compactifications on negatively
  curved manifolds},'' \href{https://dx.doi.org/10.1007/JHEP09(2025)090}{{\em
  JHEP} {\bfseries 09} (2025) 090},
  \href{https://arxiv.org/abs/2501.00093}{{\ttfamily arXiv:2501.00093
  [hep-th]}}.

\bibitem{CBootML1}
G.~K\'antor, V.~Niarchos, and C.~Papageorgakis, ``{Solving Conformal Field
  Theories with Artificial Intelligence},''
  \href{https://dx.doi.org/10.1103/PhysRevLett.128.041601}{{\em Phys. Rev.
  Lett.} {\bfseries 128} no.~4, (2022) 041601},
  \href{https://arxiv.org/abs/2108.08859}{{\ttfamily arXiv:2108.08859
  [hep-th]}}.

\bibitem{CBootML2}
G.~K\'antor, V.~Niarchos, and C.~Papageorgakis, ``{Conformal bootstrap with
  reinforcement learning},''
  \href{https://dx.doi.org/10.1103/PhysRevD.105.025018}{{\em Phys. Rev. D}
  {\bfseries 105} no.~2, (2022) 025018},
  \href{https://arxiv.org/abs/2108.09330}{{\ttfamily arXiv:2108.09330
  [hep-th]}}.

\bibitem{CBootML3}
G.~K\'antor, V.~Niarchos, C.~Papageorgakis, and P.~Richmond, ``{6D (2,0)
  bootstrap with the soft-actor-critic algorithm},''
  \href{https://dx.doi.org/10.1103/PhysRevD.107.025005}{{\em Phys. Rev. D}
  {\bfseries 107} no.~2, (2023) 025005},
  \href{https://arxiv.org/abs/2209.02801}{{\ttfamily arXiv:2209.02801
  [hep-th]}}.

\bibitem{CBootML4}
V.~Niarchos, C.~Papageorgakis, P.~Richmond, A.~G. Stapleton, and M.~Woolley,
  ``{Bootstrability in line-defect CFTs with improved truncation methods},''
  \href{https://dx.doi.org/10.1103/PhysRevD.108.105027}{{\em Phys. Rev. D}
  {\bfseries 108} no.~10, (2023) 105027},
  \href{https://arxiv.org/abs/2306.15730}{{\ttfamily arXiv:2306.15730
  [hep-th]}}.

\bibitem{CBootML5}
T.~Cai, G.~W. Merz, F.~Charton, N.~Nolte, M.~Wilhelm, K.~Cranmer, and L.~J.
  Dixon, ``{Transforming the bootstrap: using transformers to compute
  scattering amplitudes in planar $\mathcal{N} = 4$ super
  Yang\textendash{}Mills theory},''
  \href{https://dx.doi.org/10.1088/2632-2153/ad743e}{{\em Mach. Learn. Sci.
  Tech.} {\bfseries 5} no.~3, (2024) 035073},
  \href{https://arxiv.org/abs/2405.06107}{{\ttfamily arXiv:2405.06107
  [cs.LG]}}.

\bibitem{Ghosh:2026xnp}
K.~Ghosh, S.~Kumar, V.~Niarchos, and A.~Stergiou, ``{Neural Spectral Bias and
  Conformal Correlators I: Introduction and Applications},''
  \href{https://arxiv.org/abs/2604.18686}{{\ttfamily arXiv:2604.18686
  [hep-th]}}.

\bibitem{Ghosh:2026jbw}
K.~Ghosh, S.~Kumar, A.~Stergiou, and V.~Niarchos, ``{Neural Networks Reveal a
  Universal Bias in Conformal Correlators},''
  \href{https://arxiv.org/abs/2604.18673}{{\ttfamily arXiv:2604.18673
  [hep-th]}}.

\bibitem{Ghosh:2026efw}
K.~Ghosh, S.~Kumar, V.~Niarchos, and A.~Stergiou, ``{Neural Spectral Bias and
  Conformal Correlators II: Modular and Annulus Bootstrap},''
  \href{https://arxiv.org/abs/2607.24913}{{\ttfamily arXiv:2607.24913
  [hep-th]}}.

\bibitem{QMML}
S.~Mizera, ``{Scattering with neural operators},''
  \href{https://dx.doi.org/10.1103/PhysRevD.108.L101701}{{\em Phys. Rev. D}
  {\bfseries 108} no.~10, (2023) L101701},
  \href{https://arxiv.org/abs/2308.14789}{{\ttfamily arXiv:2308.14789
  [hep-th]}}.

\bibitem{SbootML1}
A.~Dersy, M.~D. Schwartz, and A.~Zhiboedov, ``{Reconstructing S-matrix Phases
  with Machine Learning},''
  \href{https://dx.doi.org/10.1007/JHEP05(2024)200}{{\em JHEP} {\bfseries 05}
  (2024) 200}, \href{https://arxiv.org/abs/2308.09451}{{\ttfamily
  arXiv:2308.09451 [hep-th]}}.

\bibitem{SbootML2}
V.~Niarchos and C.~Papageorgakis, ``{Learning S-matrix phases with neural
  operators},'' \href{https://dx.doi.org/10.1103/PhysRevD.110.045020}{{\em
  Phys. Rev. D} {\bfseries 110} no.~4, (2024) 045020},
  \href{https://arxiv.org/abs/2404.14551}{{\ttfamily arXiv:2404.14551
  [hep-th]}}.

\bibitem{Gumus:2024lmj}
M.~A. Gumus, D.~Leflot, P.~Tourkine, and A.~Zhiboedov, ``{The S-matrix
  bootstrap with neural optimizers. Part I. Zero double discontinuity},''
  \href{https://dx.doi.org/10.1007/JHEP07(2025)210}{{\em JHEP} {\bfseries 07}
  (2025) 210}, \href{https://arxiv.org/abs/2412.09610}{{\ttfamily
  arXiv:2412.09610 [hep-th]}}.

\bibitem{Gumus:2026mhb}
M.~A. Gumus, D.~Leflot, P.~Tourkine, and A.~Zhiboedov, ``{Neural S-matrix
  bootstrap II: solvable 4d amplitudes with particle production},''
  \href{https://arxiv.org/abs/2601.22145}{{\ttfamily arXiv:2601.22145
  [hep-th]}}.

\bibitem{Arkani-Hamed:2022gsa}
N.~Arkani-Hamed, L.~Eberhardt, Y.-t. Huang, and S.~Mizera, ``{On unitarity of
  tree-level string amplitudes},''
  \href{https://dx.doi.org/10.1007/JHEP02(2022)197}{{\em JHEP} {\bfseries 02}
  (2022) 197}, \href{https://arxiv.org/abs/2201.11575}{{\ttfamily
  arXiv:2201.11575 [hep-th]}}.

\bibitem{HORNIK1989359}
K.~Hornik, M.~Stinchcombe, and H.~White, ``Multilayer feedforward networks are
  universal approximators,'' {\em Neural Networks} {\bfseries 2} no.~5, (1989)
  359--366.

\bibitem{RAISSI2019686}
M.~Raissi, P.~Perdikaris, and G.~Karniadakis, ``Physics-informed neural
  networks: A deep learning framework for solving forward and inverse problems
  involving nonlinear partial differential equations,'' {\em Journal of
  Computational Physics} {\bfseries 378} (2019) 686--707.

\bibitem{Paszke:2019xhz}
A.~Paszke {\em et~al.}, ``{PyTorch: An Imperative Style, High-Performance Deep
  Learning Library},'' \href{https://arxiv.org/abs/1912.01703}{{\ttfamily
  arXiv:1912.01703 [cs.LG]}}.

\bibitem{Kingma:2014vow}
D.~P. Kingma and J.~Ba, ``{Adam: A Method for Stochastic Optimization},'' in
  {\em {International Conference on Learning Representations}}.
\newblock 12, 2014.
\newblock \href{https://arxiv.org/abs/1412.6980}{{\ttfamily arXiv:1412.6980
  [cs.LG]}}.

\bibitem{Liu:1989esw}
D.~C. Liu and J.~Nocedal, ``{On the limited memory BFGS method for large scale
  optimization},'' \href{https://dx.doi.org/10.1007/BF01589116}{{\em Math.
  Programming} {\bfseries 45} no.~1, (1989) 503--528}.

\bibitem{sdpb}
D.~Simmons-Duffin, ``{A Semidefinite Program Solver for the Conformal
  Bootstrap},'' \href{https://dx.doi.org/10.1007/JHEP06(2015)174}{{\em JHEP}
  {\bfseries 06} (2015) 174},
  \href{https://arxiv.org/abs/1502.02033}{{\ttfamily arXiv:1502.02033
  [hep-th]}}.

\bibitem{Mansfield:2024wjc}
G.~Mansfield and M.~Spradlin, ``{On Unitarity of the Hypergeometric
  Amplitude},'' \href{https://arxiv.org/abs/2409.09561}{{\ttfamily
  arXiv:2409.09561 [hep-th]}}.

\bibitem{DHS}
R.~Dolen, D.~Horn, and C.~Schmid, ``{Finite energy sum rules and their
  application to pi N charge exchange},''
  \href{https://dx.doi.org/10.1103/PhysRev.166.1768}{{\em Phys. Rev.}
  {\bfseries 166} (1968) 1768--1781}.

\bibitem{Green:1987sp}
M.~B. Green, J.~H. Schwarz, and E.~Witten, {\em {SUPERSTRING THEORY. VOL. 1:
  INTRODUCTION}}.
\newblock Cambridge Monographs on Mathematical Physics. 7, 1988.

\bibitem{Tolley:2020gtv}
A.~J. Tolley, Z.-Y. Wang, and S.-Y. Zhou, ``{New positivity bounds from full
  crossing symmetry},'' \href{https://dx.doi.org/10.1007/JHEP05(2021)255}{{\em
  JHEP} {\bfseries 05} (2021) 255},
  \href{https://arxiv.org/abs/2011.02400}{{\ttfamily arXiv:2011.02400
  [hep-th]}}.

\bibitem{Caron-Huot:2020cmc}
S.~Caron-Huot and V.~Van~Duong, ``{Extremal Effective Field Theories},''
  \href{https://dx.doi.org/10.1007/JHEP05(2021)280}{{\em JHEP} {\bfseries 05}
  (2021) 280}, \href{https://arxiv.org/abs/2011.02957}{{\ttfamily
  arXiv:2011.02957 [hep-th]}}.

\bibitem{Hamada:2018dde}
Y.~Hamada, T.~Noumi, and G.~Shiu, ``{Weak Gravity Conjecture from Unitarity and
  Causality},'' \href{https://dx.doi.org/10.1103/PhysRevLett.123.051601}{{\em
  Phys. Rev. Lett.} {\bfseries 123} no.~5, (2019) 051601},
  \href{https://arxiv.org/abs/1810.03637}{{\ttfamily arXiv:1810.03637
  [hep-th]}}.

\bibitem{Alberte:2020jsk}
L.~Alberte, C.~de~Rham, S.~Jaitly, and A.~J. Tolley, ``{Positivity Bounds and
  the Massless Spin-2 Pole},''
  \href{https://dx.doi.org/10.1103/PhysRevD.102.125023}{{\em Phys. Rev. D}
  {\bfseries 102} no.~12, (2020) 125023},
  \href{https://arxiv.org/abs/2007.12667}{{\ttfamily arXiv:2007.12667
  [hep-th]}}.

\bibitem{Tokuda:2020mlf}
J.~Tokuda, K.~Aoki, and S.~Hirano, ``{Gravitational positivity bounds},''
  \href{https://dx.doi.org/10.1007/JHEP11(2020)054}{{\em JHEP} {\bfseries 11}
  (2020) 054}, \href{https://arxiv.org/abs/2007.15009}{{\ttfamily
  arXiv:2007.15009 [hep-th]}}.

\bibitem{Caron-Huot:2021rmr}
S.~Caron-Huot, D.~Mazac, L.~Rastelli, and D.~Simmons-Duffin, ``{Sharp
  boundaries for the swampland},''
  \href{https://dx.doi.org/10.1007/JHEP07(2021)110}{{\em JHEP} {\bfseries 07}
  (2021) 110}, \href{https://arxiv.org/abs/2102.08951}{{\ttfamily
  arXiv:2102.08951 [hep-th]}}.

\bibitem{polchinskistrassler}
J.~Polchinski and M.~J. Strassler, ``{Hard scattering and gauge / string
  duality},'' \href{https://dx.doi.org/10.1103/PhysRevLett.88.031601}{{\em
  Phys. Rev. Lett.} {\bfseries 88} (2002) 031601},
  \href{https://arxiv.org/abs/hep-th/0109174}{{\ttfamily
  arXiv:hep-th/0109174}}.

\bibitem{Caron-Huot:2016icg}
S.~Caron-Huot, Z.~Komargodski, A.~Sever, and A.~Zhiboedov, ``{Strings from
  Massive Higher Spins: The Asymptotic Uniqueness of the Veneziano
  Amplitude},'' \href{https://dx.doi.org/10.1007/JHEP10(2017)026}{{\em JHEP}
  {\bfseries 10} (2017) 026},
  \href{https://arxiv.org/abs/1607.04253}{{\ttfamily arXiv:1607.04253
  [hep-th]}}.

\bibitem{Paulos:2017fhb}
M.~F. Paulos, J.~Penedones, J.~Toledo, B.~C. van Rees, and P.~Vieira, ``{The
  S-matrix bootstrap. Part III: higher dimensional amplitudes},''
  \href{https://dx.doi.org/10.1007/JHEP12(2019)040}{{\em JHEP} {\bfseries 12}
  (2019) 040}, \href{https://arxiv.org/abs/1708.06765}{{\ttfamily
  arXiv:1708.06765 [hep-th]}}.

\end{thebibliography}\endgroup

\end{document}